\documentclass[letterpaper,11pt]{article}
\pdfoutput=1

\usepackage{hyperref}
\hypersetup{
    colorlinks=true,       
    linkcolor=blue,          
    citecolor=blue,        
    filecolor=blue,      
    urlcolor=blue           
}

\usepackage{float}
                     
\usepackage{braket,slashed,bm}
\usepackage{array,multirow}
\usepackage[normalem]{ulem}
\usepackage{xcolor,cancel,youngtab}
\usepackage{verbatim}

\usepackage[T1]{fontenc} 
\usepackage[font=small,labelfont=bf,tableposition=top]{caption}

\DeclareCaptionLabelFormat{andtable}{#1~#2  \&  \tablename~\thetable}

\usepackage{mathrsfs}
\usepackage{booktabs}
\usepackage{adjustbox}
\usepackage{mathtools}

\usepackage{soul}

\usepackage{tikz}
\usetikzlibrary{arrows,decorations.pathmorphing,backgrounds,positioning,fit,petri,automata,shadows,calendar,mindmap,
decorations.markings,calc}

\definecolor{labelkey}{rgb}{0,0.5,0.0}

\usepackage{xspace}
\usepackage{slashed}
\usepackage{array,multirow}
\usepackage{graphicx}
\usepackage{graphics}
\usepackage{epsfig}
\usepackage{amsfonts}
\usepackage{amsmath}
\usepackage{amssymb}
\usepackage{dsfont}
\usepackage{color}
\usepackage{xcolor}
\usepackage{cite}
\usepackage{pdflscape}
\usepackage{pifont}
\usepackage{pgfplots}
\usepackage{tikz} 
\usepackage{pgfplotstable}
\usepackage{bbold}
\usepackage{subcaption}

\newcommand{\al}{\alpha}
\newcommand{\bt}{\beta}

\newcommand{\dt}{\delta}

\usepackage{bm}  %

\newcommand{\beq}{\begin{equation}}
\newcommand{\eeq}{\end{equation}}
\newcommand{\be}{\begin{equation}}
\newcommand{\ee}{\end{equation}}
\newcommand{\bea}{\begin{eqnarray}}
\newcommand{\eea}{\end{eqnarray}}
\newcommand{\ben}{\begin{eqnarray*}}
\newcommand{\een}{\end{eqnarray*}}

\newcommand{\boldsigma}{\mbox{\boldmath $\sigma$}}

\newcommand{\bma}{\begin{pmatrix}}
\newcommand{\ema}{\end{pmatrix}}

\def\lixo#1{}

\def\slashchar#1{\setbox0=\hbox{$#1$}           
  \dimen0=\wd0                                    
  \setbox1=\hbox{/} \dimen1=\wd1                  
  \ifdim\dimen0>\dimen1                           
    \rlap{\hbox to \dimen0{\hfil/\hfil}}            
    #1                                             
  \else                                          
    \rlap{\hbox to \dimen1{\hfil$#1$\hfil}}        
    /                                           
 \fi}                                           %

\newcommand{\Or}{\mathcal O}

\newcommand{\dslash}[1]{#1 \llap{/\kern-0.5pt}}
\newcommand{\Dslash}[1]{#1 \llap{/\kern+1.5pt}}
\newcommand{\DDslash}[1]{#1 \llap{/\kern+2.3pt}}
\newcommand{\dslashh}[1]{#1 \llap{/\kern+1pt}}

\newcommand{\nn}{\nonumber}

\newcommand{\NLDBD}{$0 \nu \beta \beta$}
\newcommand{\textoverline}[1]{$\overline{\mbox{#1}}$}

\definecolor{cadmiumgreen}{rgb}{0.0, 0.42, 0.24}
\definecolor{darkpastelgreen}{rgb}{0.01, 0.75, 0.24}
\definecolor{darkspringgreen}{rgb}{0.09, 0.45, 0.27}
\definecolor{forestgreen(web)}{rgb}{0.13, 0.55, 0.13}
\definecolor{forestgreen(traditional)}{rgb}{0.0, 0.27, 0.13}
\definecolor{cobalt}{rgb}{0.0, 0.28, 0.67}
\definecolor{darkblue}{rgb}{0.0, 0.0, 0.75}
\definecolor{darkred}{rgb}{0.55, 0.0, 0.0}
\definecolor{palatinatepurple}{rgb}{0.41, 0.16, 0.38}
\definecolor{burntorange}{rgb}{0.8, 0.33, 0.0}

\begin{document}

\begin{titlepage}

\begin{flushright}
 LA-UR-24-21117\\
INT-PUB-24-007
\\
\end{flushright}

\vspace{1.8cm}

\begin{center}
{\LARGE  \bf 
Neutrinoless double beta decay rates in \\the presence of light sterile neutrinos}
\vspace{2cm}

{\large \bf  W. Dekens$^{a}$,  J. de Vries$^{b,c}$, D. Castillo$^{d}$, J. Men\'endez$^{d,e}$,\\E. Mereghetti$^f$, V. Plakkot$^{b,c}$,   P.~Soriano$^{d,e}$, G. Zhou$^{g,h}$} 
\vspace{0.5cm}

\vspace{0.25cm}

\vspace{0.25cm}
{\large 
$^a$ 
{\it 
Institute for Nuclear Theory, University of Washington, Seattle WA 98195-1550, USA}}

\vspace{0.25cm}
{\large 
$^b$ 
{\it 
Institute of Physics Amsterdam and Delta Institute for Theoretical Physics, University of Amsterdam, Science Park 904, 1098 XH Amsterdam, The Netherlands}}

\vspace{0.25cm}
{\large 
$^c$ 
{\it 
Nikhef, Theory Group, Science Park 105, 1098 XG, Amsterdam, The Netherlands}}

\vspace{0.25cm}
{\large 
$^d$ 
{\it Departament de F\'isica Qu\`antica i Astrof\'isica, Universitat de Barcelona, Mart\'i i Franqu\`es 1, 08028, Barcelona, Spain}}

\vspace{0.25cm}
{\large 
$^e$ 
{\it Institut de Ci\`encies del Cosmos, Universitat de Barcelona, Mart\'i i Franqu\`es 1, 08028, Barcelona, Spain}}

\vspace{0.25cm}
{\large 
$^f$ 
{\it Theoretical Division, Los Alamos National Laboratory,
Los Alamos, NM 87545, USA}}

\vspace{0.25cm}
{\large 
	$^g$ 
	{\it Shenzhen Audaque Data Technology Co., Ltd., Shenzhen 518057, China}}

\vspace{0.25cm}
{\large 
	$^h$ 
	{\it School of Computer Science and Technology, Dalian University of Technology, Dalian 116024, China}}

\end{center}

\vspace{0.2cm}

\begin{abstract}
\vspace{0.1cm}

We investigate neutrinoless double-beta decay ($0\nu\beta\beta$) in minimal extensions of the Standard Model of particle physics where gauge-singlet right-handed neutrinos give rise to Dirac and Majorana neutrino mass terms. We 
argue that the standard treatment of these scenarios, based on mass-dependent nuclear matrix elements, is missing important contributions to the $0\nu\beta\beta$ amplitude. First, new effects arise from the exchange of neutrinos with very small (ultrasoft) momenta, for which we compute the associated nuclear matrix elements for the decays of ${}^{76}$Ge and ${}^{136}$Xe. These contributions can dominate the $0\nu\beta\beta$ rate in cases with light sterile neutrinos. The ultrasoft terms are also relevant in the more standard scenario of just three light Majorana neutrinos where they lead to a $10\%$ reduction of the total $0\nu\beta\beta$ amplitude. Secondly, we highlight the importance of short-range terms associated with medium-heavy sterile neutrinos and provide explicit formulae that can be used in phenomenological analyses. As examples we discuss impact of these new effects in several explicit scenarios, including a realistic $3+2$ model with two right-handed gauge-singlet neutrinos. 

\end{abstract}

\vfill
\end{titlepage}

\tableofcontents

\section{Introduction}
Understanding the origin and nature of neutrino masses is one of the most important problems in particle physics, 
which could have ramifications for other pressing open questions, ranging from the generation of the baryon matter-antimatter asymmetry in the Universe \cite{Fukugita:1986hr,Asaka:2005pn} to the nature of dark matter   \cite{Dodelson:1993je,Asaka:2005pn}.
The Standard Model of particle physics (SM) in its original form \cite{Glashow:1959wxa,Salam:1959zz,Weinberg:1967tq}
contains a left-handed neutrino field ($\nu_L$), as part of the lepton $SU(2)$ doublet. As such, the SM cannot generate a Dirac mass term for the neutrino, lacking a right handed neutrino field ($\nu_R$), while a renormalizable Majorana mass term is forbidden by the $SU(2)$ charge of the $\nu_L$ field. 
The prediction of massless neutrinos is however convincingly ruled out by neutrino oscillation experiments \cite{Superkamiokande1998,SNO:2001kpb,KamLAND:2002uet,Workman:2022ynf}. 

Without extending the SM field content, a $\nu_L$ Majorana mass term can be included in the SM as a non-renormalizable dimension-5 operator \cite{Weinberg:1979sa}.
This operator is suppressed by one power of a high-energy scale $\Lambda \gg v$, where $v \sim 246$ GeV denotes the electroweak scale, thus pointing to a high-energy origin of neutrino masses. 
Alternatively, a minimal extension of the SM, sometimes called the $\nu$SM \cite{Asaka:2005pn}, involves the addition of two or more $\nu_R$ fields, which are singlets under the SM gauge group and have only renormalizable interactions.  For $m_{\nu_R}\gg {\rm eV}$, the $\nu_R$ fall in the category of heavy neutral leptons (HNLs) \cite{Abdullahi:2022jlv}, but we will use the term sterile neutrino in this work.
At the renormalizable level, apart from a kinetic term, the sterile neutrinos have a Majorana mass term
which for $\nu_R$ is not forbidden by any symmetry,  and a Dirac term connecting sterile neutrinos to the SM left-handed lepton doublet  and the Higgs field. 

This model has several intriguing features: 1) neutrinos generally become Majorana particles, leading to the violation of lepton number, 2) it is possible to account for the baryon asymmetry of the universe (BAU) \cite{Asaka:2005pn,Shaposhnikov:2006nn,Shaposhnikov:2008pf,Canetti:2012kh,Canetti:2012vf,Drewes:2016jae,Drewes:2017zyw,Drewes:2021nqr}, 3) a very light sterile neutrino can be a dark matter candidate \cite{Asaka:2005pn,Shaposhnikov:2008pf,Boyarsky:2018tvu}. Unfortunately, only the connection to dark matter specifies a mass range, while neutrino masses and the BAU can be accounted for with sterile neutrinos in essentially any mass range. Although cosmological and big-bang-nucleosynthesis considerations typically require sterile neutrinos to have masses heavier than 10-100 MeV, these limits depend on the thermal history of the universe. We will therefore consider a broader range
of masses, which we generically denote by $M$ (although clearly not all sterile neutrinos have to fall in the same range).

Depending on their masses, sterile neutrinos can be probed by different experiments  (see Ref.\ \cite{Bolton:2022tds} for a comprehensive review).
For $M$ between  $\sim 10$ GeV and the electroweak scale, sterile neutrinos are constrained by direct collider searches at the LHC, LEP and, indirectly, by electroweak or low-energy precision observables. For $M$ between the $B$ meson mass and the kaon and pion masses, $\nu_R$ can be produced in meson decays and receive strong constraints from experiments such as Belle, NA62 and PIENU \cite{PIENU:2011aa,Belle:2013ytx,NA62:2020mcv,Bryman:2019bjg}.
Searches for kinks in the spectra of $\beta$ decays of various nuclei probe the region from the $M \sim$ MeV, via isotopes such as $^{20}$F with relatively high $Q$-value, all the way down to $M \sim$ eV,
thanks to isotopes with keV-scale $Q$-value such as  $^3$H \cite{KATRIN:2022ith}. At this scale, neutrino oscillation experiments provide additional constraints. 
However, for all mass ranges the search for lepton number violation (LNV) through neutrinoless double beta decay (0$\nu\beta\beta$) plays an important role. Limits on 0$\nu\beta\beta$ half lives provide the most sensitive probe of LNV. Current limits exceed $10^{26}$ years \cite{KamLAND-Zen:2022tow} and can be improved by one or two orders of magnitude in future experiments \cite{Abgrall:2017syy,Albert:2017hjq,Agostini:2022zub,Adams:2022jwx}. 
For $M\gg \mathcal O (\mathrm{GeV})$, $0\nu\beta\beta$ is dominated by the exchange of light active neutrinos. For lighter sterile neutrinos there appear additional contributions that can both speed up or slow down the decay rate.

While the effect of sterile neutrinos on $0\nu\beta\beta$ rates have been studied before \cite{Blennow:2010th,Mitra:2011qr,Li:2011ss,deGouvea:2011zz,Faessler:2014kka,Barea:2015zfa,Giunti:2015kza,Asaka:2005pn,Asaka:2011pb,Asaka:2013jfa,Asaka:2016zib}, these works only consider a subset of the leading contributions. In fact, they effectively replace the usual massless neutrino propagator, $1/\mathbf k^{\,2}$, valid for three active neutrinos, by a massive neutrino propagator, $1/(\mathbf k^{\,2} + M^2)$, to describe the $\nu_R$ contributions. The resulting LNV potential is then inserted into nuclear many-body computations. While these contributions are relevant, 
because of the interplay between $M$, the mass scale of quantum chromodynamics (QCD), $\Lambda_\chi$,  and the typical scales of nuclear physics, 
they only capture one part of the full $M$ dependence of the $0\nu\beta\beta$ amplitude. The full $M$-dependence involves several new effects which, in turn, require new nonperturbative input from lattice QCD (LQCD) or from models of the strong interaction, and the calculation of new sets of nuclear matrix elements. 
In what follows, we dissect these contributions using effective field theory (EFT) techniques and determine their scaling with $M$. This allows us to understand the $M$ dependence of numerical NME computations and describe $0\nu\bt\bt$ rates for a wide range of neutrino masses, from $M\ll m_\pi^2/\Lambda_\chi$ to $M\gg \Lambda_\chi$. 

Considering the great appeal of the $\nu$SM, due to its minimality and the potential resolution of major SM problems, it is important to obtain state-of-the-art predictions for key observables such as $0\nu\beta\beta$ rates. The main purpose of this work is a complete calculation of the $0\nu\beta\beta$ amplitude as function of (sterile) neutrino masses and mixing angles. A shorter version of this work was recently published in Ref.~\cite{Dekens:2023iyc} which only presented the main result in the form of a parametrization of the $0\nu\beta\beta$ amplitude as a function of the sterile neutrino mass. In this work, we present the theoretical foundations of this result. In particular, we discuss in depth the EFT power counting in the various regimes of sterile neutrino masses and show how the respective neutrino modes (hard, potential, ultrasoft, soft, and perturbative) contribute at which order in the power counting. In addition, we perform nuclear shell model calculations of the required NMEs for ${}^{136}$Xe and ${}^{76}$Ge to incorporate the ultrasoft modes and, for the first time, compute their effects on $0\nu\beta\beta$ decay rate for the standard mechanism through the exchange of three light Majorana neutrinos. The ultrasoft modes become much more important for models with light sterile neutrinos and we discuss the phenomenological for various models, including the minimal $3+2$ model. This paper is organized as follows.
After discussing the general setup in Sec.~\ref{setup} we derive the $0\nu\beta\beta$ amplitude as a function of $M$ in Sec.~\ref{Sec:contr}. Our expressions involve several (new) hadronic and nuclear matrix elements and we discuss their sizes and uncertainties in Sec.~\ref{sec:LECsNMEs}. We then apply our results to several models of phenomenological interest in Sec.~\ref{sec:pheno} and conclude in Sec.~\ref{sec:conclusion}.

\section{The $\nu$SM}\label{setup}

We consider the SM Lagrangian supplemented by $n$ gauge-singlet neutrino fields 
\begin{eqnarray}\label{eq:smeft}
\mathcal L &=&  \mathcal L_{SM} - \left[ \frac{1}{2} \bar \nu^c_{R} \, M_R \nu_{R} +\bar L \tilde H Y_\nu \nu_R + \rm{h.c.}\right]\,.
\end{eqnarray}
in terms of the lepton doublet $L=(\nu_L,\, e_L)^T$, while $\tilde H = i \tau_2 H^*$ with $H$ the Higgs doublet. In the unitary gauge
\begin{equation}
H = \frac{v}{\sqrt{2}} \left(\begin{array}{c}
0 \\
1 + \frac{h(x)}{v}
\end{array} \right)\,,
\end{equation}
where $v=246$ GeV is the Higgs vacuum expectation value (vev) and  $h(x)$ is the Higgs field. $\nu_{R}$ is a column vector of $n$ right-handed sterile neutrinos. $Y_\nu$ is a $3\times n$ matrix of Yukawa couplings and $ M_R$ is a symmetric $n \times n$ matrix.
We define charged-conjugated fields as  $\Psi^c = C \bar \Psi^T$, in terms of the charge conjugation matrix $C = - C^{-1} = -C^T = - C^\dagger$ and $\Psi_{L,R}^c = (\Psi_{L,R})^c =  C \overline{\Psi_{L,R}}^T= P_{R,L} \Psi^c$, with $P_{R,L}=(1\pm\gamma_5)/2$. Without loss of generality we will work in the basis where the charged leptons $e^i_{L,R}$ and quarks $u^i_{L,R}$ and $d^i_R$ are mass eigenstates ($i=1,2,3$). The relation between the mass and weak eigenstates for the neutrinos will be discussed below. 

After electroweak symmetry breaking the mass terms can be written as
  \bea\label{eq:mass}
 \mathcal L_m = -\frac{1}{2} \bar N^c M_\nu N +{\rm h.c.}\,,\qquad M_\nu = \bma 0 &M_D^*\\M_D^\dagger&M_R^\dagger \ema \,,
 \eea
where $N = (\nu_L,\, \nu_R^c)^T$, $M_D = \frac{v}{\sqrt{2}}Y_\nu^\dagger$.  $M_\nu$ is a symmetric matrix that can be diagonalized by a unitary matrix  $U$ 
\bea\label{Mdiag}
U^T M_\nu U =m_\nu = {\rm diag}(m_1,\dots , m_{3+n})\,, \qquad N = U N_m\,,
\eea
where $U$ is the neutrino mixing matrix and $m_i$ are real and positive. The kinetic and mass terms of the neutrinos can be written as
\bea
\mathcal L_\nu = \frac{1}{2} \bar \nu i\slashed \partial \nu -\frac{1}{2} \bar \nu^{ } m_\nu \nu\,,
\eea
in terms of the Majorana mass eigenstates $\nu = N_m +N_m^c = \nu^c$. 
A consequence of this scenario is that the following combination vanishes \cite{deGouvea:2005er,Blennow:2010th}
\bea \label{eq:cancel}
\sum_{i=1}^{n+3} U_{ei}^2 m_i = \left(M_\nu\right)_{ee}^* = 0\,,
\eea
which is important for the computation of $0\nu\bt\bt$ rates.

Eq.\  \eqref{eq:mass} is minimal in the sense that the mass spectrum of light and heavy neutrinos is purely determined by renormalizable interactions,
at the price of the introduction of new, non-SM fields.
For comparison we will consider two other scenarios. In the first, we consider a theory with the same field content as the SM, and we will thus refer to this scenario as ``SM''. In this case, neutrino masses only arise at dimension 5, via the Weinberg operator \cite{Weinberg:1979sa}. After electroweak symmetry breaking, the neutrino mass Lagrangian reads
\begin{equation}\label{eq:nuSM}
 \mathcal L_m = - \frac{1}{2} \bar\nu_L^c M_L \nu_L+{\rm h.c.}\,.
\end{equation}
$M_L$ leads to three neutrino masses, and can be fitted to reproduce the observed mass splitting and the neutrino mixing PMNS matrix. In this case the price to pay
is the introduction of non-renormalizable interactions, which parameterize beyond-the-SM physics at energy scales much larger than the electroweak one.

The final scenario is a natural combination of Eq.\ \eqref{eq:mass} and \eqref{eq:nuSM},
and yields a mass matrix of the form
\bea\label{eq:mass2}
 M_\nu = \bma M_L &M_D^*\\M_D^\dagger&M_R^\dagger \ema \,.
 \eea
Such a mass matrix arises, for example, when one or more sterile neutrinos have masses much larger than electroweak scale. $M_\nu$ then takes the form of Eq.\ \eqref{eq:mass2}, after the heavy fields are integrated out. The main difference with Eq.\ \eqref{eq:mass} is that the cancellation condition no longer holds
\bea \label{eq:nocancel}
\sum_{i=1}^{n+3} U_{ei}^2 m_i = \left(M_\nu\right)_{ee}^* \neq 0\,. 
\eea
We will thus  refer to this as a ``no cancellation'' scenario.

While Eqs.~\eqref{eq:mass}  and \eqref{eq:mass2} are general, in what follows we will consider a few concrete mass models.
The baseline to which we will compare our result is the ``standard mechanism'' in which $0\nu\beta\beta$ is mediated by the exchange of three light Majorana neutrinos
(3+0),
corresponding to Eq.~\eqref{eq:nuSM}.
The simplest scenario with sterile neutrinos is the one in which the SM is extended by one gauge-singlet neutrino, the so called 3+1 model.
In the case of Eq.\ \eqref{eq:mass}, this model leads to two massless neutrinos, $m_1 =m_2 =0$, and thus is not realistic. It will however be useful to illustrate
the impact of the new contributions identified in this paper. 
The observed neutrino spectrum can be reproduced  in  3+1 models if $M_L\neq 0$. In this case 
$0\nu\beta\beta$ predictions can be significantly affected even by very light sterile neutrinos with masses in the eV range.
As a more realistic scenario, we will examine a minimal 3+2 model with two sterile neutrinos which can reproduce all oscillation data, if the mass of the lightest active neutrino vanishes.
Finally, we will consider pseudo-Dirac scenarios in which lepton number is an approximate symmetry. These correspond to a subset of Eq.\ \eqref{eq:mass},
with a specific structure for $M_D$ and $M_R$. We will focus on a 1+2 case, with one light neutrino with sub-eV mass, and two nearly degenerate sterile neutrinos 
with mass between 1 MeV and 10 GeV.

\section{Contributions to $0\nu\bt\bt$}\label{Sec:contr}
The amplitude for \NLDBD\ arises from the exchange of the light and heavy neutrinos and is second order in the Fermi constant, $G_F=\frac{1}{\sqrt{2}v^2}$.
The general expression below the electroweak scale can be written as  
\begin{align}\label{eq:quark}
\langle e_1 e_2 h_f|S^{\Delta L=2}_{eff}|h_i\rangle& =\frac{8 G_F^2 V^2_{ud} \sum_{i=1}^{n} U^2_{ei}m_{i}}{2!} \int d^4 x   d^4 r \langle e_1 e_2 | T \Big( \bar{e}_L (x+r/2)\,\gamma^\mu \gamma^\nu   e_L^c (x-r/2) \Big)| 0 \rangle \nonumber  \\
& \times \int \frac{d^4 k}{(2\pi)^4}\frac{e^{i k\cdot r} }{k^2 -m_i^2 +i\epsilon}  \langle   0^+_f|  T\Big(J_\mu(x+r/2)J_\nu (x-r/2) \Big) |0^+_i\rangle\,,
\end{align}
where $V_{ud}\simeq 0.97$ is the $u$-$d$ element of the quark mixing CKM matrix, $J_\mu = \bar u_L\gamma_\mu d_L$, $e_{1,2}$ stand for the final state electrons, and $h_{i,f}$ are the initial and final nuclear states, which we describe by their total angular momentum and parity quantum numbers, $0^+_{i,f}$.
The evaluation of Eq.\ \eqref{eq:quark} involves several steps, the first being the matching of the quark-level Lagrangian onto Chiral EFT ($\chi$EFT), which requires knowledge of hadronic matrix elements. After evaluating the LNV amplitude at the nucleon level, many-body computations are needed to describe the physical processes inside nuclei. 

The application of the EFT framework can essentially be seen as dividing up the neutrino-momentum integral in Eq.\ \eqref{eq:quark} into various regions.  The integral receives contributions from several momentum regions, which are set by the relevant scales in the problem. 
For each momentum region, we make use of the hierarchy between the different scales and the method of regions \cite{Beneke:1997zp} to expand the integrand in small ratios of $k/\Lambda\ll1$ or $m/k\ll 1$, where $\Lambda$ ($m$) is a large (small) scale. The EFT approach thereby allows one to consider one scale at a time. In our case Eq.\ \eqref{eq:quark} receives contributions from the following momentum regions:
\begin{itemize}
\item Hard neutrinos with momenta $k_0\sim |\mathbf {k}| \sim \Lambda_\chi$. As we will see below, this region leads  to short-distance interactions in the $\chi$EFT Lagrangian.
\item Soft neutrinos whose momenta scale as $k_0\sim |\mathbf {k}| \sim m_\pi$. These neutrinos contribute to loop diagrams involving nucleons and pions within $\chi$EFT.
\item Potential neutrinos with momenta, $k_0\sim |\mathbf {k}|^2/m_N \sim k_F^2/m_N$, where $k_F\sim m_\pi$ is the Fermi momentum of the nucleons and $m_N$ the nucleon mass. This region can be described by the LNV potentials between nucleons that are usually considered in calculations of $0\nu\bt\bt$.
\item Ultrasoft neutrinos with momenta that scale as $k_0\sim |\mathbf {k}| \sim k_F^2/m_N$. These neutrinos can be described as coupling to the nucleus as a whole, instead of individual nucleons.
\end{itemize}

We start by briefly recalling the contributions due to the exchange of the light SM neutrinos that appear in this EFT framework, before discussing the effects from sterile neutrinos as a function of their mass.

\subsection{Contributions from active Majorana neutrinos}
We first discuss the effects due to the usual active neutrinos.
After matching of the quark-level interactions onto $\chi$EFT, there are two types of leading-order (LO) contributions due to the SM neutrinos. We first have the exchange of neutrinos between nucleons, which results from hadronizing the quark-level weak currents. At LO this contribution is captured by replacing 
\be
J_\mu \to\mathcal J_\mu = \frac{1}{2} \bar N \tau^+ \left( g_V v_\mu-2g_A S_\mu \right)N\,,
\ee
 in Eq.\ \eqref{eq:quark}, where 
$N=(p\,n)^T$ is the nucleon doublet, $\tau^+ = \frac{\tau_1+i \tau_2}{2}$ is an isospin ladder operator with $\tau_i$ the Pauli matrices, and $v^\mu = (1,\vec{0})$ and $S^\mu = (0, \boldsigma/2)$ are the nucleon velocity and spin, while $g_{V}$ $(g_A)$ is the (axial-)vector charge of the nucleon. We will use $g_V \simeq 1$ up to tiny corrections and  $g_A = 1.2754\pm 0.0013$ \cite{Workman:2022ynf}. 
After including the effects of pions and neglecting the difference between the lepton momenta, this leads to the well-known long-range potential \cite{Doi:1985dx,Agostini:2022zub} arising from the exchange of three neutrinos with masses $m_i \ll k_F$
\begin{eqnarray}\label{eq:potential}
V_{\nu}^{(\rm pot )}(\mathbf{k}) &= &\tau^{(a)+}\tau^{(b)+} \times (4 G^2_F V^2_{ud})  \sum_{i=1}^3 \frac{ U^2_{ei}m_{i}}{\mathbf{k}^2 }\nonumber\\
&&\times  \Bigg[
1-g_A^2\left(\boldsigma^{(a)}\cdot \boldsigma^{(b)}-\frac{2m_\pi^2+\mathbf{k}^2}{(\mathbf{k}^2+m_\pi^2)^2}\boldsigma^{(a)} \cdot \mathbf{k}\,\boldsigma^{(b)} \cdot \mathbf{k}\right)
\Bigg] \bar{u}(p_1)  P_R{u}^c(p_2)\,,
\end{eqnarray}
where $(a,b)$ label the nucleons and $u(p_{1,2})$ are the electron spinors.

Apart from the exchange of `potential' neutrinos, with three-momenta similar to the Fermi momentum, $k_0\ll |\mathbf k| \sim k_F\sim m_\pi$,  substantial contributions arise from the exchange of `hard' neutrinos with momenta $k_0\sim |\mathbf {k}| \sim \Lambda_\chi$ \cite{Cirigliano:2018hja,Cirigliano:2019vdj}. The latter induces a contact interaction between nucleons and electrons in the chiral Lagrangian which generates the following potential 
\begin{eqnarray}\label{eq:potentialhard}
V_{\nu}^{(\rm hard)}({\mathbf k}) & = &\tau^{(a)+}\tau^{(b)+} \times \left(4 G_F^2 V_{ud}^2 \right)\times  \sum_{i=1}^3 \left[ -2g^{NN}_\nu U^2_{ei}m_{i}\right ]\times  \bar{u}(p_1)  P_R{u}^c(p_2)\,.
\end{eqnarray}
Here $ g^{NN}_\nu $ is a low-energy constant (LEC) that is in principle amenable to LQCD determinations \cite{Cirigliano:2022rmf,Davoudi:2020gxs,Cirigliano2020May,Davoudi2022May}, however, so far only phenomenological estimates are available \cite{Cirigliano:2021qko,Cirigliano:2020dmx,Richardson:2021xiu}, giving $g_\nu^{NN} = \Or(1/k_F^2)$.

The inverse $0\nu\bt\bt$ half life arising from light neutrino exchange can then be expressed as 
\bea\label{eq:rate}
\left(T_{1/2}^{0\nu}\right)^{-1}
= g_A^4G_{01}  \Big |\sum_{i=1}^3 V_{ud}^2\frac{U_{ei}^2 m_i}{m_e}A_\nu \Big| ^2\,,
\eea
where $G_{01}$ is a phase-space factor arising from the integral over the electron momenta. We use $G_{01} = 1.5\cdot 10^{-14}$ yr$^{-1}$ for ${}^{136}$Xe and $G_{01} = 2.2\cdot 10^{-15}$ yr$^{-1}$ for ${}^{76}$Ge  \cite{Kotila:2012zza,Horoi:2017gmj}. We define\footnote{Here $A_\nu$ is related to the potential by
\bea
\left[\frac{g_A^2 V_{ud}^2G_F^2}{\pi R_A}m_i U_{ei}^2  \bar u(p_1)P_R u^c(p_2)\right]\left[A_\nu^{\rm (pot)}+A_\nu^{\rm (hard)} ]\right] = \langle 0^+_f|\sum_{a,b}\int \frac{d^3 k}{(2\pi)^3}e^{i\mathbf{k\cdot r}}\left[V_{\nu}^{(\rm pot)} + V_{\nu}^{(\rm hard)}\right] | 0^+_i\rangle \,.\nn
\eea}
\bea\label{eq:AnuPot}
A_\nu &= A_\nu^{\rm (pot)}+A_\nu^{\rm (hard)} =  \frac{\mathcal{M}_F}{g_A^2}-\mathcal{M}_{GT}-\mathcal{M}_T-2 g_\nu^{NN} m_\pi^2\frac{\mathcal{M}_{F,sd}}{g_A^2} = \mathcal{M}_\text{long} + \mathcal{M}_\text{short} \,,
\eea
where $\mathcal{M}_{i}$ are the total Fermi (F), Gamow-Teller (GT) and tensor (T)  nuclear matrix elements (NMEs). These combine into the usual long-range NME, $\mathcal{M}_\text{long}$, whereas $\mathcal{M}_{F,sd}$, which is normalized such that it is $\Or(1)$ \cite{Cirigliano:2017djv}, drives the short-range NME, $\mathcal{M}_\text{short}$. See e.g.\ Ref.\ \cite{Agostini:2022zub} for an overview~\footnote{$\mathcal{M}_{F,sd}$ corresponds to $\mathcal{M}_{F,h}$ in the notation of Ref.\ \cite{Agostini:2022zub}, as {\it short-distance} NMEs are sometimes identified with {\it heavy} particle exchange. Likewise, the same correspondence between the notation for subindices ``$sd$'' in this work and ``$h$'' in Ref.\ \cite{Agostini:2022zub} holds across the manuscript.}. 
In these expressions we have neglected the dependence of the NMEs on the mass of the exchanged active neutrino which is a minuscule effect.

At next-to-next-to-leading order (N$^2$LO) one encounters  
$\pi\pi$, $\pi N N$ and momentum-dependent $NN$ counterterms \cite{Cirigliano:2017tvr,Cirigliano:2019vdj}, corrections to the axial and vector currents $\mathcal J_\mu$, and loops involving `soft' neutrinos with momenta $k_0\sim \mathbf k\sim k_F$ \cite{Cirigliano:2017tvr}, which lead to a correction to the potential amplitude, $A_\nu^{\rm (pot,2)}$. At the same order, there are contributions which depend on the intermediate nuclear states  due to the exchange of `ultrasoft' neutrinos, with momenta $k_0\sim |\mathbf k|\sim k_F^2/m_N$. This additional contribution to $A_\nu$ can be obtained from the chiral version of Eq.\ \eqref{eq:quark} by replacing the quark currents with their hadronic counterparts and expanding in $|\boldsymbol{k}|/k_F\ll1$, 
\bea\label{eq:ampusoftMS}
A_\nu^{\rm (usoft)}(m_i) &=& 8
\frac{\pi R_A}{g_A^2}\sum_n \langle 0^+_f | \mathcal J_\mu | 1^+_n\rangle\langle 1^+_n| \mathcal J^\mu | 0^+_i\rangle\int \frac{d^{d-1}k}{(2\pi)^{d-1}} \frac{1}{E_\nu\left[E_\nu +\Delta E_1-i\epsilon\right]} \nn\\ 
&&+(\Delta E_{1}\to \Delta E_{2})\,,
\eea
where $| 1^+_n\rangle$ indicates a complete set of intermediate states denoted by their total angular momentum and parity as $1^+_n$. $E_\nu = \sqrt{\mathbf k^2+m_i^2} \simeq |\mathbf k|$, $\Delta E_{1,2} = E_{1,2}+E_n-E_i$, with $E_{i}$ and $E_{n}$ denoting the energies of the initial and intermediate states ($E_f$ indicates the energy of the final state), while $E_{1,2}$ stand for the electron energies. The nuclear radius $R_A \simeq 1.2\,A^{1/3}\,\mathrm{fm}$ appears due to the conventional normalization of $G_{01}$ and the NMEs in Eq.\ \eqref{eq:rate}. Because these terms explicitly depend on the intermediate states, they represent the first corrections to the so-called ``closure'' approximation in $\chi$EFT \cite{Cirigliano:2017tvr}.
The integral in Eq.\ \eqref{eq:ampusoftMS} is UV divergent, but the dependence on the ultrasoft cut-off scale is cancelled  by a term in the
N$^2$LO potential \cite{Cirigliano:2017tvr}.

Scenarios involving sterile neutrinos generate the same types of contributions as those discussed above. Several of these have not been considered in the literature before. 
First, it has been convincingly demonstrated that the usual long-range $0\nu\beta\beta$ potential (Eq.\ \eqref{eq:potential}) has to be supplemented by the additional short-range interaction in Eq.\ \eqref{eq:potentialhard} that arises from the exchange of light neutrinos with large virtual momenta. In the standard mechanism (the exchange of three light Majorana neutrinos) these contributions have been found to enhance $0\nu\beta\beta$ rates by a factor two-to-three for light (e.g.\ ${}^{12}$Be), medium (${}^{48}$Ca), to heavy isotopes (e.g. ${}^{76}$Ge and ${}^{136}$Xe) \cite{Pastore:2017ofx,Wirth:2021pij,Weiss:2021rig,Belley:2023btr}. Although the uncertainties are still sizable, mainly due to poor knowledge of the associated QCD matrix element $g_\nu^{NN}$, LQCD efforts can pave the way towards more reliable predictions \cite{Cirigliano:2022rmf,Davoudi:2020gxs,Cirigliano2020May,Davoudi2022May}. In case of sterile neutrinos with $M\lesssim \Lambda_\chi\sim$ GeV, similar contributions proportional to this matrix element appear, but now with an additional $M$ dependence, $g_\nu^{NN}(M)$ \cite{Dekens:2020ttz}. These effects appear at leading order.
Second, 
several of the effects that appear at N$^{2}$LO in the $\chi$EFT expansion in $\epsilon_\chi = m_\pi/\Lambda_\chi$ in the standard mechanism, can become important in specific scenarios. Here we highlight the case when all sterile neutrinos have masses below the pion mass in the minimal extension of the SM. The leading contributions are then strongly suppressed \cite{Blennow:2010th} due to the cancellation in Eq.\ \eqref{eq:cancel} and the formally subleading terms become dominant. In particular, as we discuss below, contributions from the ultrasoft region become significant. 

\subsection{Contributions from sterile neutrinos}
\begin{figure}[!t]
	\center
	\begin{subfigure}[b]{0.49\textwidth}
		\includegraphics[trim={1.5cm 0 0 0},clip ,width=\textwidth]{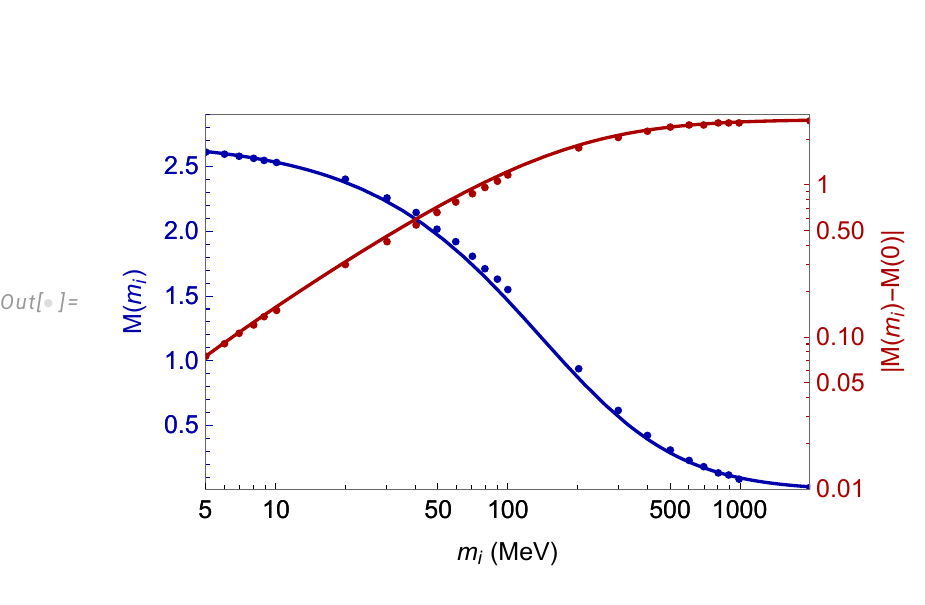}
		\hfill
	\end{subfigure}
	\begin{subfigure}[b]{0.49\textwidth}
		\includegraphics[width = \textwidth]{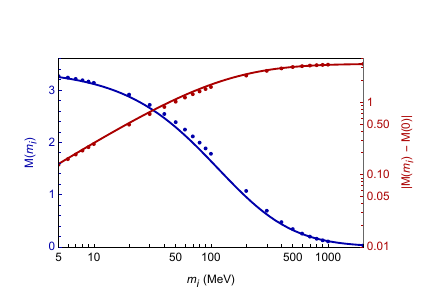}
		\hfill
	\end{subfigure}
	\caption{NMEs in Eq.\ \eqref{eq:LDNME} for $^{136}$Xe (left panel) and $^{76}$Ge (right panel) as a function of the neutrino mass (in blue), as well as the difference $\mathcal{M}(0)-\mathcal{M}(m_i)$ (in red). The circles show the numerical results of the nuclear shell model calculation described in Sec.\ \ref{sec:shell}, while the solid lines depict the interpolation formula of Eq.\ \eqref{eq:fitM}. 
	}
	\label{fig:NME}
\end{figure}

In this section we discuss the contributions from sterile neutrinos for different mass ranges. Before describing our approach, we briefly discuss what is presently done in most of the literature. 
The standard procedure is to modify the neutrino potential in Eq.~\eqref{eq:potential} to include heavier states by effectively replacing
\begin{equation}
 \sum_{i=1}^3 \frac{ U^2_{ei}m_{i}}{\mathbf{k}^2 } \rightarrow  \sum_{i=1}^3 \frac{ U^2_{ei}m_{i}}{\mathbf{k}^2 } + \sum_{i=4}^{n+3} \frac{ U^2_{ei}m_{i}}{\mathbf{k}^2 + m_i^2 }\,,
\end{equation}
for essentially any value of $m_i$ and where $n$ denotes the number of sterile neutrinos. The contribution from hard neutrinos, which appears at LO, is not considered but would lead to a mass dependent LEC $g_\nu^{NN}(m_i)$. The modified potential is then evaluated between the initial and final nuclear states leading to mass-dependent NMEs. 
Figure~\ref{fig:NME} shows an example of this $m_i$ dependence for $^{136}$Xe  and $^{76}$Ge, where  the blue circles correspond to shell-model results for
the amplitude
\begin{equation}\label{eq:LDNME}
\mathcal{M}(m_i) \equiv -\left(\frac{\mathcal{M}_F(m_i}{g_A^2}-\mathcal{M}_{GT}(m_i)-\mathcal{M}_T(m_i)\right)
\end{equation}
 for a range of $m_i$. 
The main features of this line are easy to understand. For light $m_i$ the NMEs are almost mass independent, whereas for heavy $m_i$ the NMEs scale as $m_i^{-2}$ due to the massive neutrino propagator. These arguments have led to interpolation formulae that are used in effectively all analyses in the literature, see e.g. Refs.~\cite{Faessler:2014kka,Asaka:2016zib,Bolton:2019pcu,Fang:2021jfv,Bolton:2022tds}. These formulae take the form
\begin{equation}\label{eq:naiveNME}
\mathcal{M}(m_i) = \mathcal{M}(0) \frac{\langle p^2\rangle}{\langle p^2\rangle +m_i^2}\,,
\end{equation} 
ensuring the appropriate scalings at $m^2_i \ll \langle p^2 \rangle$ and $m^2_i \gg \langle p^2 \rangle$. A fit to the nuclear shell model results
discussed in Sec.\ \ref{sec:shell}, and shown  in Fig.~\ref{fig:NME}, gives $\langle p^2 \rangle \simeq (175\,\mathrm{MeV})^2$
for $^{136}$Xe and $\langle p^2 \rangle \simeq (160.5\,\mathrm{MeV})^2$ for $^{76}$Ge.
Thus $\sim m_\pi^2 \sim k_F^2$ is of the expected size.

\begin{figure}[t!]
\center
\includegraphics[width=0.48\textwidth]{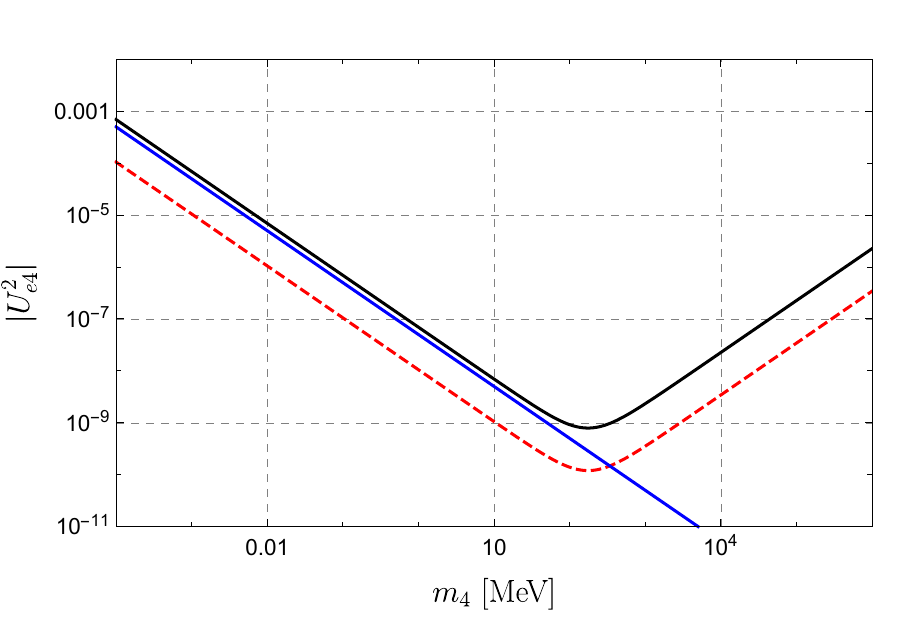}
\includegraphics[width=0.48\textwidth]{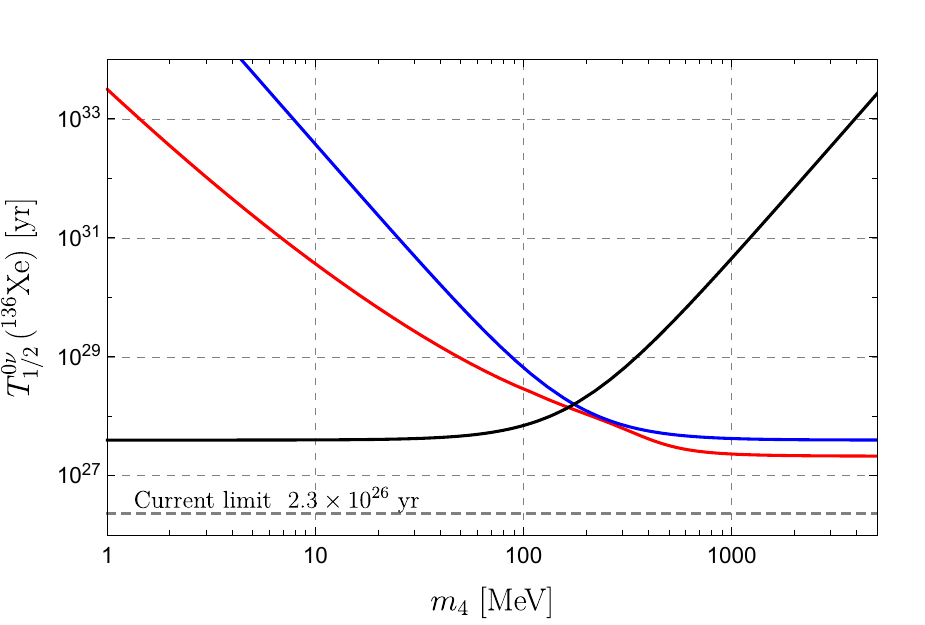}
\caption{ Left: Limits on $U_{e4}^2$ as function of $m_4$ assuming the $0\nu\beta\beta$ rate is dominated by a single sterile neutrino, for current (solid black line) and future (dashed red line) experimental limits, compared to the naive seesaw expectation (blue line). Right: ${}^{136}$Xe half-life considering  contributions only  from $m_4$  (black) and  all neutrinos (blue) using Eq.~\eqref{eq:naiveNME} in the 3+1 model. The red line denotes the half-life by using Eq.\ \eqref{eq:fullint} including the new ultrasoft contributions. }
\label{naiveUlim}
\end{figure}

With this description it is possible to compute the contribution of a sterile neutrino with mass $m_i$ and mixing angle $U_{ei}$. If we assume no cancellation with other contributions, as we expect for example in the scenario decribed by Eq.\ \eqref{eq:mass2},  the above description allows us to constrain the mixing angle $U_{ei}$ as a function of $m_i$. The resulting constraints are depicted in the left panel of Fig.~\ref{naiveUlim} using the current KamLAND-Zen \cite{KamLAND-Zen:2022tow} limit for $^{136}$Xe, $T_{1/2} > 2.3 \times 10^{26}$ yr, and the reach of next-generation experiments assuming a lower bound $T_{1/2} > 10^{28}$ yr. The blue line depicts the naive seesaw expectation for the size of the mixing angle, $U_{e4} = \sqrt{m_3/m_4}$, where we set $m_3=0.05$ eV. These limits are in good agreement with Ref.~\cite{Bolton:2019pcu}.

While this approach seems reasonable, the above description of $0\nu\beta\beta$ rates from massive sterile neutrinos comes with several shortcomings:
\begin{itemize}
\item The meaning of the NMEs becomes unclear for $m_i > m_N \simeq 1\,\mathrm{GeV}$ ($m_N$ is the nucleon mass) as the $\chi$EFT expansion, used to obtain Eq.~\eqref{eq:potential}, no longer converges when $m_i/\Lambda_\chi\gtrsim 1$. A more appropriate description for sterile neutrinos in this mass range is to integrate them out at the quark level, before hadronization. This leads to a local LNV dimension-nine operator containing four quarks and two charged leptons which, after renormalization-group evolution from $m_i$ to $m_N$, can be matched to LNV hadronic operators. This procedure has been worked out in Ref.~\cite{Dekens:2020ttz}. 
\item For sterile neutrino masses $m_i \lesssim \Lambda_\chi$ there are leading-order contributions from hard neutrino exchange. These are captured by the mass-dependent $g_\nu^{NN}(m_i)$ LEC. We will argue that these terms can have sizeable impact on $0\nu\beta\beta$ rates in the $\nu$SM. 
\item In minimal models in which Eq.\ \eqref{eq:cancel} holds and all sterile neutrinos have masses $m_i \ll k_F$ the total $0\nu\beta\beta$ rate is strongly suppressed. The commonly used parametrization in the light-$m_i$ regime gives
\begin{equation}
\mathcal{M}(m_i) \simeq \mathcal{M}(0)\left( 1 - \frac{m_i^2}{\langle p^2 \rangle} + \dots \right)\,.
\end{equation}
Considering the simple $3+1$ model then leads to (ignoring again the $g_\nu^{NN}$ contributions which are affected by the same cancellation) 
\begin{eqnarray}
\left(T_{1/2}^{0\nu}\right)^{-1}\Bigg|_{m_4 \ll k_F} &=&  g_A^4G_{01} \Big |\sum_{i=1}^4 V_{ud}^2\frac{U_{ei}^2 m_i}{m_e}\mathcal{M}(0) \left( 1 - \frac{m_i^2}{\langle p^2 \rangle} \right)  \Big| ^2\nonumber\\
&=& g_A^4G_{01} \Big |\sum_{i=1}^4 V_{ud}^2\frac{U_{ei}^2 m_i^3}{m_e \langle p^2 \rangle}\mathcal{M}(0) \Big|^2\,,
\end{eqnarray} 
where in the second equality we applied the identity in Eq.~\eqref{eq:cancel}. In the right panel of Fig.~\ref{naiveUlim} we demonstrate this cancellation by computing the ${}^{136}$Xe half-life by considering only the contribution from $m_4$ (this contribution becomes mass independent at small energies because $m_4 U_{e4}^2 \sim m_3$) and by the sum of all contributions. For $m_4 < k_F$ the cancellation is severe, leading to extremely suppressed decay rates. Within this approach the first corrections to the amplitude scale as $U_{ei}^2 m_i^3$, but, as we will show in more detail below, a more careful analysis of the various contributions leads to new terms related to ultrasoft neutrino exchange that scale more favourably, as $U_{ei}^2 m_i^2$ or $U_{ei}^2 m_i^3  \log m_i$. We anticipate these findings by depicting the solid red line in the right panel of Fig.~\ref{naiveUlim}, that includes these corrections. Clearly, the commonly used parametrization in Eq.~\eqref{eq:naiveNME} is unable to capture the correct $m_i$ dependence in these scenarios.
\end{itemize}

In what follows we discuss how one can improve upon the method described in this section.

\subsubsection{Region 1: $m_i > \Lambda_\chi$ }\label{heavy}
An EFT approach to this region has been extensively discussed in the literature, starting from Refs.\ \cite{Prezeau:2003xn,Graesser:2016bpz}. Here the heavy neutrino can be integrated out at the quark-gluon level, which gives rise to a LNV dimension-nine operator containing four quarks and two electrons. At the scale  $\mu_0\simeq 2$ GeV we have
\begin{equation}\label{eq:Ldim9}
\mathcal{L}^{(9)} = C_L (\mu_0) \bar{u}_L \gamma^\mu d_L \bar{u}_L \gamma_\mu d_L \bar{e}_L e^c_L\,,  
\end{equation}  
with $C_L (\mu_0) =\eta(\mu_0,m_i) C_L (\mu=m_i)=-\eta(\mu_0,m_i)\frac{4 V^2_{ud} G^2_F}{m_i} U^2_{ei}$. Here $\eta(\mu_0,m_i)$ takes into account the QCD renormalization-group evolution from the scale $m_i$ to the QCD scale \cite{Buras:2000if,Buras:2001ra,Cirigliano:2017djv}, 
\bea\label{QCDrunning}
\eta(\mu_0,m_i) =\left\{
  \begin{array}{@{}ll@{}}
   \left(\frac{\al_s(m_i)}{\al_s(\mu_0)}\right)^{6/25} & m_i\leq m_{\text{bottom}} \\
\left(\frac{\al_s(m_{\text{bottom}})}{\al_s(\mu_0)}\right)^{6/25}\left(\frac{\al_s(m_i)}{\al_s(m_{\text{bottom}})}\right)^{6/23} & m_{\text{bottom}} \leq m_i\leq m_{\text{top}}\\
\left(\frac{\al_s(m_{\text{bottom}})}{\al_s(\mu_0)}\right)^{6/25}\left(\frac{\al_s(m_{\text{top}})}{\al_s(m_{\text{bottom}})}\right)^{6/23} \left(\frac{\al_s(m_i)}{\al_s(m_{\text{top}})}\right)^{2/7} & m_i \geq m_{\text{top}}
  \end{array}\right. \,,
\eea
in terms of the bottom and top quark masses $m_{\text{bottom}}$ and $m_{\text{top}}$, and where, at one loop, the strong coupling can be written as $\al_s(\mu) = \frac{2\pi}{\bt_0\log(\mu/\Lambda^{(n_f)})}$, with $\bt_0 = 11-\frac{2}{3}n_f$. Together with $\alpha_s(m_Z) = 0.1179$ \cite{Workman:2022ynf}, this gives $\Lambda^{(4,5,6)} \simeq \{119,\, 87,\, 43 \}$ MeV.

Matching the interaction in Eq.\ \eqref{eq:Ldim9} onto $\chi$EFT leads to LNV $\pi\pi \bar ee^c$, $\pi (\bar p n) \bar ee^c$, and $(\bar p n)^2 \bar ee^c$ vertices \cite{Prezeau:2003xn,Cirigliano:2017djv,Cirigliano:2018yza}.
The resulting neutrino potential is 
\bea\label{V9}
V_9(\mathbf{k})&=& -8\tau^{(a)+}\tau^{(b)+}\eta(\mu_0,m_i) g_A^2 G_F^2\frac{V_{ud}^2 U_{ei}^2}{m_i}\bar u(p_1)P_R u^c(p_2)\nn\\
&&\times \left[  \boldsigma^{(a)} \cdot \mathbf k \,\boldsigma^{(b)} \cdot \mathbf k \, \left(\frac{5}{6}g_1^{\pi\pi}
\frac{\mathbf k^2}{(\mathbf k^2+m_\pi^2)^2}- \frac{g_1^{\pi N}}{\mathbf k^2+m_\pi^2}\right)-\frac{2}{g_A^2}g_1^{NN}
\right]\,,
\eea
where $g_1^{\pi\pi}$, $g_1^{N\pi}$, and $g_1^{NN}$ denote the LECs, evaluated at the scale $\mu=2$ GeV, corresponding to the $\pi\pi$, $\pi N$, and $NN$ interactions, which are expected to be $\Or(1)$. So far only the pionic coupling has been determined using LQCD calculations \cite{Nicholson:2018mwc,Detmold:2020jqv,Detmold:2022jwu}, which give 
$g_1^{\pi\pi} = 0.36\pm 0.019$ \cite{Nicholson:2018mwc} and $g_1^{\pi\pi} = 0.17\pm 0.016$ \cite{Detmold:2022jwu}. 
Taking the results at face value there is a disagreement between the two determinations, however, they both confirm the $\Or(1)$ expectation from naive dimensional analysis (NDA).
They also show that QCD corrections cause significant deviations from the naive factorization results, $g^{\pi\pi}_{1} = 0.6$.
We expect similar $\mathcal O(1)$ deviations for $g_{1}^{\pi N}$ and $g_{1}^{NN}$.
The potential in Eq.~\eqref{V9} leads to the amplitude 
\bea\label{A9}
A_\nu^{(9)} = -2\eta(\mu_0,m_i) \frac{m_\pi^2}{m_i^2}\left[\frac{5}{6}g_1^{\pi\pi}\left(\mathcal{M}_{GT,sd}^{PP}+\mathcal{M}_{T,sd}^{PP}\right)+\frac{g_1^{\pi N}}{2}\left(\mathcal{M}_{GT,sd}^{AP}+\mathcal{M}_{T,sd}^{AP}\right)
-\frac{2}{g_A^2}g_1^{NN}\mathcal{M}_{F,sd}\right],
\eea
where the NMEs, Ref.~\cite{Cirigliano:2017djv} and Ref.~\cite{Agostini:2022zub}, are also normalized such that they are expected to be $\Or(1)$. We discuss their values in Sec.~\ref{sec:LECsNMEs}.

\subsubsection{Region 2: $k_F<m_i < \Lambda_\chi$}\label{sec:below_lambda_chi}
In this mass region the sterile neutrino cannot be integrated out at the quark level, so that its contributions can no longer be described by $A_\nu^{(9)}$. The sterile neutrino now has to be kept as an explicit degree of freedom in $\chi$EFT. Its effects are more similar to that of the light SM neutrinos and can partially be captured by including the $m_i$ dependence of $V_\nu^{\rm (pot)}$ and $V_\nu^{\rm (hard)}$
\begin{eqnarray}\label{eq:potential_usoft}
& & V_{\nu}^{(\rm pot )} + V_{\nu}^{(\rm hard )} = \tau^{(a)+}\tau^{(b)+} \times (4 G^2_F V^2_{ud})  \sum_{i=1}^n  U^2_{ei}m_{i} \bar{u}(p_1)  P_R{u}^c(p_2)\\
&&\times \Bigg\{  \frac{1}{\mathbf{k}^2  +m_i^2}\Bigg[
1-g_A^2\left(\boldsigma^{(a)}\cdot \boldsigma^{(b)}-\frac{2m_\pi^2+\mathbf{k}^2}{(\mathbf{k}^2+m_\pi^2)^2}\boldsigma^{(a)} \cdot \mathbf{k}\,\boldsigma^{(b)} \cdot \mathbf{k}\right) \Bigg] -  2 g_\nu^{\rm NN}(m_i)
\Bigg\} \,. \nonumber
\end{eqnarray}
Similar to Eq.\ \eqref{eq:AnuPot} this gives,
\bea \label{eq:pot>}
A_\nu^{\rm (pot)}(m_i)+A_\nu^{\rm (hard)}(m_i) =  \frac{\mathcal{M}_F(m_i)}{g_A^2}-\mathcal{M}_{GT}(m_i)-\mathcal{M}_T(m_i)-2 g_\nu^{NN}(m_i) m_\pi^2\frac{\mathcal{M}_{F,sd}}{g_A^2}\,,
\eea
where the $m_i$ dependence of these amplitudes now becomes significant and is expected to scale as $\frac{k_F^2}{m_i^2}$. With the exception of the
contributions proportional to $g_\nu^{NN}$, these terms are similar those captured by the literature approach in Eq.\ \eqref{eq:naiveNME}.
Evaluating these contributions requires knowledge of the NMEs and the LEC $g_\nu^{NN}$ as a function of $m_i$, motivating nuclear-structure and LQCD determinations.

In addition, there are contributions from loops involving soft sterile neutrinos, leading to a correction to the potential, $A_\nu^{\rm (pot,2)}$. Although such contributions appear at N$^2$LO for light neutrinos, they can give rise to terms scaling as $m_i^2/\Lambda_\chi^2$ for the sterile neutrinos. These effects lead to a breakdown of the $\chi$EFT expansion when $m_i$ approaches the QCD scale, so that our estimates become unreliable for $m_i\sim \Lambda_\chi$. Finally, similar to the region $m_i>\Lambda_\chi$, there are no contributions from ultrasoft sterile neutrinos since $m_i\gg k$, so that the integrals vanish in dimensional regularization once we expand the integrand of Eq.\ \eqref{eq:ampusoftMS} in terms of $k/m_i$ and $k/k_F$.

\subsubsection{Region 3: $m_i < k_F$}\label{sec:below_kf}
Sterile neutrinos in this regime look even more similar to the usual SM neutrinos and contribute to $A_\nu^{\rm (pot)}$, $A_\nu^{\rm (hard)}$, $A_\nu^{\rm (usoft)}$, $A_\nu^{\rm (pot,2)}$, although with different relative importance than the SM neutrinos. We organize the discussion by these different momentum regions. 
\\\\\textbf{Hard and potential neutrinos}\\
As before, the evaluation of the $m_i$ dependence for the $A_\nu^{\rm (pot)}$ and $A_\nu^{\rm (hard)}$ terms requires non-perturbative many-body and LQCD methods.
Naively, the neutrino potential is again given by Eq.\ \eqref{eq:potential_usoft}.
However, to be able to treat the different momentum regions separately and avoid double counting, we employ the method of regions \cite{Beneke:1997zp} and expand the $k$ integrand whenever small ratios of scales appear. This implies that Eq.\ \eqref{eq:potential_usoft}
should be expanded in $m_i^2/{\mathbf{k}^2}$
whenever $m_i$ is in the ultrasoft domain, $m_i\sim k_F^2/m_N$. This procedure of expanding can be seen as a matching calculation between theories with and without potential neutrinos. This matching can be computed by  subtracting the terms from the low-energy theory, which only involves ultrasoft neutrinos, from the contributions from the full theory (including potential and ultrasoft neutrinos).
The contributions from potential neutrinos then lead to
\begin{eqnarray}\label{eq:potential_usoft2}
V_{\nu}^{(\rm pot )}({\bf k}) &= &\tau^{(a)+}\tau^{(b)+} \times (4 G^2_F V^2_{ud})  \sum_{i=1}^n  U^2_{ei}m_{i} \bar{u}(p_1)  P_R{u}^c(p_2) \nonumber\\
&&\times\left( \frac{1}{\mathbf{k}^2}
- \frac{m_i^2}{(\mathbf{k}^2)^2}  \right)
\Bigg[
1-g_A^2\left(\boldsigma^{(a)}\cdot \boldsigma^{(b)}-\frac{2m_\pi^2+\mathbf{k}^2}{(\mathbf{k}^2+m_\pi^2)^2}\boldsigma^{(a)} \cdot \mathbf{k}\,\boldsigma^{(b)} \cdot \mathbf{k}\right) \Bigg]  . 
\end{eqnarray}
The corrections from the potential thus scale as $m_i^2/k_F^2$, for $m_i\ll k_F$. Traditionally, the potential is not expanded in $m_i$ in many-body calculations, leading to a mismatch between the definition of the potential contributions used here --- employing Eq.\ \eqref{eq:potential_usoft2} --- and the NME results in the literature --- using Eq.\ \eqref{eq:potential_usoft}. 
The difference is that NMEs obtained from the unexpanded potential have an $m_i^1$ term, which is absent when using Eq.\ \eqref{eq:potential_usoft2}. This is most easily seen from the Fourier transforms of Eqs.\ \eqref{eq:potential_usoft} and \eqref{eq:potential_usoft2}, where $\Or(m_i^3)$, the difference of the unexpanded and expanded potentials in coordinate space behaves like
\begin{equation}
V-V_{\rm exp.} \sim \left(1-g_A^2 \boldsigma^{(a)}\cdot \boldsigma^{(b)}\right)\left(\frac{e^{-m_i r}}{r} - \left[\frac{1}{r}+\frac{1}{2}m_i^2 r\right]\right)\simeq -m_i \left(1-g_A^2 \boldsigma^{(a)}\cdot \boldsigma^{(b)}\right).
\end{equation}
To connect our definition of the potential contributions to the usually determined NMEs, we thus have to correct for the additional linear term. We have
\bea\label{eq:Apot_expand}
A_\nu^{(\rm pot,<)}(m_i) = -\left[\mathcal{M}(m_i) - m_i \left[\frac{d}{d m_i} \mathcal{M}(m_i)\right]_{m_i=0}\right]\,,
\eea
where the $<$ label of the amplitude denotes that it applies in the $m_i<k_F$ region and $\mathcal{M}(m_i)$, defined in Eq.\ \eqref{eq:LDNME},
correspond to the NMEs computed from the unexpanded potential in Eq.\ \eqref{eq:potential_usoft}. The derivative term in Eq.\ \eqref{eq:Apot_expand} removes the $m_i^1$ term, which does not appear when starting from Eq.\ \eqref{eq:potential_usoft2}.

After expanding the hard-neutrino contributions we have~\footnote{In principle, the LEC $g_\nu^{NN}$ could have a linear dependence on $m_i$. However, $g_\nu^{NN}$ arises from contributions to Eq.\ \eqref{eq:quark} in the $k_0\sim \boldsymbol{k}\sim\Lambda_\chi$ region, where $\frac{1}{k^2-m_i^2}\simeq \frac{1}{k^2}\left[1+\frac{m_i^2}{k^2}\right]$ is a good approximation and no linear $m_i$ dependence should appear.}
\bea
 V_{\nu}^{(\rm hard )} &= &-2\tau^{(a)+}\tau^{(b)+} \times (4 G^2_F V^2_{ud})  \sum_{i=1}^n  U^2_{ei}m_{i} \bar{u}(p_1)  P_R{u}^c(p_2) \times \left( g_\nu^{\rm NN}(0)  + m_i^2 \frac{d}{d m_i^2 } g_{\nu}^{\rm NN} \right) 
\, ,
\eea
which leads to 
\begin{equation}
A_\nu^{\rm (hard)} = -2 \left( g_\nu^{\rm NN}(0)  + m_i^2 \frac{d}{d m_i^2 } g_{\nu}^{\rm NN} \right) m_\pi^2\frac{M_{F,sd}}{g_A^2},
\end{equation}
where only the hadronic matrix element, and not the NME, is $m_i$ dependent.
The resulting $m_i^2$ term is hard to compute from first principles, but we will argue below that it scales as $m_i^2/\Lambda_\chi^2$ and thus only provides a small next-to-next-to-leading-order correction. 
\\\\\textbf{Ultrasoft neutrinos}\\
A new effect appears due to ultrasoft sterile neutrinos. When $k_F>m_i$, we can evaluate the usoft terms by performing the integrals of Eq.\ \eqref{eq:ampusoftMS}  in the \textoverline{MS} scheme, which leads to the following expression 
\begin{equation}\label{eq:ampusoftMSregion3}
\begin{aligned}
A_{\nu}^{\rm (usoft)} &=& 
2\frac{R_A}{\pi g_A^2}   \sum_{n}   \langle 0^+_f|\mathcal J^\mu |1^+_n\rangle \langle 1^+_n| \mathcal J_\mu |0^+_i\rangle  \Big(f(m_i, \Delta E_1) +f(m_i,\Delta E_2)\Big)  \,,
\end{aligned}
\end{equation}
with 
\begin{eqnarray}\label{eq:Fusoft}
f(m,E) = \begin{cases}
-2\left[ E\left(1+ \log \frac{\mu_{us}}{m} \right)  +\sqrt{m^2 - E^2} \left(\frac{\pi}{2}-\tan^{-1}\, \frac{ E}{\sqrt{m^2 -E^2}}\right) \right]\,,
&{\rm if}\,  m > E\,, \\
-2\left[ E \left(1+\log \frac{\mu_{us}}{m} \right)  -\sqrt{ E^2-m^2} \log \frac{ E+\sqrt{ E^2-m^2}}{m}  \right]\,,& {\rm if}\,  m \le E\,.
\end{cases}
\end{eqnarray}
Eqs.~\eqref{eq:ampusoftMSregion3} and \eqref{eq:Fusoft} 
depend on the ultrasoft renormalization scale
$\mu_{us}$ and require knowledge of the intermediate state energies, $E_{n}$, as well as the  nuclear matrix elements, $\langle 0^+_f | \mathcal J_\mu | 1^+_n\rangle$, which appear in first-order single-$\beta$ decays and only involve one-body operators. The $m_i$-dependent contributions scale as 
\be
A_\nu^{(\rm usoft)}\sim \frac{ m_i^2}{4\pi \Delta E_{1,2}k_F} \log\frac{m_i}{\Delta E_{1,2}}\,,\qquad \mathrm{for}\,\,m_i<\Delta E_{1,2}\ll k_F\,.
\ee
A power-counting estimate $\Delta E_{1,2}\sim k_F/4\pi$  gives $A_\nu^{(\rm usoft)}\sim  m_i^2/k_F^2 \log\frac{m_i}{\Delta E_{1,2}}$, similar to the scaling of the potential-neutrino contributions. However, this underestimates the contributions from the lowest-lying states with $\Delta E_{1,2}\lesssim 1 \,{\rm MeV}\ll k_F/4\pi$, see Table \ref{tab:overlapNME}, and in practice the ultrasoft contributions dominate. 
For $k_F>m_i>\Delta E_{1,2}$, there is a region where ultrasoft neutrinos induce contributions that are linear in $m_i$ and scale as $A_\nu\sim m_i/k_F$. Comparing the $m_i$-dependent terms to those resulting from the potential and hard contributions, one finds that, in this region, the ultrasoft contributions can give LO contributions in minimal scenarios where Eq.\ \eqref{eq:cancel} holds.

As mentioned above, the unexpanded and expanded potentials in Eqs.\ \eqref{eq:potential_usoft} and \eqref{eq:potential_usoft2} differ by a term $\sim m_i^1$. 
This linear term is related to ultrasoft neutrinos as it appears when we do not subtract the contributions of the ultrasoft neutrinos in the matching calculation. As a result, the ultrasoft contributions in Eq.\ \eqref{eq:Fusoft} involve an $m_i^1$ dependence, which can be shown to be related to the linear term that would result from Eq.\ \eqref{eq:potential_usoft}. 
It may seem surprising that the potential NMEs are related to the ultrasoft contributions, as only the latter involve excited state information. However, in the $m_i > \Delta E$ regime the $m_i^1$ term in Eq.\ \eqref{eq:Fusoft} is independent of $E_n$ and we can perform the sum over the complete set of intermediate states, $\sum_{n}   \langle 0^+_f|\mathcal J^\mu |1^+_n\rangle \langle 1^+_n| \mathcal J_\mu |0^+_i\rangle  \sim \langle 0^+_f|\tau^+ \tau^+(1-g_A^2 \boldsigma  \cdot  \boldsigma | 0^+_i \rangle $,
leaving just the dependence on the initial and final states. Likewise, in coordinate space, the $m_i^1$ term arising from Eq.\ \eqref{eq:potential_usoft} multiplies an $r$-independent potential, which leads to the same NME.
The correspondence between the sum of the product of the two first-order matrix elements and the linear term in the NMEs (computed with the unexpanded potential) allows for a consistency check, which we discuss in more detail in Sec.~\ref{sec:LECsNMEs}. 
\\\\\textbf{Soft neutrinos}\\
Finally, there are again effects due to loops involving soft sterile neutrinos. In fact, it can be shown that the dependence on $\mu_{us}$ above cancels in the total amplitude when taking into account loop contributions to the potential \cite{Cirigliano:2017tvr}. Including these effects would require the computation of NMEs due to $V_{\nu,2}$ of Ref.\ \cite{Cirigliano:2017tvr}, modified to include the $m_i$ dependence. As the relevant NMEs have only been estimated in light nuclei \cite{Pastore:2017ofx}, here we will estimate part of the terms due to $V_{\nu,2}$ by setting $\mu_{us}=m_\pi$ in Eq.\ \eqref{eq:ampusoftMSregion3} and neglecting the remaining contributions.

\subsubsection{Summary}
\begin{table*}[t]
\small
\renewcommand{\arraystretch}{1.7}    \centering
    \begin{tabular}{|lc|c|c|c|c|}
 \hline
\multicolumn{2}{|c|}{} & $m_i\ll k_F^2/m_N$ & $k_F^2/m_N<m_i < k_F$ & $k_F<m_i<\Lambda_\chi$ & $\Lambda_\chi <m_i$ \\ 
 \hline
${\rm hard}$&$ k_0\sim |\boldsymbol{k}|\sim\Lambda_\chi$ & $\boldsymbol{ \color{darkred}1}$ & $\boldsymbol{\color{darkred}1}$ &$\boldsymbol {\color{darkred}\frac{k_F^2}{m_i^2}} $& -\\
${\rm potential}$&$ k_0\sim \frac{\boldsymbol{k}^2}{m_N}\sim\frac{k_F^2}{m_N} $ & $\boldsymbol{ \color{darkred}1}$ & $\boldsymbol{\color{darkred}1}$ &$\boldsymbol {\color{darkred}\frac{k_F^2}{m_i^2}} $& -\\
${\rm ultrasoft} $&$k_0\sim  |\boldsymbol{k}|\sim \frac{k_F^2}{m_N}$ & $\frac{\Delta E_{}}{4\pi k_F}\sim \frac{1}{(4\pi)^2}$ & $\frac{m_i}{ k_F}$ & - & -\\
${\rm soft}$&$ k_0\sim |\boldsymbol{k}|\sim k_F$ &  $\frac{1}{(4\pi)^2}$ & $ \frac{1}{(4\pi)^2} $  & $\boldsymbol{ \color{darkred} \frac{1}{(4\pi)^2}\frac{m_i^2}{k_F^2}} $  & -\\
${\rm perturbative}$&$ k_0\sim |\boldsymbol{k}|\gg \Lambda_\chi$  & - & - & - & $\boldsymbol{ \color{darkred}\frac{k_F^2}{m_i^2}}$\\
\hline\hline
 $\sum_{i=1}^n m_i U_{ei}^2 = 0$ && $m_i\ll k_F^2/m_N$ & $k_F^2/m_N<m_i < k_F$ & $k_F<m_i<\Lambda_\chi$ & $\Lambda_\chi <m_i$ \\ 
 \hline
 ${\rm hard}$& & ${\color{gray}1+}\frac{m_i^2}{\Lambda_\chi^2}$ & {\color{gray}1+}$\frac{m_i^2}{\Lambda_\chi^2}$ & $\boldsymbol{\color{darkred}\frac{k_F^2}{m_i^2}}$ & -\\
 ${\rm potential}$& & ${\color{gray}1+}\boldsymbol{\color{darkred}\frac{m_i^2}{k_F^2}}$ & ${\color{gray}1+}\frac{m_i^2}{k_F^2}$ & $\boldsymbol{\color{darkred}\frac{k_F^2}{m_i^2}}$ & -\\
${\rm ultrasoft} $&& ${\color{gray}\frac{1}{(4\pi)^2}+}\boldsymbol{\color{darkred}\frac{m_i^2 \log \frac{m_i}{\Delta E_{}}}{4\pi \Delta E_{}k_F}}$ & $\boldsymbol{\color{darkred}\frac{m_i}{ k_F}}$ & - & -\\
${\rm soft}$& & $ {\color{gray}\frac{1}{(4\pi)^2}+}\frac{1}{(4\pi)^2}\frac{m_i^2}{k_F^2} $ & $ {\color{gray}\frac{1}{(4\pi)^2}+}\frac{1}{(4\pi)^2}\frac{m_i^2}{k_F^2} $  & $\boldsymbol {\color{darkred}\frac{1}{(4\pi)^2}\frac{m_i^2}{k_F^2}} $  & -\\
${\rm perturbative}$& & - & - & - & $\boldsymbol{\color{darkred}\frac{k_F^2}{m_i^2}}$\\
 \hline
    \end{tabular}
\vspace{.2cm}
    \caption{ The expected scaling of the contributions of a neutrino with mass $m_i$ to $A_\nu(m_i)$.
The contributions are shown separately for different ranges of $m_i$ (the columns) and are organized by the neutrino momentum regions that induced them (the rows).    
   The lower panel shows the same information as the top panel, but assumes a minimal scenario in which Eq.\ \eqref{eq:cancel} holds. For each $m_i$ region, the parts of the amplitude that are expected to be leading are shown in red. The order-of-magnitude estimates for terms that cancel in the total amplitude, due to $\sum_i m_i U_{ei}^2=0$, are shown in gray.}  %
    \label{tab:scalings}
\end{table*}

Table \ref{tab:scalings} summarizes the scaling of the contributions of a neutrino with mass $m_i$, induced by the different momentum regions.
The top panel shows the scenario assuming no cancellations, while the bottom panel focuses on the terms that survive when taking into account $\sum_{i=1}^n m_i U_{ei}^2 = 0$. Here the columns show the different ranges of $m_i$, while the rows show the contributions due to a particular neutrino-momentum region.
We first focus on the top panel for which the usual expressions dominate, namely, $A_\nu^{(\rm pot,hard)}$, while the ultrasoft contributions provide comparable corrections if $m_i$ lies somewhat below $k_F$. This picture changes drastically once we assume the minimal ultraviolet completion of the $\nu$SM for which Eq.\ \eqref{eq:cancel} holds. Now the usually dominant terms are only nonzero after taking into account the $m_i$ dependence of the NMEs and the LEC $g_\nu^{NN}(m_i)$.
Other contributions to the amplitude, which would otherwise appear at sub-leading orders, now give leading contributions. In particular, the contributions due to ultrasoft neutrinos become significant or even dominant in the range $m_i<k_F$, while loop diagrams involving soft neutrinos can be relevant for $m_i<\Lambda_\chi$. Although the latter have not been computed so far, and would be hard to control for $m_i\sim \Lambda_\chi$, the ultrasoft contributions can be estimated reliably.

\section{Nuclear and hadronic matrix elements}\label{sec:LECsNMEs}

 \begin{table}
	\center
	\begin{tabular}[b]{|c|ccccccccc|}    
		\hline
		$ m_i \,{\rm [MeV]}$
		&   5 & 
		6 & 
		7 & 
		8 & 
		9 & 
		10 & 
		20 & 
		30 & 
		40  \\\hline
		$\mathcal M(m_i)$ &
		2.62 & 2.60&2.59  & 2.57  & 2.55    & 2.54  & 2.4  & 2.3  & 2.1   \\\hline
		$ m_i \,{\rm [MeV]}$ &50 &60
		&   70 & 
		80 & 
		90 & 
		100 & 
		200 & 
		300 & 
		400 \\\hline
		$\mathcal M(m_i)$   & 2.0 &1.9  &
		1.8 & 1.7  & 1.6  & 1.5  & 0.94  &  0.61  & 0.42  \\\hline
		
		$ m_i \,{\rm [MeV]}$
		& 
		500 & 
		600 &700 &800 & 
		900 & 
		1000 & 
		2000 & 
		& 
		\\\hline
		$\mathcal M(m_i)$  & 0.31  & 0.23 & 0.18 &0.14   & 0.11  &  0.094 & 0.025  &   &    \\\hline
	\end{tabular}
	\caption{Shell-model $0\nu\bt\bt$ NMEs for ${}^{136}$Xe as a function of the neutrino mass.}
	\label{tab:NME}
\end{table}

\begin{table}
	\center
	\begin{tabular}[b]{|c|ccccccccc|}    
		\hline
		$ m_i \,{\rm [MeV]}$
		&   5 & 
		6 & 
		7 & 
		8 & 
		9 & 
		10 & 
		20 & 
		30 & 
		40  \\\hline
		$\mathcal M(m_i)$ &
		3.26 & 3.24 & 3.21 & 3.19 & 3.16 & 3.14 & 2.9 & 2.7 & 2.5   \\\hline
		
		$ m_i \,{\rm [MeV]}$ &50 &60
		&   70 & 
		80 & 
		90 & 
		100 & 
		200 & 
		300 & 
		400 \\\hline
		$\mathcal M(m_i)$   & 
		2.4 & 2.2 & 2.1 & 2.0 & 1.9 & 1.8 & 1.1 & 0.69 & 0.47  \\\hline
		
		$ m_i \,{\rm [MeV]}$
		& 
		500 & 
		600 &700 &800 & 
		900 & 
		1000 & 
		2000 & 
		& 
		\\\hline
		$\mathcal M(m_i)$  &  0.34 & 0.25 & 0.20 & 0.15 & 0.13 & 0.10 & 0.027 &  &   \\\hline
	\end{tabular}
	\caption{Shell-model $0\nu\bt\bt$ NMEs for ${}^{76}$Ge as a function of the neutrino mass.}
	\label{tab:NME_Ge}
\end{table}

\begin{table*}[t]
\setlength{\tabcolsep}{2pt}
\renewcommand{\arraystretch}{1.3}    \centering
\footnotesize
    \begin{tabular}{|c|c|c|}
    \hline
    $\frac{E_n-E_i}{\rm MeV}$&  $\langle 1^+_n| \boldsigma\tau^+ | 0^+_i\rangle$& $\langle 0^+_f| \boldsigma\tau^+ | 1^+_n\rangle$\\\hline
 0.17 & 1. & 0.13 \\
 0.63 & -0.19 & -0.0063 \\
 0.89 & -0.25 & -0.016 \\
 1.02 & 0.3 & 0.036 \\
 1.05 & 0.23 & 0.025 \\
 1.1 & -0.13 & -0.00076 \\
 1.2 & 0.12 & -0.0052 \\
 1.3 & 0.16 & -0.0028 \\
 1.4 & -0.23 & -0.0098 \\
 1.5 & 0.2 & -0.012 \\
 1.6 & -0.36 & 0.0084 \\
 1.7 & -0.24 & 0.00058 \\
 1.9 & 0.22 & 0.011 \\
 2.0 & 0.34 & 0.007 \\
 2.2 & 0.35 & 0.006 \\
 2.3 & -0.49 & -0.0086 \\
 2.6 & 0.62 & 0.021 \\
 2.7 & -0.91 & -0.024 \\
 2.9 & 0.37 & 0.0064 \\
 3.1 & 0.3 & 0.0013
  \\\hline
    \end{tabular}
\hfill    
        \begin{tabular}{|c|c|c|}
    \hline
    $\frac{E_n-E_i}{\rm MeV}$&  $\langle 1^+_n| \boldsigma\tau^+ | 0^+_i\rangle$& $\langle 0^+_f| \boldsigma\tau^+ | 1^+_n\rangle$\\\hline
 3.3 & 0.39 & -0.0013 \\
 3.6 & 0.39 & 0.0021 \\
 3.8 & 0.45 & -0.013 \\
 4.0 & -0.44 & -0.0032 \\
 4.3 & -0.35 & -0.0038 \\
 4.6 & -0.36 & -0.0067 \\
 4.8 & 0.44 & 0.0083 \\
 5.1 & 0.44 & 0.0066 \\
 5.4 & -0.55 & -0.0093 \\
 5.7 & 0.63 & 0.012 \\
 6.1 & 0.85 & 0.013 \\
 6.3 & -1.2 & -0.016 \\
 6.7 & -1.3 & -0.014 \\
 7.0 & -1.9 & -0.016 \\
 7.3 & 3.1 & 0.023 \\
 7.5 & -4. & -0.028 \\
 7.7 & 2.6 & 0.017 \\
 8.1 & 1.4 & 0.0091 \\
 8.4 & -1. & -0.0057 \\
 8.8 & -0.93 & -0.0064 
  \\\hline
    \end{tabular}
\hfill
        \begin{tabular}{|c|c|c|}
    \hline
 $\frac{E_n-E_i}{\rm MeV}$&  $\langle 1^+_n| \boldsigma\tau^+ | 0^+_i\rangle$& $\langle 0^+_f| \boldsigma\tau^+ | 1^+_n\rangle$\\\hline
 9.1 & 0.8 & 0.0038 \\
 9.4 & 0.59 & 0.0014 \\
 9.8 & -0.5 & 0.0027 \\
 10.1 & 0.35 & -0.0027 \\
 10.5 & 0.26 & -0.00053 \\
 10.9 & -0.22 & -0.00021 \\
 11.3 & 0.17 & -0.00037 \\
 11.7 & -0.16 & -0.00054 \\
 12.0 & -0.16 & -0.001 \\
 12.4 & 0.14 & 0.00092 \\
 12.8 & 0.12 & -0.00014 \\
 13.1 & 0.092 & -0.0004 \\
 13.5 & -0.079 & -0.00019 \\
 13.9 & 0.071 & -0.00026 \\
 14.2 & -0.07 & 0.000031 \\
 14.6 & -0.035 & 0.00021 \\
 15.1 & -0.051 & -0.00015 \\
 16.2 & -0.039 & 0.00011 \\
 17.3 & -0.043 & -0.000091 \\
 17.7 & 0.11 & -0.000029 
  \\\hline
    \end{tabular}    
    \caption{Values of the first-order nuclear matrix elements in Eq.\ \eqref{eq:overlapNMEs}, that enter the $0\nu\bt\bt$ of $^{136}$Xe.}
    \label{tab:overlapNME}
\end{table*}

\begin{table*}[t]
	\setlength{\tabcolsep}{2pt}
	\renewcommand{\arraystretch}{1.2}    \centering
	\footnotesize
	\begin{tabular}[t]{|c|c|c|}
		\hline
		$\frac{E_n-E_i}{\rm MeV}$&  $\langle 1^+_n| \boldsigma\tau^+ | 0^+_i\rangle$& $\langle 0^+_f| \boldsigma\tau^+ | 1^+_n\rangle$\\\hline
		0.5 & -0.33 & -0.11 \\
		0.7 & 0.67 & 0.29 \\
		0.8 & -0.024 & -0.052 \\
		0.8 & -0.5 & -0.14 \\
		1.2 & 0.075 & 0.017 \\
		1.3 & -0.14 & -0.22 \\
		1.5 & 0.55 & 0.18 \\
		1.8 & 0.18 & 0.016 \\
		2.1 & -0.28 & -0.088 \\
		2.3 & 0.35 & 0.019 \\
		2.8 & 0.65 & 0.11 \\
		3.0 & 1.12 & 0.091 \\
		3.3 & 0.9 & 0.098 \\
		3.7 & 1.059 & 0.078 \\
		4.0 & 1.005 & 0.075 \\
		4.4 & 1.19 & 0.078 \\
		4.9 & 1.29 & 0.055 \\
		5.4 & 1.4 & 0.05 \\
		5.9 & 1.45 & 0.023 \\
		6.3 & 1.25 & 0.0065 \\
		6.9 & 1.09 & -0.0026 \\
		\hline
	\end{tabular}
	\hfill
	\begin{tabular}[t]{|c|c|c|}
		\hline
		$\frac{E_n-E_i}{\rm MeV}$&  $\langle 1^+_n| \boldsigma\tau^+ | 0^+_i\rangle$& $\langle 0^+_f| \boldsigma\tau^+ | 1^+_n\rangle$\\\hline
		7.4 & 0.92 & -0.0089 \\
		8.0 &0.81 & -0.011 \\
		8.6 & 0.71 & -0.0075 \\
		9.3 & 0.6 & -0.0075 \\
		9.9 & 0.52 & -0.0046 \\
		10.6 & 0.46 & -0.0059 \\
		11.3 & 0.41 & -0.0032 \\
		12.1 & 0.34 & -0.0033 \\
		12.9 & 0.32 & -0.000093 \\
		13.6 & 0.28 & -0.0017 \\
		14.4 & 0.24 & -0.0002 \\
		15.2 & 0.2 & -0.0004 \\
		16.0 & 0.17 & 0.00021 \\
		16.9 & 0.15 & -0.00016 \\
		17.7 & 0.14 & 0.00022 \\
		18.6 & 0.12 & -0.00019 \\
		19.5 & 0.1 & 0.00018 \\
		20.4 & 0.086 & -0.00024 \\
		21.3 & 0.07 & 0.00017 \\
		22.2 & 0.058 & -0.0002 \\
		23.1 & 0.049 & 0.00017 \\
		24.1 & 0.037 & -0.00018 \\
		\hline
	\end{tabular}
	\hfill
	\begin{tabular}[t]{|c|c|c|}
		\hline
		$\frac{E_n-E_i}{\rm MeV}$&  $\langle 1^+_n| \boldsigma\tau^+ | 0^+_i\rangle$& $\langle 0^+_f| \boldsigma\tau^+ | 1^+_n\rangle$\\\hline
		25.1 & 0.031 & 0.00016 \\
		26.0 & 0.025 & -0.00017 \\
		27.0 &0.022 & 0.00014 \\
		28.0 & 0.017 & -0.00015 \\
		29.0 & 0.014 & 0.00014 \\
		29.9 & 0.0099 & -0.00015 \\
		31.0 & 0.0073 & 0.00014 \\
		32.0 & 0.0057 & -0.00013 \\
		33.0 & 0.0044 & 0.000125 \\
		34.0 & 0.0032 & -0.00013 \\
		35.1 & -0.0024 & -0.00012 \\
		36.1 & -0.0018 & 0.00011 \\
		37.1 & -0.0013 & -0.00011 \\
		38.2 & -0.00096 & 0.00011 \\
		39.2 & 0.00071 & 0.000105 \\
		40.3 & 0.00052 & -0.0001 \\
		41.3 & 0.00038 & 0.0001 \\
		42.4 & -0.00028 & 0.000097 \\
		43.4 & 0.00021 & 0.000094 \\
		44.5 & -0.00017 & 0.00009 \\
		45.5 & 0.00014 & 0.000088 \\
		\hline
	\end{tabular}    
	\caption{Values of the first-order nuclear matrix elements in Eq.\ \eqref{eq:overlapNMEs}, that enter the $0\nu\bt\bt$ of $^{76}$Ge.}
	\label{tab:overlapNME_Ge}
\end{table*}

Computing all the contributions identified above requires knowledge of various  hadronic and nuclear matrix elements. For relatively light sterile neutrinos, $M\lesssim\Lambda_\chi$, these matrix elements are non-trivial functions of the sterile neutrino mass. In this work, we use a single nuclear framework in which we can compute all NMEs consistently, the nuclear shell model~\cite{MPinedo}. Note that only many-body methods which can calculate $\beta\beta$-decay NMEs beyond the closure approximation, such as the nuclear shell model or the quasiparticle random-phase approximation~\cite{Engel:2016xgb,Jokiniemi:2022yfr}, can provide ultrasoft NMEs.  
We further focus on $0\nu\bt\bt$ in ${}^{136}$Xe and ${}^{76}$Ge, which presently give some of the most stringent limits on neutrino Majorana masses, and are also expected to do so for next-generation experiments~\cite{Agostini:2022zub}. It is worthwhile to consider other nuclear many-body methods and isotopes, but our main goal here is to assess the newly identified contributions with respect to traditional contributions for representative experimentally relevant isotopes. 
\\\\\textbf{Nuclear shell model NME calculations}\label{sec:shell}\\
We perform a nuclear shell-model study of the decays of the ground-state to ground-state transition of ${}^{136}$Xe into ${}^{136}$Ba and ${}^{76}$Ge into ${}^{76}$Se. For the ultrasoft NMEs, in addition to these initial and final states, we also need to calculate a set of states of the intermediate nuclei ${}^{136}$Cs and ${}^{76}$As. The initial and final states are well converged by the diagonalization of the Hamiltonian in the configuration spaces given below. However, for the set of intermediate states, we use the Lanczos strength function method~\cite{MPinedo}, which gives a set of approximate eigenstates. Nonetheless, we have checked that with the $\sim$ sixty approximate eigenstates kept, for which we give results in Tables~\ref{tab:overlapNME} and~\ref{tab:overlapNME_Ge}, the ultrasoft NMEs for both isotopes are well converged.

For the germanium decay, we use a configuration space consisting of the $\rm1p_{3/2}$, $\rm0f_{5/2}$, $\rm1p_{1/2}$ and $\rm0g_{9/2}$ single-particle orbitals for protons and neutrons, with a ${}^{56}$Ni inert core. As in previous shell-model studies~\cite{Menendez:2017fdf}, we use  the GCN2850 effective Hamiltonian~\cite{Caurier08}. For the decay of xenon, the configuration space of our calculations comprises the $\rm1d_{5/2}$, $\rm0g_{7/2}$, $\rm2s_{1/2}$, $\rm1d_{3/2}$ and $\rm0h_{11/2}$ single-particle orbitals for protons and neutrons, on top of a ${}^{100}$Sn core. Here we use the GCN5082 shell-model interaction~\cite{Caurier08}, also in line with previous works~\cite{Menendez:2017fdf}. We use the shell-model codes ANTOINE~\cite{FNowacki,MPinedo} and NATHAN~\cite{MPinedo} to obtain the nuclear states and to evaluate the NMEs.
\\\\\textbf{LECs and NMEs from potential and hard neutrinos}\\
Let us begin by discussing the required matrix elements induced by potential neutrinos, which appear in the linear combination
$\mathcal{M}(m_i)$ defined in Eq.\ \eqref{eq:LDNME}.
For very light masses, $m_i \ll k_F$, these NMEs have been calculated for many isotopes with a broad range of nuclear many-body methods \cite{Agostini:2022zub}, but the explicit mass dependence has only been considered in a handful of works \cite{Blennow:2010th,Barea:2015zfa}. Here we use the shell-model results depicted in Fig.~\ref{fig:NME}, with the numerical values given in Tables~\ref{tab:NME} and~\ref{tab:NME_Ge}. Note that for lighter neutrino masses $m_i\lesssim5$~MeV the difference between the NMEs shown in Fig.~\ref{fig:NME} requires very precise calculation of the corresponding neutrino potentials.

Not surprisingly, Fig.~\ref{fig:NME} shows that for light $m_i \ll k_F$ the NMEs become roughly constant, while they scale as $m_i^{-2}$ for heavy $m_i\sim \Lambda_\chi$. The description in terms of NMEs no longer applies for masses $m_i \geq \mu_0 = 2$ GeV, as we integrate out the heavy neutrinos at the quark level in this case, see Sec.~\ref{heavy}. 
In practice, it is useful to have an interpolation formula that describes the shell-model results for $m_i\leq 2$ GeV. As they include a linear $m_i$ dependence in the light mass regime, see the discussions below Eqs.~\eqref{eq:potential_usoft2} and \eqref{eq:Fusoft}, we use the functional form
\begin{equation}\label{eq:fitM}
\mathcal{M}_{\mathrm{int}}(m_i) =\mathcal{M}(0)\frac{1}{1+m_i/m_a+(m_i/m_b)^2}\,,
\end{equation}
where $\mathcal{M}(0)=2.7$ for $^{136}$Xe and 3.4 for $^{76}$Ge. For these decays, we set $m_a = 157 $ MeV (117 MeV) for $^{136}$Xe ($^{76}$Ge) which is the prediction of the linear slope from ultrasoft corrections as explained in more detail below, and perform a $\chi^2$-fit to the NMEs in the mass range of $2~\text{MeV}\leq m_i\leq2000~\text{MeV}$  to obtain $m_b = 221$ MeV (218 MeV) for $^{136}$Xe ($^{76}$Ge). 
The resulting curve fits the data points in this range to about $5\%$ accuracy, well within the expected theoretical uncertainty~\footnote{
 If we perform a $\chi^2$-fit to the NMEs keeping both $m_{a,b}$ as fit parameters we obtain 
$m_a = 192 (157)$ MeV and $m_b = 208 (202)$ MeV for $^{136}$Xe ($^{76}$Ge). Assuming a conservative flat $10\%$ theoretical uncertainty on the NMEs, the total $\chi^2$ of the two fits are very similar.\label{fn:chisq}}.

The potential contributions always appear in combination with hard contributions
\be\label{eq:totNME}
\mathcal{M}^{\mathrm{tot}}(m_i) \equiv    \mathcal{M}_{\mathrm{int}}(m_i)+ \frac{ 2 m_\pi^2 g_\nu^{NN}(m_i)}{g_A^2} \mathcal{M}_{F,\mathrm{sd}}\,.
\ee
The hard contributions depend on a hadronic and a nuclear matrix element $\mathcal{M}_{F,\mathrm{sd}} \times g_\nu^{NN}(m_i)$ and, in fact, it is only the combination with $\mathcal{M}_{\mathrm{int}}(m_i)$ in Eq.\ \eqref{eq:totNME} that is independent of regulators used in nuclear computations \cite{Cirigliano:2018hja}. 
As $g_\nu^{NN}$ has not been determined using LQCD methods yet, only model-dependent estimates are available. It was pointed out in Ref.~\cite{Cirigliano:2019vdj} that $g_\nu^{NN}(0)$ is connected to charge-indepedence-breaking (CIB) nucleon-nucleon interactions that are known up to $N_c$-suppressed corrections \cite{Richardson:2021xiu}, where $N_c=3$ denotes the number of colors in QCD. The value provided by the CIB strategy is in reasonable agreement with a model estimate of $g_{\nu}^{NN}(0)$ \cite{Cirigliano:2020dmx,Cirigliano:2021qko}. As we are using the nuclear shell model results for the NMEs, here we take advantage of the connection to CIB and follow Ref.~\cite{Jokiniemi:2021qqv} which gives a range of values for $g_{\nu}^{NN}(0)$ based on various nucleon-nucleon potentials. For this work, we pick the intermediate value 
\begin{equation}\label{eq:gnuvalue}
g_{\nu}^{NN}(0) = -1.01\,\mathrm{fm}^2\,,
\end{equation}
with the corresponding NMEs shown in Table \ref{tab:sdNME}.

 \begin{table}
	\center
		\renewcommand{\arraystretch}{1.2}    
	\begin{tabular}[b]{|c|ccccc|}    
		\hline		
		&  $\mathcal M_{F,sd}$ & $\mathcal M_{GT,sd}^{AP}$ &$\mathcal M_{GT,sd}^{PP} $&$\mathcal M_{T,sd}^{AP} $&$\mathcal M_{T,sd}^{PP}$\\\hline
		$^{76}$Ge&-2.21 & -2.26 &0.82&-0.05&0.02\\
		$^{136}$Xe&-1.94 &-1.99 &0.74 &0.05 &-0.02\\\hline
	\end{tabular}
	\caption{Shell-model determinations from Refs.\ \cite{Menendez:2017fdf,Jokiniemi:2021qqv}  of the relevant short-distance NMEs for ${}^{76}$Ge and ${}^{136}$Xe.}
	\label{tab:sdNME}
\end{table}

The hadronic matrix element $g_\nu^{NN}(m_i)$ has a non-trivial mass dependence of which little is known. Around $m_i \sim \Lambda_\chi$ the sum of the potential and hard contributions in Eq.~\eqref{eq:totNME} should match to the description provided by Eq.~\eqref{A9} which scales as $m_i^{-2}$. Because $\mathcal M_{\mathrm{}}(m_i)$ has the same scaling, this requires $g_{\nu}^{NN}(m_i \sim \Lambda_\chi) \sim m_i^{-2}$ as well. 
In the opposite limit,  $m_i\ll k_F$, we would expect the form $g_{\nu}^{NN}(m_i)\simeq g_{\nu}^{NN}(0)+g_{\nu,2}^{NN}m_i^2$, see Sec.\ \ref{sec:below_kf}.
Although the renormalization-group invariance of the $nn \rightarrow pp + ee$ amplitude requires $g_{\nu}^{NN}(0)=\Or(F_\pi^{-2})$,
where $F_\pi$ is the pion decay constant,  to appear at leading order  \cite{Cirigliano:2018hja}, no such argument exists for the enhancement of $m_i^2$-dependent LECs, leading to an 
estimate of $g_{\nu,2}^{NN}=\Or(F_\pi^{-2}\Lambda_\chi^{-2})$ (see App.~\ref{app:gnuNN}).
We therefore assume the functional form
\begin{equation}\label{eq:gnu_int}
g_{\nu}^{NN}(m_i) = g_{\nu}^{NN}(0) \frac{1+ (m_i/m_c)^2\,{\rm sign}(m_d^2)}{1 + (m_i/m_c)^2(m_i/|m_d|)^2}\,,
\end{equation}
where sign$(m_d^2) = m_d^2/|m_d^2|$, while $|m_d|$ appears in the denominator in order to avoid possible poles.
This generalizes the interpolation constructed in Ref.~\cite{Dekens:2020ttz}, which overestimates the $m_i$ dependence in the small $m_i$ regime. 
We set $m_c = 1$ GeV as expected from the NDA estimate of  $g_{\nu,2}^{NN}$. We subsequently tune $m_d$ such that Eq.~\eqref{eq:totNME} matches Eq.~\eqref{A9} at a scale $m_i = \mu_0 = 2$ GeV. 
This last step requires values of $g_1^{\pi\pi}$, $g_1^{\pi N}$, and $g_1^{NN}$, the latter two of which are currently poorly known. To get a reasonable estimate we only consider the contributions from $g_1^{NN}$ and $g^{\pi\pi}_1$, for which we use  $g_1^{NN}= (1+3 g_A^2)/4 $, inspired by the factorization estimate, and $g^{\pi\pi}_1=0.36$ \cite{Nicholson:2018mwc}. The corresponding short-distance NMEs we use are collected in Table \ref{tab:sdNME}.
We then obtain $m_d \simeq 146 (139)$ MeV for $^{136}$Xe ($^{76}$Ge). It is encouraging that the value of $m_d$ obtained in this way is mostly independent of the applied nucleus as $g_\nu^{NN}(m_i)$ is related to a two-nucleon matrix element. 
Clearly, our estimates of these LECs and their $m_i$ dependence come with sizable uncertainties. 
Future LQCD determinations of $g_{1}^{\pi N, NN}$ and $g_\nu^{NN}(m_i)$ will allow one to verify the functional form of Eq.\ \eqref{eq:gnu_int} and reduce the current uncertainties.
\\
\\
\textbf{NMEs from ultrasoft neutrinos}\\
For the ultrasoft contributions in Eqs.~\eqref{eq:ampusoftMS} and \eqref{eq:ampusoftMSregion3} we require the intermediate-state energies, $E_n$, and first-order NMEs of the form
\begin{eqnarray}\label{eq:overlapNMEs}
 A^{(\rm usoft)}_\nu &\propto& \frac{1}{4} \left[ g_V^2  \langle 0^+_f |  \tau^+  | 1^+_n\rangle \langle 1^+_n |  \tau^+  | 0^+_i \rangle - g_A^2  \langle 0 ^+_f |  \tau^+ \boldsigma  | 1^+_n\rangle \cdot  \langle 1^+_n |  \tau^+  \boldsigma | 0^+_i \rangle \right]\nonumber\\
&\simeq& -\frac{g_A^2}{4}  \langle 0 ^+_f |  \tau^+ \boldsigma  | 1^+_n\rangle \cdot  \langle 1^+_n |  \tau^+  \boldsigma | 0^+_i \rangle\,,
\end{eqnarray} 
where we neglected Fermi transitions because they vanish up to tiny isospin-breaking corrections. We include excited states up to $E_n - E_i = 18$ MeV as higher-states provide negligible contributions to the first-order matrix elements in our shell-model calculations. We approximate the electron energies by $E_1\simeq E_2\simeq Q_{\bt\bt}/2 +m_e $, where the $Q$-value is $Q_{\bt\bt}=E_i-E_f-2m_e\simeq 2.5$ MeV for ${}^{136}$Xe and likewise $Q_{\bt\bt}\simeq 2.0$ MeV for ${}^{76}$Ge~\footnote{Since $A_\nu^{\rm (usoft)}$ is even in $\Delta E_1\leftrightarrow \Delta E_2$, corrections to the approximation $E_1=E_2$ scale like $\sim \frac{\delta ^2}{\Delta E_{1,2}^2}$, where $\delta = \frac{E_2-E_1}{2}$ and $|\delta| \leq Q_{\bt\bt}/2$. For typical intermediate states with $E_n-E_i = (5-10)$ MeV, such corrections are at the percent level at most.}.
We tabulate the corresponding first-order matrix elements for the decay of $^{136}$Xe and ${}^{76}$Ge in, respectively, Tables~\ref{tab:overlapNME} and \ref{tab:overlapNME_Ge}. The results for $^{136}$Xe were presented earlier in Ref.~\cite{Dekens:2023iyc}.

As mentioned in Sec.\ \ref{sec:below_kf}, the linear $m_i$ dependence appearing in the NMEs, $\mathcal{M}(m_i)$, should correspond to the linear term in $A_\nu^{\rm (usoft)}$, allowing for a consistency check. In the regime $\Delta E < m_i < k_F$, the linear term in the ultrasoft expression in Eq.~\eqref{eq:Fusoft} is given by
\be\label{eq:usoftCheck}
A^{(\rm usoft)}_\nu\Big|_{m_i^1}= R_A m_i    \sum_{n}   \langle 0^+_f|\mathcal \tau^+ \boldsigma |1^+_n\rangle  \cdot  \langle 1^+_n |  \tau^+  \boldsigma | 0^+_i \rangle  + \mathcal O(\Delta E/m_i) \simeq 
 \begin{cases}
 \frac{m_i}{58\,\mathrm{MeV}}\,,&^{136}{\rm Xe}\\
\frac{m_i}{35\,\mathrm{MeV}}\,,& ^{76}{\rm Ge}
 \end{cases}
\ee
where the last result sums over the shell-model contributions for $^{136}$Xe and $^{76}$Ge presented in Tables \ref{tab:overlapNME} and \ref{tab:overlapNME_Ge}, which are the values used for $m_a$ in Eq.~\eqref{eq:fitM}.  These values are in pretty good agreement with fits of $m_a$ to the NMEs in Fig.\ \ref{fig:NME} in the same regime, $m_a/\mathcal{M}(0) \simeq 71\, \mathrm{MeV}$ and $m_a/\mathcal{M}(0) \simeq 46\, \mathrm{MeV}$ for $^{136}$Xe and $^{76}$Ge, respectively (see footnote \ref{fn:chisq}). This confirms that the usual definition of the NMEs includes part of the ultrasoft contributions. 

\subsection{A practical formula}
Having discussed all contributions, we finally need to construct an effective formula that connects the various regions. A very useful parametrization for the contribution from a neutrino of mass $m_i$ to the $0\nu\beta\beta$ amplitude is given by
\begin{eqnarray}\label{eq:fullint}
A_\nu (m_i) = \begin{cases}
A_\nu^{\rm (pot,<)}(m_i)+A_\nu^{\text{(hard)}}(m_i)+A_\nu^{\text{(usoft)}}(m_i) \,,&  m_i <  100 \text{ MeV}\,, \\
A_\nu^{\rm (pot)}(m_i)+A_\nu^{\text{(hard)}}(m_i)\,,& 100 \text{ MeV} \le m_i <  2 \text{ GeV}\,,\\
A_\nu^{\text{(9)}}(m_i)\,,& {\rm }\,  2 \text{ GeV} \le m_i \,.
\end{cases}
\end{eqnarray}
The needed input to these amplitudes is given by 
\begin{itemize}
\item The potential contributions require the interpolation of the NMEs Eq.\ \eqref{eq:fitM}.  For $m_i>k_F$ the relation of the NMEs to $A_\nu^{\rm (pot)}$ is given by  Eq.\ \eqref{eq:pot>}, while in the $m_i<k_F$ region the expression for $A_\nu^{\rm (pot,<)}$ is given by Eq.\ \eqref{eq:Apot_expand}. The latter subtracts the derivative with respect to $m_i$ to avoid double counting the linear terms that appear both in the usual definition of the mass dependent NMEs and the ultrasoft expression. 
\item The hard contributions require the interpolation formula for $g_\nu^{NN}$ in Eq.\ \eqref{eq:gnu_int} and the NME $\mathcal{M}_{F,sd}$ in Table \ref{tab:sdNME}.
\item The ultrasoft contributions involve the first-order NMEs and intermediate-state energies listed in Tables \ref{tab:overlapNME} and \ref{tab:overlapNME_Ge}.
\item Finally, $A_\nu^{\text{(9)}}$ requires knowledge of the LECs $g_1^{\pi\pi}$, $g_1^{\pi N}$, $g_1^{NN}$, and several short-distance NMEs. Since several of the LECs are currently unknown, we approximate this region by using $g_1^{NN}=(1+3 g_A^2)/4$, $g^{\pi\pi}_1=0.36$, combined with the short-distance NMEs from Table \eqref{tab:sdNME}. The QCD evolution factors are given by Eq.~\eqref{QCDrunning}. 
\end{itemize}

A Mathematica notebook implementing Eq.~\eqref{eq:fullint} is provided as supplementary material attached to the paper, and is available from the authors upon request.
\begin{figure}[t!]
\center
\includegraphics[width=0.48\textwidth]{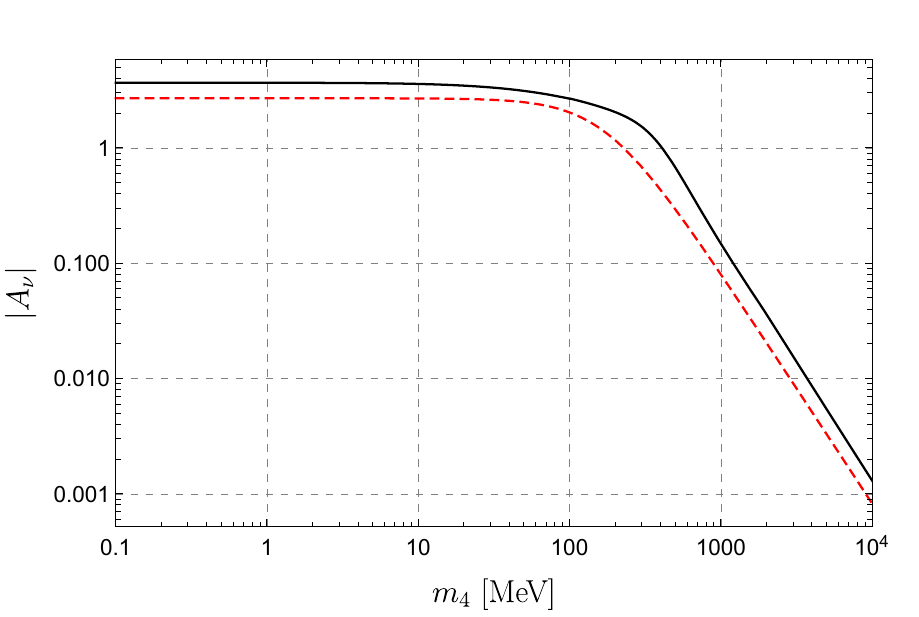}
\includegraphics[width=0.48\textwidth]{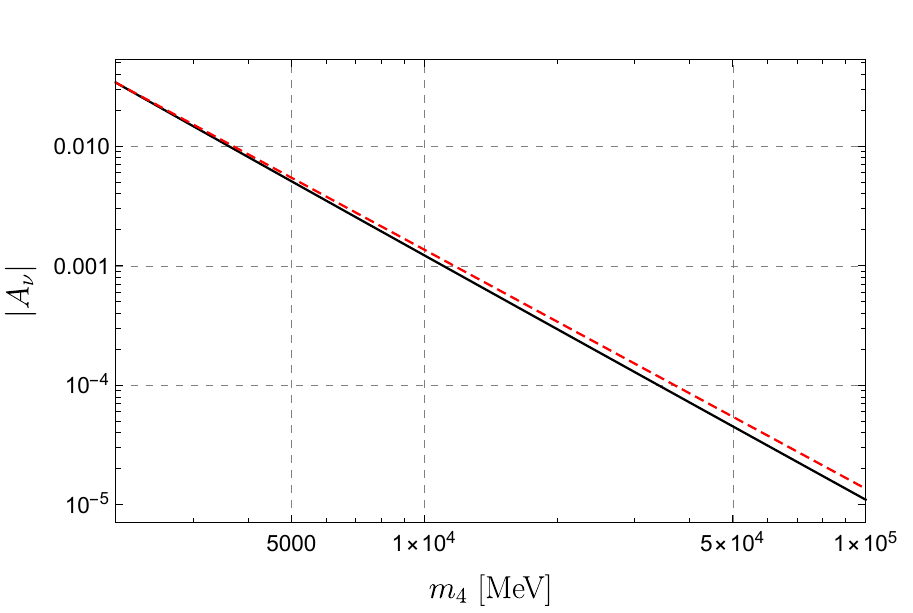}
\includegraphics[width=0.48\textwidth]{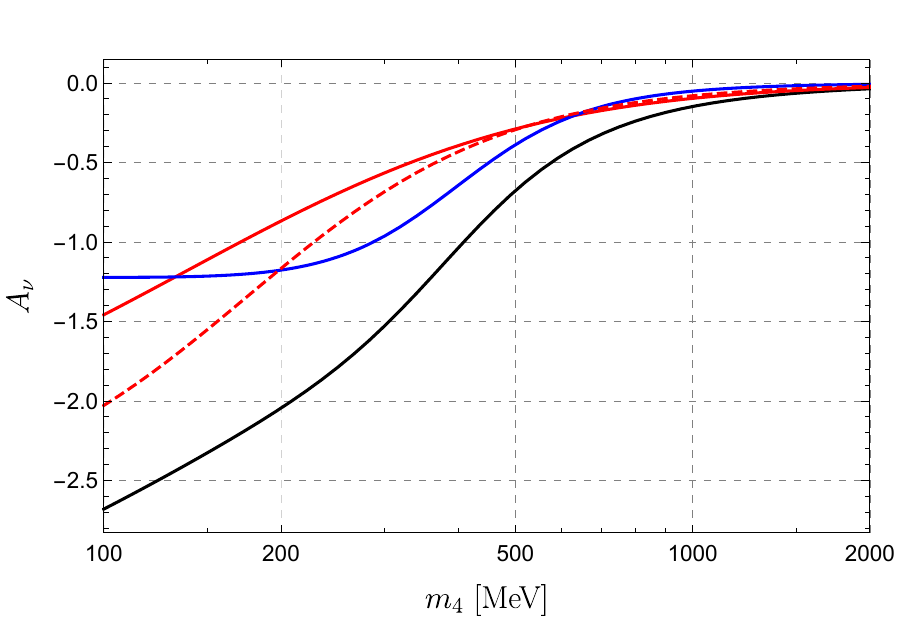}
\includegraphics[width=0.48\textwidth]{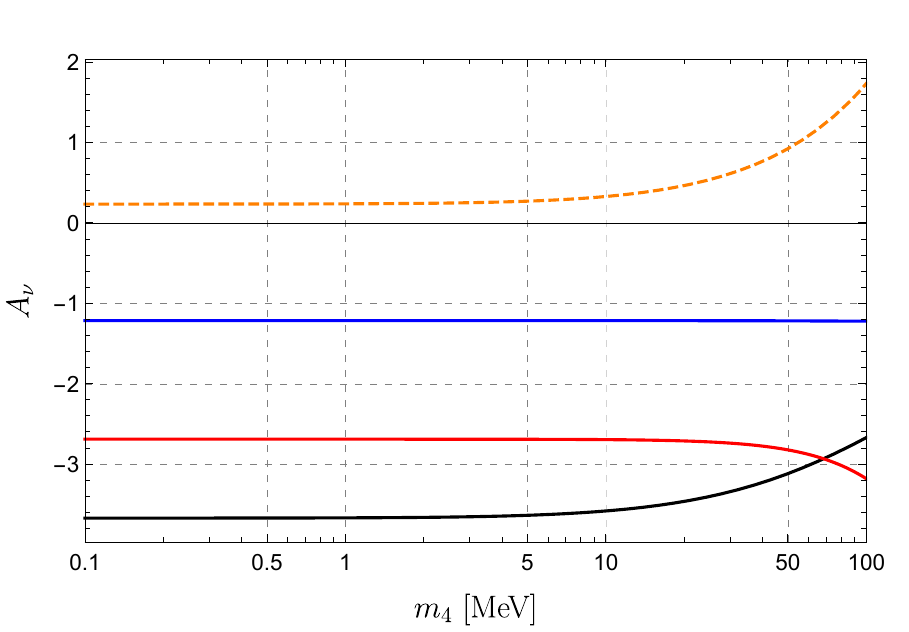}
\caption{ Contributions to the $0\nu\beta\beta$ amplitude from a neutrino with mass $m_4$. Top-left: total contribution derived here (black) compared to the literature result (dashed red). Top-right: amplitude in the heavy mass regime with (black) and without (dashed red) QCD renormalization group evolution. Bottom-left: total (black) amplitude in the intermediate-mass regime arising from potential (red) and hard (blue) neutrino exchange. The literature result for the total amplitude is shown by the dashed red curve. Bottom-right: total (black) amplitude in the light-mass regime consisting of potential (red), hard (blue), and ultrasoft (dashed orange) contributions.}
\label{Anu}
\end{figure}

To get a sense of the behavior of the amplitude in various $m_i$ regions, Fig.~\ref{Anu} shows $|A_\nu|$ for ${}^{136}$Xe, as induced by a single sterile neutrino. The top-left panel shows our result for the amplitude over a wide range of neutrino masses in solid black, compared to the commonly used parametrization of Eq.~\eqref{eq:naiveNME} in  red (dashed). Overall the shape is similar, but, as we will discuss below, the differences are important in specific scenarios. The top-right panel illustrates $A_\nu$ in the heavy $m_i$-region ($m_i \geq \mu_0 =2$ GeV) with (solid black) and without (dotted red) QCD renormalization-group evolution which is a minor effect. For instance, the ratio of $A_\nu(100\,\mathrm{GeV})/A^{\mathrm{no}\,\mathrm{RGE}}_\nu(100\,\mathrm{GeV})= 0.81$ implying a $20\%$ reduction of the amplitude. For $m_i=1$ TeV, the reduction grows to $25\%$. 
 
The bottom-left panel depicts the $100\,\mathrm{MeV}<m_i<\mu_0$ regime. The solid black line again denotes the total amplitude whereas the red  and blue  line denote, respectively, the potential and hard contribution. We also show the red dashed line for the commonly used parametrization. In this window of neutrino masses, the hard regime provides $\mathcal O(100\%)$ contributions, with the same sign, with respect to the usually considered potential contributions leading to faster $0\nu\beta\beta$ rates.
 
In the bottom-right panel we zoom in on the small $m_i$-regime, $m_i \leq 100$ MeV. Here we show in black the total amplitude, in red  the potential contribution, in blue  the hard contributions (which are essentially mass independent in this regime), and in orange (dashed) the ultrasoft contributions. 
We see that the latter are relatively small for small $m_4$, as predicted by power counting, and add destructively to the hard and potential regime. Despite being subleading, when considering contributions from a single $\nu_R$, they will play an important role when we consider the minimal $\nu$SM in which Eq.\ \eqref{eq:cancel} holds.
 
 \subsection{Current uncertainties and future improvements}
 \begin{figure}[t!]
	\centering
	\begin{subfigure}[b]{0.495\textwidth}
		\raggedleft
		\includegraphics[width=\textwidth]{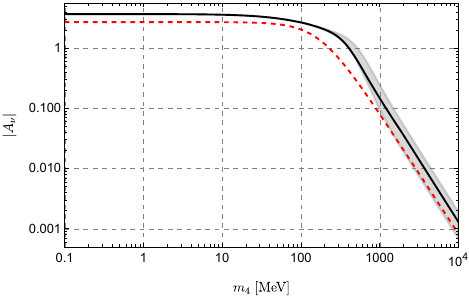}
	\end{subfigure}
	\hfill
	\begin{subfigure}[b]{0.495\textwidth}
		\raggedright
		\includegraphics[width=\textwidth]{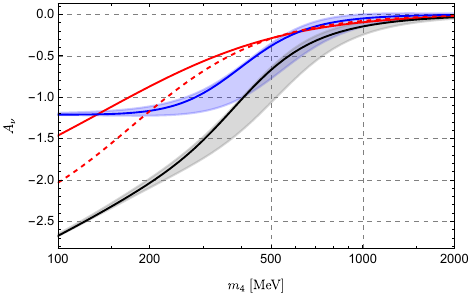}
	\end{subfigure}
	\caption{Uncertainties in $|A_\nu|$ and its constituents obtained by varying $g_1^{NN}$. The lines are the same as in the left panel of Fig.~\ref{Anu}, and the bands represent their variation when $g_1^{NN}$ is varied between 50\% and 150\% of $(1 + 3 g_A^2)/4$.}
	\label{fig:unc}
\end{figure}
 
While it correctly describes the leading contributions to $0\nu\beta\beta$ rates, Eq.\ \eqref{eq:fullint} involves several sources of uncertainty. In all mass regions, the amplitude depends on both hadronic and nuclear matrix elements (LECs and NMEs). Starting with the former, so far the only LEC that has been determined on the lattice is $g_1^{\pi\pi}$ \cite{Nicholson:2018mwc,Detmold:2020jqv,Detmold:2022jwu}, which appears in the $m_i>2$ GeV region. In the same mass range there appear two other LECs, $g_1^{\pi N}$ and $g_1^{N N}$. Instead, below $m_i<2$ GeV, the amplitude depends on a single LEC, $g_\nu^{NN}$, which has a nontrivial $m_i$ dependence unlike $g_1^{\pi \pi}$, $g_1^{\pi N}$, and $g_1^{N N}$. Although there are model- and NDA-estimates of $g_\nu^{NN}(0)$, $g_1^{\pi N}$, and $g_1^{N N}$, these LECs currently come with an $\Or(1)$ uncertainty. As a subset of the LECs appears for any $m_i$, these hadronic uncertainties in principle affect all mass regions. However, in cases where Eq.\ \eqref{eq:cancel} holds, we see from Table \ref{tab:scalings} that these poorly known matrix elements have a smaller impact for $m_i<k_F$, as the hard-neutrino contributions are expected to be subleading in this mass range.
Future LQCD determinations could significantly reduce the hadronic uncertainties, especially for $m_i>2$ GeV, where all contributions come with LECs, and $k_F<m_i<2$ GeV where the hard-neutrino contributions are a leading effect. 
Although the $m_i$-independent LECs $g_1^{\pi N}$, $g_1^{NN}$, and $g_\nu^{NN}(0)$ are not yet within reach of LQCD computations, their determination is part of a large ongoing effort, see e.g.\ Refs.\ \cite{Cirigliano:2019jig,Cirigliano:2022oqy} for an overview. Recent lattice studies have started to consider the $m_i$ dependence of LECs as well, although, so far, only in the meson sector \cite{Tuo:2022hft}.

Consider, as a demonstration, the effect of uncertainties on the amplitude $A_\nu$ by varying $g_1^{NN}$. 
Using Eq.\  \eqref{eq:gnu_int}, we vary $g_1^{NN}$ between $0.5(1+ 3 g_A^2)/4$ and $1.5 (1+3 g_A^2)/4$ while keeping all other parameters unchanged. The results shown in Fig.~\ref{fig:unc} may now be compared with the left panel of Fig.~\ref{Anu}. The variation affects only the dim-9 and hard (via $m_d$) contributions. The lines show the values considered previously, and the bands show the ``uncertainty'' in the amplitudes when $g_1^{NN}$ is varied within the range mentioned above. 

Likewise, NMEs are necessary in all mass regions. It is well known that many-body determinations of, for example the NMEs in the `standard scenario' of light-neutrino exchange, $\mathcal{M}(m_i=0)$, can vary by a factor of $\Or(1)$ between different methods, see e.g.\ Refs.~\cite{Engel:2016xgb,Agostini:2022zub}, and estimated uncertainties within particular many-body methods are at least about 50\%~\cite{Jokiniemi:2022ayc}. The nuclear uncertainties are similar for $m_i$-dependent NMEs~\cite{Agostini:2022zub} and therefore affect our results at the same level. Recently, several \textit{ab-inito} determinations have been able to reach the heavy isotopes that are used in experiments \cite{Yao2019Aug,Belley2020Aug,Novario2020Aug,Wirth:2021pij,Belley:2023btr,Belley:2023lec}. Further developments in this direction could provide calculations of NMEs with controlled error estimates \cite{Agostini:2022zub,Cirigliano:2022oqy}. Given the current uncertainties, such results would significantly improve the accuracy of our estimates in all mass regions.

\begin{figure}[t!]
	\centering
		\includegraphics[width=.65\textwidth]{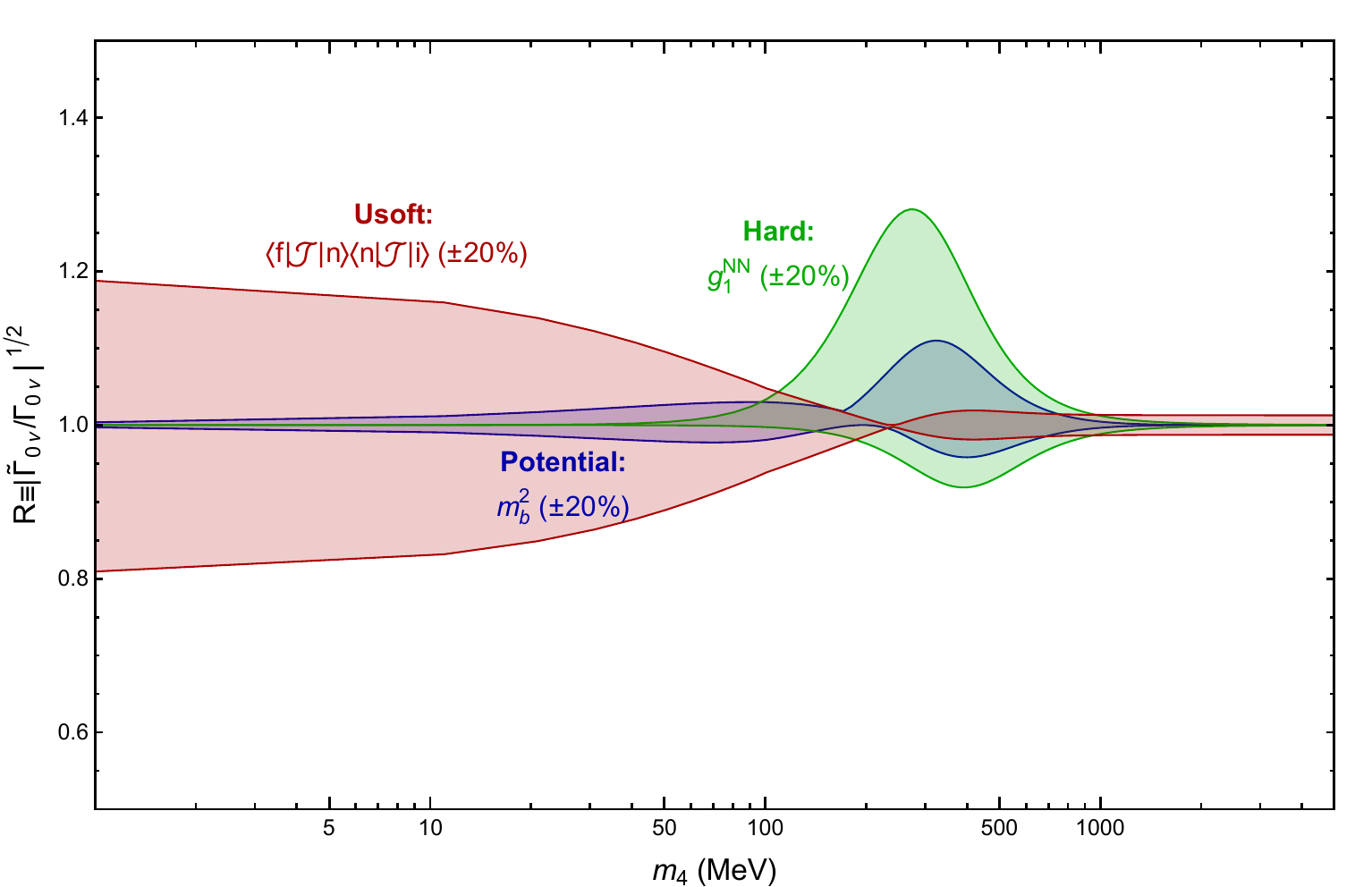}
	\caption{The relative change in the amplitude, $R$,  obtained by varying several hadronic and nuclear inputs in the $3+1$ scenario discussed in Section \ref{sec:3+1}. Here $R\equiv |\tilde\Gamma_{0\nu}/\Gamma_{0\nu}|^{1/2}$ where  $\Gamma^{}_{0\nu}$ is the original decay rate, while $\tilde\Gamma_{0\nu}$ is obtained by   varying the indicated hadronic or nuclear parameter.	
	The green, blue, and red bands are related to uncertainties in the hard, potential, and ultrasoft contributions. They are obtained by varying the LEC $g_1^{NN}$ in Eq.\ \eqref{A9}, $m_b^2$ in Eq.\ \eqref{eq:fitM}, and the NMEs $\langle 0^+_f|\mathcal J^\mu |1^+_n\rangle \langle 1^+_n| \mathcal J_\mu |0^+_i\rangle$  in Eq.\ \eqref{eq:ampusoftMSregion3}, by $\sim 20\%$ respectively. }
	\label{fig:unc2}
\end{figure}

As another example, we consider the impact of varying several hadronic and nuclear matrix elements in the case of a $3+1$ scenario, discussed in Section \ref{sec:3+1} below. This scenario implements the cancellation in Eq.\ \eqref{eq:cancel}, which allows us to study the impact of uncertainties in a minimal extension of the SM. We again vary the LEC $g_1^{NN}$, related to the dim-9 and hard regions, the parameter $m_b$, related to the potential contributions in Eq.\ \eqref{eq:fitM}, and the first-order nuclear matrix elements, needed for the ultrasoft contributions in Eq.\ \eqref{eq:ampusoftMSregion3}. To illustrate the impact of uncertainties from the different neutrino-momentum regions we show the relative change in the  amplitude for $^{136}$Xe after varying a parameter
as a function of the sterile neutrinos mass, $m_4$, in Fig.\ \ref{fig:unc2}. The y-axis shows the total modified amplitude, $\propto |\tilde \Gamma_{0\nu}|^{1/2}$, obtained after varying the input parameters, relative to the original amplitude, $\propto | \Gamma_{0\nu}|^{1/2}$, which can be written as $R\equiv|\tilde  \Gamma_{0\nu}/ \Gamma_{0\nu}|^{1/2}$.
The impact of varying $g_1^{NN}$, $m_b^2$, and the first-order NMEs by $20\%$ is shown in green, blue, and red, respectively. For illustration, all parameters were varied by the same amount. While the uncertainty on $g_1^{NN}$ is most likely larger than $20\%$, an estimate of the uncertainties in the potential and ultrasoft regions can be obtained from the consistency check discussed around Eq.\ \eqref{eq:usoftCheck}, which is consistent with an $\Or(20\%)$ uncertainty. As one might expect, these variations lead to a  $\sim 20\%$ effect at the level of the amplitude and peak in the region between $k_F<m_4<\Lambda_\chi$ for the uncertainties related to $m_b$ and $g_1^{NN}$, while the uncertainties in the first-order NMEs are most noticeable for small $m_4\lesssim k_F$.
The main purpose of Fig.\ \ref{fig:unc2} is to illustrate which hadronic or nuclear matrix elements are important in which mass region. We therefore stress that it does not  capture the significant theoretical uncertainties related to the NMEs and LECs required for the case of three active Majorana neutrinos, namely $\mathcal M(0)$ and $g_{\nu}^{NN}(0)$.

Finally, there are in principle errors due to missing higher orders in the $\chi$EFT expansion. In particular, the current work does not include the contributions from $A_\nu^{\rm (pot,2)}$, 
which involves additional NMEs and LECs. The induced corrections to the potential are known for $m_i=0$ \cite{Cirigliano:2017tvr} and have so far been estimated only in light nuclei \cite{Pastore:2017ofx}. From Table \ref{tab:scalings} we expect these contributions to be most relevant when $\Lambda_\chi\gtrsim m_i\gtrsim k_F$ and Eq.\ \eqref{eq:cancel} holds.

The main purpose of this work has been the systematic derivation of Eq.~\eqref{eq:fullint}  which describes the largest $0\nu\beta\beta$ contributions in the $\nu$SM in terms of well-defined QCD and nuclear matrix elements. It will be straightforward to update the expressions once more refined calculations of these matrix elements exist.

\section{Phenomenology}\label{sec:pheno}
\subsection{3+0}\label{sec:3+0}
We begin with the standard mechanism through the exchange of three light Majorana neutrinos (see Eq.\ \eqref{eq:nuSM}). 
Unlike the $\nu$SM, this scenario assumes that the masses of the neutrinos are generated by heavy beyond-the-SM fields that have been integrated out. This leads to Majorana masses for the active neutrinos without a cancellation mechanism as in Eq.\ \eqref{eq:cancel}.
The total amplitude is then given by the sum of Eqs.~\eqref{eq:AnuPot} and \eqref{eq:ampusoftMS} 
\begin{equation}
A_\nu = A_\nu^{\text{(pot)}}(0)+A_\nu^{\text{(hard)}}(0)+A_\nu^{\text{(usoft)}}(0)\,,
\end{equation}
and we can safely neglect the mass dependence of $A_\nu$. The three terms can be read from the bottom-right panel of Fig.~\ref{Anu}. The potential contribution, usually the only term considered in the literature, indeed provides the largest piece, $A_\nu^{\text{(pot)}}(0) = -2.7$ and $-3.4$ for $^{136}$Xe and $^{76}$Ge, respectively. The hard-neutrino exchange mechanism \cite{Cirigliano:2018hja, Cirigliano:2019vdj} provides a constructive $45(41)\%$ correction in $^{136}$Xe ($^{76}$Ge), given by $A_\nu^{\text{(hard)}}(0) = -1.2 (-1.4)$, and has been considered in various modern $0\nu\beta\beta$ computations \cite{Wirth:2021pij,Weiss:2021rig,Jokiniemi:2021qqv, Jokiniemi:2022ayc}.

The last term is new and provides a smaller, destructive, $10\%$ correction $A_\nu^{\text{(usoft)}}(0)=0.23(0.28)$  for $^{136}$Xe ($^{76}$Ge). While small, the contribution is somewhat larger than expected from power counting (see Table \ref{tab:scalings}) due to the large logarithms in Eq.\ \eqref{eq:Fusoft}, $\log (m_\pi/(2\Delta E_{1,2}))$, which are responsible for $\sim 70\%$ of the ultrasoft amplitude. Nonetheless, this contribution has the same sign and is of similar size as usual contributions to $0\nu\beta\beta$ beyond the closure approximation~\cite{Muto94,Senkov13,Senkov16}, which are related to the ultrasoft term as discussed in Sec.~\ref{Sec:contr}. As far as we are aware, this is the first calculation and analysis of the ultrasoft contributions to $0\nu\beta\beta$. 

\begin{figure}[t!]
\center
\begin{subfigure}[b]{0.49\textwidth}
\centering
\includegraphics[width=\textwidth]{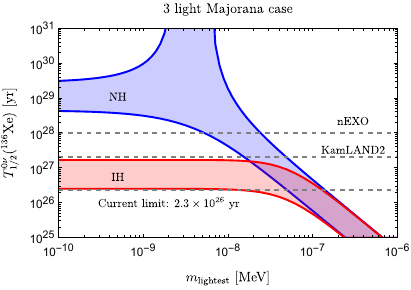}
\end{subfigure}
\begin{subfigure}[b]{0.49\textwidth}
\centering
\includegraphics[width=\textwidth]{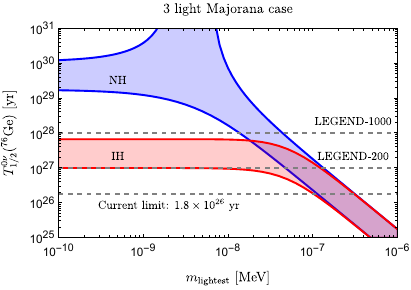}
\end{subfigure}
\caption{Predicted ${}^{136}$Xe (left panel) and ${}^{76}$Ge (right panel) $0\nu\beta\beta$ half-life in the normal (blue) and inverted (red) neutrino-mass hierarchy as function of the lightest neutrino mass,  marginalized with respect to the Majorana phases. The half-life includes contributions from potential, hard, and ultrasoft neutrino exchange calculated with the nuclear shell model. The current lower bounds on the half-life is shown in gray~\cite{KamLAND-Zen:2022tow,GERDA:2020xhi}, along with future prospects~\cite{Albert:2017hjq,Ichimura:2022kvl,Abgrall:2017syy}.}
\label{tau3p0}
\end{figure}

Using the usual parametrization of the PMNS matrix,
\bea 
U_{\rm PMNS} = 
\bma 
1 & 0 & 0\\
0 & c_{23} & s_{23}\\
0 & -s_{23} & c_{23}\\ 
\ema \cdot
\bma 
c_{13} & 0 & s_{13}e^{-i\dt}\\
0 & 1 & 0\\
-s_{13}e^{i\dt} & 0 & c_{13}\\ 
\ema\cdot
\bma 
 c_{12} & s_{12} & 0\\
-s_{12} & c_{12} & 0\\ 
0 & 0 & 1
\ema\cdot
\bma 
 1 & 0 & 0\\
0 & e^{i \al_1} & 0\\ 
0 & 0 & e^{i\al_2}
\ema\,,
\eea
where $s_{ij}=\sin\theta_{ij}$ and $c_{ij}=\cos \theta_{ij}$, and the PDG determinations of the mixing angles $\theta_{ij}$ and CP-violating phase $\delta$ \cite{Workman:2022ynf}, we obtain predictions for the half-life of ${}^{136}$Xe and ${}^{76}$Ge.
Figure~\ref{tau3p0} shows the results for the normal (NH) and inverted (IH) neutrino-mass hierarchy in blue and red, respectively. The width of the bands mostly arises from the variation of the Majorana phases, $\al_{1,2}$. Searches involving ${}^{136}$Xe are currently more sensitive; the figure shows that, assuming our calculated shell-model NMEs and neglecting theory uncertainties, the most recent KamLAND-Zen measurement is approaching the lower edge of the IH band.
Future bounds promise to probe the entirety of the IH band for both isotopes.

\subsection{Limits on $U_{e4}^2$}
\begin{figure}[t!]
\center
\includegraphics[width=0.8\textwidth]{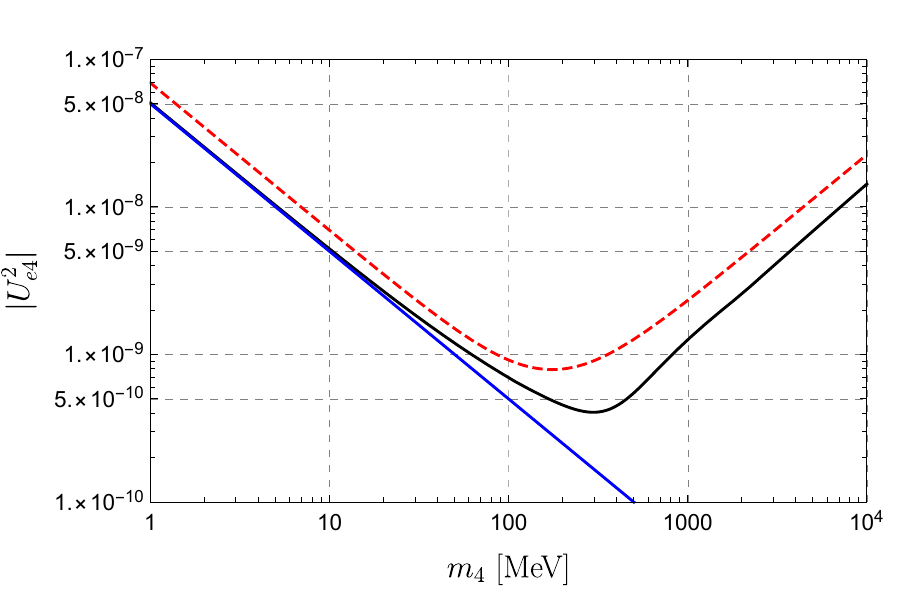}
\caption{ Limits on $U_{e4}^2$ as function of $m_4$ assuming the $0\nu\beta\beta$ rate is dominated by a single sterile neutrino. The black line corresponds to our result, while the dashed line corresponds to the usual procedure in the literature following Eq.~\eqref{eq:naiveNME}. The blue line denotes the naive seesaw relation $U_{e4}^2 = (0.05\,\mathrm{eV})/m_4$. }
\label{Ulim}
\end{figure}

Before moving on to the $\nu$SM, we consider a 4th sterile neutrino and assume its contribution saturates the $0\nu\beta\beta$ amplitude, i.e.\ we assume there are no cancellations with the active neutrinos. Such scenarios, without the cancellation mechanism in Eq.\ \eqref{eq:cancel}, are possible when going beyond minimal extensions of the SM, and are represented by Eq.\ \eqref{eq:mass2}.
The resulting limit (from ${}^{136}$Xe) on $U_{e4}^2$ as a function of $m_4$ is depicted by the solid line in Fig.~\ref{Ulim}, while the approach followed in the literature using Eq.~\eqref{eq:naiveNME} is shown by the dashed line. Our approach leads to tighter limits (up to a factor $2$ around $m_4=300$ MeV), especially in the MeV-GeV regime of $m_4$, because of hard-neutrino-exchange contributions. The limits reach the naive seesaw expectations (indicated by the blue line) for $m_4 \leq 10$ MeV. 

\begin{figure}[t!]
	\center
	\begin{subfigure}[b]{0.49\textwidth}
		\centering
		\includegraphics[width=\textwidth]{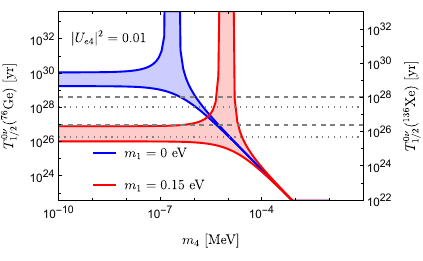}
	\end{subfigure}
	\begin{subfigure}[b]{0.49\textwidth}
		\centering
		\includegraphics[width=\textwidth]{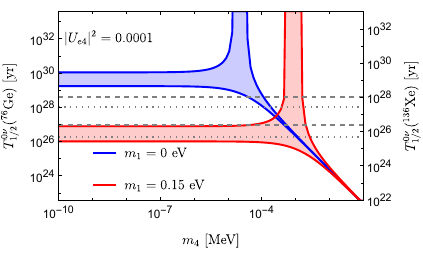}
	\end{subfigure}
	\caption{Predicted ${}^{136}$Xe and ${}^{76}$Ge $0\nu\beta\beta$ half-life in NH as a function of the light sterile neutrino mass $m_4$ for lightest active neutrino mass $m_1 = 0$ (blue) and $0.15$ eV (red), for two different values of the coupling strength $|U_{e4}|$. New funnels appear as the sterile contribution can contribute destructively to the decay rate. The current and future limits on $0\nu \beta\beta$ lifetime are shown in gray (dashed and dotted for ${}^{136}$Xe and ${}^{76}$Ge respectively)~\cite{KamLAND-Zen:2022tow,GERDA:2020xhi,Albert:2017hjq,Abgrall:2017syy}.}
	\label{fig:lightsterilefunnel}
\end{figure}

This assumption of the sterile contribution saturating the $0\nu\beta\beta$ amplitude breaks down for light sterile neutrino masses, where the sterile contribution can be of same size as that of active neutrinos. In this region, there can be cancellations between the contributions, leading to the possibility of new ``funnels'' in the $0\nu\beta\beta$ rate, depending on the values of the phases, neutrino masses, as well as possibly the nucleus in question~\cite{Abada:2018qok,Asaka:2020wfo,Asaka:2020lsx,Asaka:2021hkg}. Consider a scenario with one light sterile neutrino, and where the neutrinos acquire a mass through an unspecified UV mechanism. The relevant elements of the mixing matrix can be parametrized as 
\begin{align}
	U_{e1} = c_{12} \,c_{13} \,c_{14},\quad U_{e2} = c_{14} \,c_{13} \,s_{12} \,e^{\alpha_1},\quad U_{e3} = c_{14}\,s_{13}\,e^{\alpha_2},\quad U_{e4} = s_{14}\,e^{\alpha_3}\,, 
\end{align}
where $c_{ij}$ and $s_{ij}$ are $\cos(\theta_{ij})$ and $\sin(\theta_{ij})$ respectively. As shown in Fig.~\ref{fig:lightsterilefunnel} for NH, the decay rate can go to zero at certain values of the sterile mass which is determined by the strength of the coupling $U_{e4}$ and the mass of the lightest neutrino. The region beyond which the sterile neutrino starts to dominate the amplitude can also be clearly seen, as the bands (the width of which is given by the phases in the mixing matrix) shrink to a line. Smaller values of $|U_{e4}|$ cause the effect of the sterile contribution to kick in only at larger $m_4$. We also see that the $m_1 = 0.15$~eV case is ruled out (as expected from Fig.~\ref{tau3p0}) for either value of $|U_{e4}|$ unless there is a significant destructive contribution from the sterile neutrino, that is, until a funnel appears.

\begin{figure}[h!]
	\center
		\centering
		\includegraphics[width=0.6\textwidth,trim={1.2in 8.1in 3.2in 1in},clip]{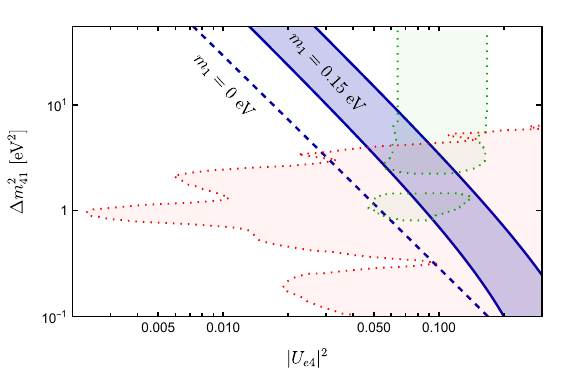}
	\caption{Bounds in the $\Delta m_{41}^2 - |U_{e4}^2|$ plane for $m_1 = 0$ (dashed line) and $m_1 = 0.15$ eV (solid lines) for light sterile neutrinos, assuming NH. The region to the right of the dashed line is excluded by current $0\nu \beta\beta$ limits for a massless lightest neutrino, while a scenario with $m_1 = 0.15$ eV is allowed by $0\nu \beta\beta$ limits only within the shaded region. In green and red are results from Ga anomaly ($2\sigma$ allowed region from a combined analysis) and DANSS (excluded region at 90\% CL)~\cite{Barinov:2021asz,DANSS:2018fnn}.}
	\label{fig:m4coupling}
\end{figure}

With this in mind, we show in Fig.~\ref{fig:m4coupling} the bounds on $|U_{e4}|^2$ as a function of $\Delta m_{41}^2 = m_4^2 - m_1^2$, assuming NH, for $m_1 = 0$ (dashed line) as well as $m_1 = 0.15$ eV (solid lines). The bounds are obtained by assuming that the active neutrino contributions to the decay rate add up with the same phase, while the sterile neutrino contributes with an opposite sign, thus making the upper bounds conservative. In case of $m_1 = 0.15$ eV we also obtain a lower bound on top of an upper limit (thus restricting the allowed region to the shaded area in Fig.~\ref{fig:m4coupling}) since the $0\nu \beta \beta$ rate gets oversaturated unless there is a partial cancellation from the sterile neutrino contribution, while only an upper bound can be drawn for $m_1 = 0$. This plane is often studied in the context of neutrino oscillation anomalies~\cite{Gariazzo:2018mwd,Dentler:2018sju,Giunti:2019aiy}. It is seen that with the assumption that neutrinos are Majorana particles, current $0\nu \beta \beta$ limits can rule out chunks of the preferred parameter region outlined in Refs.~\cite{Gariazzo:2018mwd,Dentler:2018sju}. For reference, we also show the $2\sigma$ contours from a combined analysis of Ga anomaly~\cite{Barinov:2021asz} in green, and the exclusion region from the DANSS experiment ($90\%$ CL)~\cite{DANSS:2018fnn} in red. We see that if the lightest neutrino is massless, current $0\nu\beta\beta$ limits alone exclude the region allowed by Ga anomaly almost entirely, and DANSS rules out the remaining sliver. For $m_1 = 0.15$ eV, there exists an allowed region that explains the Ga anomaly and is not excluded by DANSS, just below $\Delta m_{41}^2 = 10$ eV$^2$ and $|U_{e4}|^2 \lesssim 0.1$.

\subsection{3+1}\label{sec:3+1}
In this section we discuss the 3+1 scenario with 3 light left-handed neutrinos and one sterile neutrino. This model is not realistic as it predicts two massless neutrinos, $m_1=m_2=0$, but serves as a useful toy model to illustrate the importance of the newly identified contributions in scenarios where Eq.\ \eqref{eq:cancel} holds. We consider a mass matrix
\begin{align}
M_\nu = \bma 0 & 0 & 0 & M^*_{D,1}\\ 
		      0 & 0 & 0 & M^*_{D,2}\\
		      0 & 0 & 0 & M^*_{D,3}\\
		      M^*_{D,1} & M^*_{D,2}  & M^*_{D,3} & M_R		      
		    \ema\,,
\end{align}
and set  for simplicity $M^*_{D,1} = M^*_{D,2} = M^*_{D,3} \equiv M^*_{D}$. We diagonalize the mass matrix to obtain the PMNS matrix, which we parametrize as \cite{Giunti:2019aiy}
\bea
U &=&D_L R^{34}R^{24}R^{23}R^{14}R^{13}W^{12}D_R\,,\nn\\
 D_{L}&=&e^{i(\al_D+\al_R/2)}{\rm diag}(1,1,1,e^{-i(\al_R+\al_D)})\,\qquad D_{R}={\rm diag}(1,1,i,1)\,,
\eea 
where $\left[W^{ab}(\theta_{ab},\dt_{ab})\right]_{ij} =\dt_{ij}+ (\dt_{ia}\dt_{jb}e^{i\dt_{ab}}-\dt_{ib}\dt_{ja}e^{-i\dt_{ab}})s_{ab}+ (\dt_{ia}\dt_{ja}+\dt_{ib}\dt_{jb})(c_{ab}-1)$ and $R^{ab}(\theta_{ab}) =W^{ab} (\theta_{ab},0)$ and $\al_{D,R} = {\rm Arg}\,M_{D,R}$. We can now read off the relevant mixing angles for $0\nu\beta\beta$
\bea
U_{e3}^2=-\frac{m_4}{m_3}U_{e4}^2 =- \frac{1}{3}\frac{m_4}{m_3+m_4}e^{i(2\al_D+\al_R)}\,.
\label{eq:3plus1}
\eea
The phases $\alpha_D$ and $\alpha_R$ drop out in the decay rate and can be effectively set to zero. In this model we have
\beq
m_{\beta\beta} = m_3 U_{e3}^2 + m_4 U_{e4}^2 = 0\,,
\eeq
and we need to consider the mass dependence of the hadronic and nuclear matrix elements in order to get a non-zero $0\nu\beta\beta$ rate. 
For small $m_4 \ll k_F$ the non-vanishing combination of mixing angles and masses can be written as 
\beq
m_3 U_{e3}^2 A_\nu(0) + m_4 U_{e4}^2 A_\nu(m_4) =  m_3 U_{e3}^2 \left[A_\nu(0) - A_\nu(m_4)\right] \simeq - m_3 m_4 U_{e3}^2 A'_\nu(0)\,,
\eeq
and depends on the derivative of the amplitude as function of the neutrino mass. 

\begin{figure}[t!]
	\center
	\includegraphics[width=0.48\textwidth]{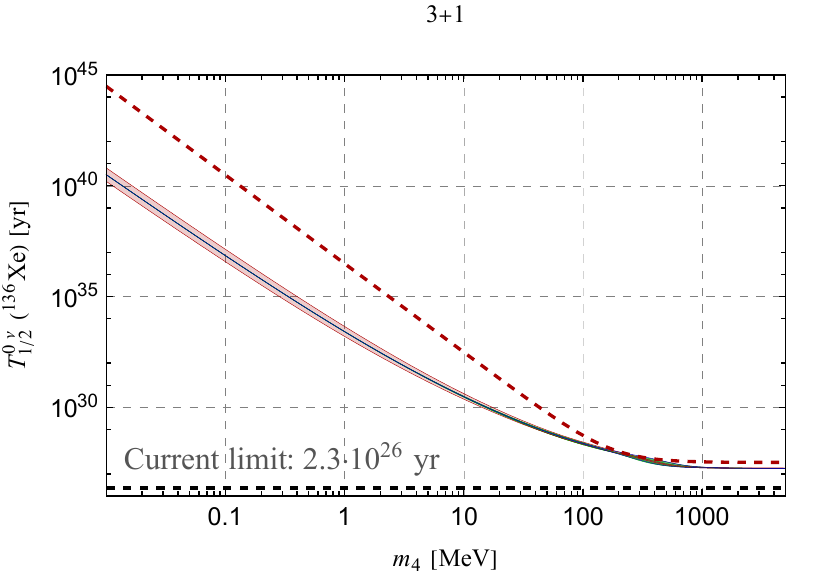}
		\includegraphics[width=0.48\textwidth]{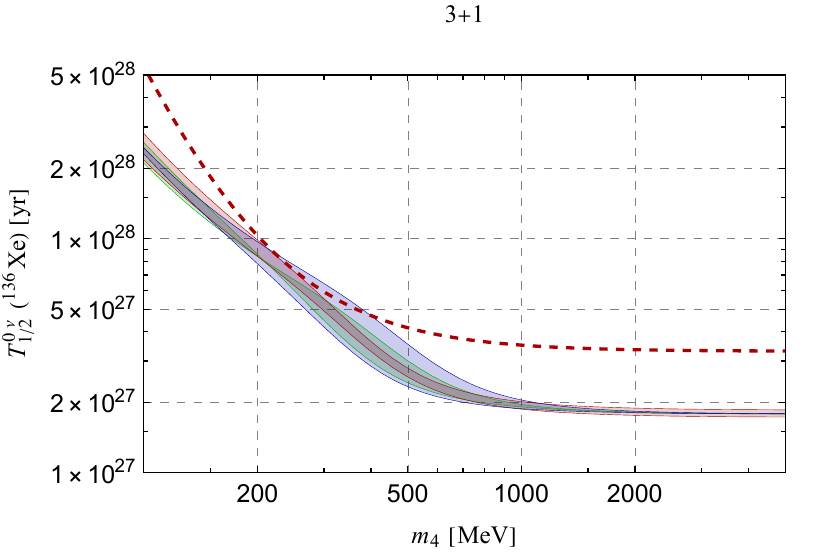}
	\caption{\NLDBD\, half-life of $^{136}{\text{Xe}}$ as a function of $m_4$ in the 3+1 model. In the left panel, the half-life is obtained by using the NMEs in  Eqs.\ \eqref{eq:naiveNME} (dashed red) and \eqref{eq:fullint} (black).  The right panel zooms in the heavy $m_4$ region. The coloured bands show uncertainty estimates similar to Fig.~\ref{fig:unc2}, but with parameters being allowed to vary up to 50\% from their central values.}
	\label{fig:Thalf3p1}
\end{figure}

We set $m_3\simeq 0.05$ eV, treating it as a light active neutrino, and vary $m_4$. The resulting lifetime is shown in Fig.~\ref{fig:Thalf3p1} where the right panel focuses on $m_4 >100$ MeV. The red dashed lines correspond to the approach in the literature using the parametrization of the NMEs in Eq.~\eqref{eq:naiveNME}. For light $m_4$ this leads to a decay rate that scales as $|m_4^3 U_{e4}^2|^2 \sim m_4^4$ such that the half-life grows as $m_4^{-4}$. The black solid line instead corresponds to the expressions obtained in this work leading to significantly shorter half-lives in the light $m_4$ regime because of the larger derivative arising from the ultrasoft term. The differences for $m_4 > 100$ MeV are less profound. In this case, $A_\nu(m_4) \ll A_\nu(0)$ and only $A_\nu(0)$ matters. The main difference then is our inclusion of hard neutrino exchange contributions leading to decay rate that is faster by about a factor of 2, as illustrated in the right panel. The coloured bands in both panels indicate the estimated uncertainties; see Fig.~\ref{fig:unc2}. The parameters here are allowed to vary by 50\% from the central values used in this work. Only uncertainties from new pieces are shown to aid comparison with the standard approach in red, which also suffers from uncertainties in NMEs similar to our approach.

\subsection{A pseudo-Dirac scenario}\label{sec:pseudo}
Another interesting way to generate neutrino masses is through the inverse seesaw mechanism \cite{Mohapatra:1986bd,Mohapatra1986Feb,Nandi1986Feb}. This model and its generalizations \cite{Dev:2012sg,BhupalDev:2012jvh,Bolton:2019pcu} add two types of singlets to the SM, leading to a neutrino mass matrix of the form,
\bea\label{eq:inv_seesaw}
M_\nu = 
\bma
0& m_D & 0\\
m_D^T & \mu_X & m_S\\
0 & m_S^T & \mu_S
\ema\,,
\eea
which is a special case of Eq.\ \eqref{eq:mass}, with the upper right part $\bma m_D & 0\ema = M_D^*$ and the lower right block with $\mu_X$, $\mu_S$, $m_S$, and $m_S^T$ forming $M_R^\dagger$.
Assuming we add the same number, $n_S$, of each type of singlet, the block matrices in $M_\nu$ become a $3\times n_S$ matrix in the case of $m_D$, while $\mu_X$, $m_S$, and $\mu_S$ are $n_S\times n_S$ matrices. 
The minimal inverse seesaw scenario additionally assumes $\mu_X=0$, which, together with the hierarchy $m_S\gg m_D,\, \mu_S$, leads to the following mass matrix for the active neutrinos
\bea
m_{\nu,{\rm light}}\simeq -  m_D(m_S^T)^{-1} \mu_S m_S^{-1}m_D^T\,.
\eea
Unlike the usual seesaw scenario, where the active neutrinos masses are inversely proportional to the Majorana mass of the sterile neutrinos, here the light neutrino masses are proportional to a small LNV parameter, $\mu_S$. 
The new singlets lead to states that can be organized in pairs; each pair consists of two Majorana neutrinos with $\Or(m_S)$ masses and an $\Or(\mu_S)$ mass splitting.  
The assumption $m_S\gg m_D,\, \mu_{X,S}$ implies that lepton number is an approximate symmetry. This case is often referred to as pseudo-Dirac, as it reduces to a scenario with purely Dirac neutrinos in the limit of $\mu_{X,S}\to 0$. Variants of these models appear in scenarios of low-scale leptogenesis, see e.g. Refs.~\cite{Canetti:2012kh,Drewes:2021nqr, Hernandez:2022ivz} and references therein. 

Given the small mass splitting between the added neutrinos in these scenarios, it is useful to note that the contributions to $0\nu\bt\bt$ simplify whenever the sterile neutrinos are nearly degenerate.
To see this, we can rewrite the effective amplitude relevant for $0\nu\beta\beta$ by using $\sum_{i=1}^{n+3} U_{ei}^2 m_i =0$ and Taylor-expanding the sterile contributions around their common mass scale.
In cases with $n$ nearly-degenerate sterile neutrinos, this leads to
\bea\label{eq:degen}
A_{\mathrm{eff}} &=& 
\sum_{i=1}^3 U_{ei}^2 m_i A_\nu(0)+\sum_{i=4}^{n+3} U_{ei}^2 m_i A_\nu(m_i)\nn\\
&=& \sum_{i=1}^3 U_{ei}^2 m_i \left[A_\nu(0)-A_\nu (M_1)\right]+M_1\sum_{i=4}^{n+3} U_{ei}^2  \Delta_i A'_\nu(M_1)+\Or(\Delta_i^2)\,,
\eea
where we introduced $M_i=m_{i+3}$ for the masses of the sterile neutrinos,  $\Delta_{i+3} = M_i - M_1 \ll M_{1}$, and $A'_\nu(M_1)$ denotes $ (d A_\nu(m)/ d m) |_{m=M_1}$.  As the first term scales with the light neutrino masses, the second term can dominate if the mixing angles are not too small. Thus, whenever the sterile neutrinos are nearly degenerate and dominate over the active contributions, the decay rate is determined by the {\it derivative} of $A_\nu$, instead of the $0\nu\bt\bt$ amplitude itself, and scales with the mass differences $\Delta_i$.

In order to assess the impact of our determination of $A_\nu$ on these scenarios, we specify to a toy model with a single active neutrino and $n_X = n_S = 1$ \cite{Bolton:2019pcu} which captures the main features of a pseudo-Dirac sterile neutrino pair. The resulting mass matrix can be parametrized as
\bea
M_\nu = U^* {\rm diag}(m_\nu,M_1,M_2)\, U^\dagger\,,
\eea
with $m_\nu$ and $M_{1,2}$ the masses of the light and sterile neutrinos, respectively, and
\bea 
U_{} = 
\bma 
1 & 0 & 0\\
0 & c_{12} & s_{12}\\
0 & -s_{12} & c_{12}\\ 
\ema \cdot
\bma 
c_{e2} & 0 & s_{e2}e^{-i\dt}\\
0 & 1 & 0\\
-s_{e2}e^{i\dt} & 0 & c_{e2}\\ 
\ema\cdot
\bma 
 c_{e1} & s_{e1} & 0\\
-s_{e1} & c_{e1} & 0\\ 
0 & 0 & 1
\ema\cdot
\bma 
 1 & 0 & 0\\
0 & e^{i \al_1} & 0\\ 
0 & 0 & e^{i(\al_2+\dt)}
\ema\,,
\eea
in terms of three mixing angles, $\theta_{12},\, \theta_{e1}$, and $\theta_{e2}$ and three phases, $\dt$, $\al_{1,2}$. 
$M_\nu$ is then made up of nine parameters (three masses with six angles and phases), which are subject to four constraints $\left(M_\nu\right)_{11}=\left(M_\nu\right)_{13,31}=\left(M_\nu\right)_{22}=0$ in the pseudo-Dirac case.
\begin{figure}[t!]
	\center
		\includegraphics[width=0.47\textwidth]{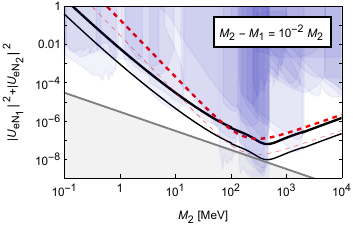}\hfill
		\includegraphics[width=0.47\textwidth]{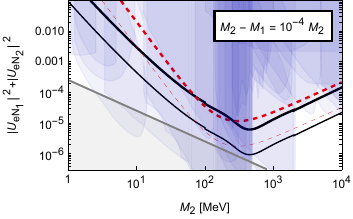}
	\caption{Limits on the mixing angles, $|U_{e1}|^2+|U_{e2}|^2$, as a function of $M_2$ in the pseudo-Dirac scenario discussed in Sec.\ \ref{sec:pseudo}. The left (right) panel shows the case in which the mass splitting between the sterile neutrinos is $\frac{M_2-M_1}{M_2}=10^{-2}\, \left(10^{-4}\right)$. The thick (thin) lines correspond to the current (future, $T_{1/2}^{0\nu}\geq 10^{28}$yr) constraints, obtained using Eq.\ \eqref{eq:naiveNME} (dashed red) or Eq.\ \eqref{eq:fullint} (black). }
	\label{fig:pseudo}
\end{figure}
Here we focus on a scenario with the benchmark values
\bea
m_\nu = 2.6 \times 10^{-3} \, {\rm eV}\,,\qquad    \al_1=0\,,\qquad  \al_2 = \pi/2\,,
\eea
where the choice of $\al_2$ induces a relative sign between the contributions of the two sterile neutrinos, allowing them to act as a pseudo-Dirac pair. The constraint $\left(M_\nu\right)_{11}=0$ can be used to eliminate the combination $s_{e1}^2-s_{e2}^2$ in favor of $|U_{e1}|^2+|U_{e2}|^2\simeq s_{e1}^2+s_{e2}^2 \equiv s_+^2$ and $M_{1,2}$, which gives $s_{e1}^2\simeq s_{e2}^2$ up to $\Or(m_\nu/M_1)$ and $\Or(\Delta/M_1)$ corrections. 
As we assume the sterile neutrinos to be nearly degenerate we can use Eq.\ \eqref{eq:degen} with $\Delta\equiv \Delta_5=M_2-M_1$, $\Delta_4=0$, and  $U_{e3}^2 \simeq -s_{e2}^2\simeq -s_+^2/2$, which allows us to set constraints on $s_+^2$ as a function of $M_1$ and $\Delta$.

Together with the current lower limit on the half life of $^{136}$Xe \cite{KamLAND-Zen:2022tow}, the above gives rise to Fig.\ \ref{fig:pseudo} which shows the constraints on $s_{e1}^2+s_{e2}^2$ as a function of the sterile neutrino mass, $M_2$. Here the left and right panels depict scenarios with different mass splittings, $\frac{M_2-M_1}{M_2}=10^{-2}$ and $\frac{M_2-M_1}{M_2}=10^{-4}$, respectively. The black lines show the current and projected limits obtained using the approach discussed in this work, while the red lines again depict the results using Eq.\ \eqref{eq:naiveNME}.
The blue regions are excluded by other laboratory constraints~\cite{Bolton:2019pcu}. Below the kaon mass, the strongest limits come from missing energy experiments which probe $|U_{\alpha}|^2$ through invisible decays of the pion or kaon~\cite{PIENU:2017wbj,NA62:2020mcv}, and inverse beta decays~\cite{Friedrich:2020nze,Borexino:2013bot}. At higher masses, the limits come from displaced vertex searches which probe long-lived sterile neutrinos via their decay to SM particles~\cite{Barouki:2022bkt,DELPHI:1996qcc}. The gray region shows the part of parameter space that does not satisfy the inverse seesaw expectation, $\left(M_\nu\right)_{22}\sim\left(M_\nu\right)_{33}=\Or(M_2-M_1)$,  and corresponds to $|\left(M_\nu\right)_{22}|=|\mu_X| > 3(M_2-M_1)$. These results can be obtained by using the constraint $\left(M_\nu\right)_{13}=0$ to determine $s_{12}$ and $\dt$, after which $\left(M_\nu\right)_{22}$ and $\left(M_\nu\right)_{33}$ are fixed for a given point in Fig.\ \ref{fig:pseudo}.
We find that the expectation $\left(M_\nu\right)_{22}\sim\left(M_\nu\right)_{33}=\Or(M_2-M_1)$, can roughly be satisfied whenever $s_+^2\gtrsim \frac{m_\nu}{M_2-M_1}$.

The parametric dependence of the limits in Fig.~\ref{fig:pseudo} can be understood from Eq.\ \eqref{eq:degen}. Assuming the $m_\nu$ terms can be neglected, we have
\bea
A_{\mathrm{eff}} &=&-\frac{s_{+}^2}{2}M_1 \Delta  A'_\nu(M_1)+\Or(\Delta^2,\, m_\nu)\,.
\eea
The limit on the $0\nu\beta\beta$ rate then sets a bound on $s_+^2$ as function of $M_1$ and $\Delta$ through
\beq
s^2_+ < \frac{2 m_e}{g_A^2V_{ud}^2 } \sqrt{\frac{ 1}{ G_{01}T_{1/2}^{0\nu} }}\times \frac{1}{M_1 \,\Delta\,|A'_\nu(M_1)|}\,,
\eeq
which agrees well with the bounds we find numerically. This form explains the $\Delta^{-1}$ scaling seen in Fig.~\ref{fig:pseudo}. Second, the limits depend on the derivative of the neutrino amplitude with respect to the sterile neutrino mass. We already saw in the $3+1$ analysis that the ultrasoft corrections identified in this work lead to a larger slope compared to the usual approach, explaining the significantly tighter limits, about an order of magnitude for $M_1 =1$ MeV. For larger masses $M_1 > 200$ MeV we again obtain stronger limits (roughly a factor 2.5 for $M_1=400$ MeV), mainly because of the hard-neutrino-exchange contributions. As mentioned above, this analysis is not specific for the $1+1+1$ toy model discussed here and the dependence on $A'_\nu$, rather than $A_\nu$, holds for more general scenarios involving pseudo-Dirac sterile neutrino pairs. The main conclusion here is that ultrasoft and hard contributions can have a significant impact on $0\nu\beta\beta$ rates in the $\nu$SM including well-motivated variants involving low-scale leptogenesis.  

\subsection{3+2 model}
\label{sec:3+2}
A more realistic model from a phenomenological point of view is a scenario with two right-handed neutrinos, such that it can explain neutrino oscillations (and thus their light masses), as well as address the matter-antimatter asymmetry of the universe via leptogenesis. However, the presence of two sterile neutrinos allows only two light neutrino masses, so that the lightest neutrino is necessarily massless in this case. One option that can reproduce the light neutrino masses is the hierarchical limit, in which the heavy neutrino masses are required to be $\gtrsim 10^{9}$~GeV~\cite{Davidson:2002qv}. Here we will instead consider scenarios where the sterile neutrinos are pseudo-degenerate in mass, which allows for much lower mass scales compatible with leptogenesis. 

The neutrino oscillation data are most easily implemented in this scenario with the Casas-Ibarra (CI) parametrisation~\cite{Casas:2001sr}, in which the Yukawa coupling matrix between the left-handed lepton doublet and the right-handed neutrino is given by
\begin{align}
	 Y_\nu^\dagger \propto U_\text{PMNS} \sqrt{m_\nu^d}\,\mathcal{R} \,\sqrt{M^d}\,,
\end{align}
where $m_\nu^d$ is a diagonal matrix of light neutrino masses and $M^d$ is a diagonal matrix of the (two) heavy neutrino masses. $\mathcal{R}$ is a complex orthogonal matrix given by
\begin{align}
	\mathcal{R}_\text{NH} = \left(\begin{array}{cc}
		0 & 0 \\
		\cos(\omega)  & \sin(\omega) \\
		-\sin(\omega) & \cos(\omega)
	\end{array}\right)\,,\quad \mathcal{R}_\text{IH} = \left(\begin{array}{cc}
		\cos(\omega)  & \sin(\omega) \\
		-\sin(\omega) & \cos(\omega) \\
		0 & 0
	\end{array}\right)\,,
\end{align}
with $\omega\in\mathbb{C}$ for normal and inverted hierarchies respectively. The PMNS matrix is parametrised as before, but with only one effective Majorana phase given by $\eta = \frac{1}{2}(\alpha_{21}-\alpha_{31})$ for NH and $\eta = \frac{1}{2}\alpha_{21}$ for IH. With this, the active-sterile mixing angles can be approximated to be, assuming NH,
\begin{align}
	|U_{ei}|^2\approx \frac{6.2 \times 10^{-10}}{M_i/\text{MeV}}\mathrm{e}^{2\text{Im}(\omega)}\,,
	\label{eq:Uei}
\end{align}
where the NuFIT central values for light neutrino masses and mixing angles have been used~\cite{Esteban:2020cvm,NuFIT}, and we have picked Re$(\omega) = \eta =0$, $\delta = \frac{3 \pi}{2}$. These parameters enter the effective amplitude as arguments of oscillating functions and thus can only play a limited role in enhancing the rate; see Eq.~\eqref{eq:effamp3+2} below.

In the pseudo-degenerate regime, we can trade $M_1$ and $M_2$ for the average mass $M_N = (M_1 + M_2)/2$ and the mass difference $\Delta M = (M_2-M_1)/2$. Since the active-sterile mixing angles depend exponentially on Im$(\omega)$, it is clear that this parameter plays a major part in the possibility of $0\nu\beta\beta$ rate enhancement. However, at $\mathcal{O}(\Delta M^0)$, Im$(\omega)$ does not enter the effective amplitude and sterile neutrino exchange only suppresses the rate. For example, the effective amplitude in NH is given by
\begin{align}
		A_\text{eff} \approx \quad & m_3 \sin^2(\theta_{13}) \mathrm{e}^{-2i(\delta + \eta)} \left[A_\nu(m_3)-A_\nu(M_N)\right] + m_2 \sin^2(\theta_{12}) \cos^2(\theta_{13}) \left[A_\nu(m_2)-A_\nu(M_N)\right] \nonumber\\
		+ &\Delta M A_\nu'(M_N) \mathrm{e}^{-2i(\delta + \eta)} \cdot \nonumber\\
		&\Big[\cosh(-2i\omega)\left(m_2 \cos^2(\theta_{13}) \sin^2(\theta_{12}) \mathrm{e}^{2i(\delta+\eta)} - m_3 \sin^2(\theta_{13})\right) \nonumber\\
		&- 2 i \mathrm{e}^{i(\delta + \eta)} \sqrt{m_2 m_3} \sin^2(\theta_{12}) \sin^2(\theta_{13}) \cos^2(\theta_{13}) \sinh(-2i\omega)\Big]\,,
		\label{eq:effamp3+2}
\end{align}
where we  explicitly show the dependence of the amplitude in Eq.~\eqref{eq:degen} on CI parameters. Clearly, Im$(\omega)$ comes into play at first order in $\Delta M$, and thus there is an interplay of these parameters that dictate the $0\nu\beta\beta$ lifetime. 

Given that there are quite stringent bounds on the mixing angles across a wide range of heavy neutrino masses, we must be careful when choosing a value for Im$(\omega)$. 
The strongest bounds are set for masses around $M_N \approx 300$ MeV, where values of Im$(\omega) \approx3$ can be excluded \cite{Bolton:2019pcu}. Nevertheless, we can safely have larger values for Im$(\omega)$ for other values of heavy neutrino masses, as the bounds sharply weaken away from $M_N\simeq 300$ MeV. 
To illustrate our results we therefore set  {Im$(\omega) \approx 4.5$}. Note that this essentially fixes the strength of $U_{ei}$ for a sterile neutrino of a given mass. 

\begin{figure}[t!]
	\center
	\includegraphics[width=0.55\textwidth]{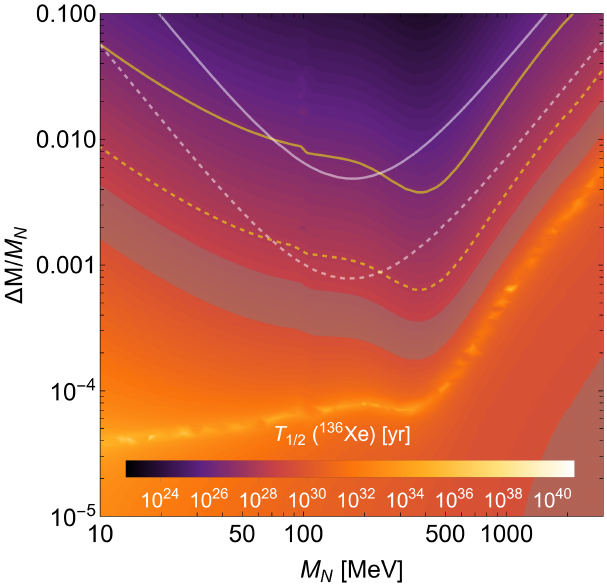}
	\caption{$0\nu\beta\beta$ lifetime using Eq.~\eqref{eq:fullint} as a function of the averaged heavy sterile mass ($M_N$) and the mass splitting $(\Delta M)$ for the pseudo-degenerate 3+2 NH scenario discussed in Sec.~\ref{sec:3+2}. The parameters chosen are $\delta = 3\pi/2,\, \eta = \text{Re}(\omega) =0,\, \text{Im}(\omega)=4.5$. The yellow lines correspond to the current and projected (solid and dashed respectively) limits on the lifetime calculated using Eq.~\eqref{eq:fullint}, and in white are the results obtained when Eq.~\eqref{eq:naiveNME} is used. The shaded regions are the bands in Fig.~\ref{tau3p0}.}
	\label{fig:contour}
\end{figure}

By fixing the parameters appearing in the $\mathcal{R}$ matrix and the unknown phases (a more comprehensive scan and the interplay with low-scale leptogenesis is studied in Ref.~ \cite{deVries:2024rfh}), we can study the $0\nu\beta\beta$ lifetime as a function of $M_N$ and $\Delta M$. Such an example for NH is shown in Fig.~\ref{fig:contour}, where we have picked $\delta = 3\pi/2,\, \eta = \text{Re}(\omega) =0,\, \text{Im}(\omega)=4.5$. The solid (dashed) yellow lines are the current (predicted) limits on the $0\nu\beta\beta$ lifetime of ${}^{136}$Xe (from KamLAND-Zen and nEXO, respectively), calculated using Eq.~\eqref{eq:fullint}. The points are color coded and show the lifetimes calculated using the same expression. For comparison, we show in white the limits one would obtain using Eq.~\eqref{eq:naiveNME} instead.  

The regions above the solid (dashed) lines are (will be) excluded by current (future) $0\nu\beta\beta$ constraints. It is seen that the new effects included in Eq.~\eqref{eq:fullint} give rather interesting differences, especially away from $M_N \sim 300$ MeV. Although the new formula enhances the decay rates at large masses, allowing us to probe smaller mass splittings, the slope is very similar to the white lines. 
Similar to Sec.\ \ref{sec:pseudo}, we find that our approach differs by about a factor of 2 from that of the literature in the large $M_N$ limit for this particular set of parameters. For smaller $M_N$ the slopes differ significantly due to the ultrasoft contributions. 

The gray-shaded regions correspond to lifetimes predicted by the 3 light Majorana-neutrino case shown in Fig.~\ref{tau3p0}. In this region, the effects of the pseudo-degenerate neutrino species on the \NLDBD~will thus be difficult to disentangle from the active neutrino contribution. In the bottom right region, for higher $M_N$ and small mass splittings, the sterile neutrinos start to decouple and the lifetime begins to fall into the aforementioned band. For the NH, the gray bands will not be probed in the near future, and any signal would thus point towards the existence of sterile neutrinos. 

Instead of either dominating (e.g., in the case of large mass splittings and Im$(\omega)$) or having a sub-leading effect (much smaller compared to the active neutrino contribution) on \NLDBD, sterile neutrinos can also contribute destructively to the \NLDBD~rate; i.e., the contribution of sterile neutrinos to $A_\text{eff}$ can be of similar size to the contribution of active neutrinos, but opposite in sign. This is emphasized by the thin curve of yellow-to-white points below the gray band, which indicate a very large half-life. This curve represent a new ``funnel'' where $0\nu\beta\beta$ is strongly suppressed even when the lightest active neutrino is massless. The amount of cancellation is highly dependent on the parameters that have been fixed in Fig.~\ref{fig:contour}, and again we point out that this is not a comprehensive study of this scenario, but rather an illustration of the importance of newly-identified contributions to \NLDBD.

\section{Conclusions}\label{sec:conclusion}
The Standard Model extended with several gauge-singlet right-handed neutrinos (the $\nu$SM) is a very promising framework that can solve several of the shortcomings of the vanilla SM. Right-handed neutrinos 
can account for neutrino masses through the seesaw mechanism while, at the same time, accommodating the universal matter/anti-matter asymmetry through leptogenesis. The possibility of low-scale leptogenesis has led to an increased interest in the search for relatively light sterile neutrinos at the GeV scale. 
One of the generic features of the $\nu$SM is the violation of lepton number which can be detected in $0\nu\beta\beta$ experiments. 
Essentially all analyses of $0\nu\bt\bt$ are based on the same computational framework, in which the effect of sterile neutrino masses is incorporated by modifying the propagator of the neutrino that is exchanged between two neutrons in a nucleus. 
As these modifications are only made in the LNV potentials used in nuclear many-body calculations, they capture the 
contributions from neutrinos with momenta of potential scaling, $k_0\ll |{\bf k}|\sim k_F$. The resulting NMEs depend on the sterile neutrino mass, which are assumed to take a simple functional form, see Eq.~\eqref{eq:naiveNME}.  

Considering the interest in detecting sterile neutrinos, in this work, we have taken a fresh look at these computations using a recently developed EFT framework for $0\nu\beta\beta$. We find several important new contributions that can significantly alter the $0\nu\beta\beta$ rates in seesaw models involving light sterile neutrinos. The most important findings and applications of our work are:
\begin{itemize}
\item Our main result is a practical, and relatively easy-to-use, formula for the $0\nu\beta\beta$ contribution from a sterile neutrino of any mass $m_i$ given in Eq.~\eqref{eq:fullint}. This formula includes the effects of potential, hard, and ultrasoft modes for $m_i \lesssim 2$ GeV, and the correct QCD renormalization-group-evolution and matching to hadronic scales for $m_i  \gtrsim $ 2 GeV. We advocate the use of this formula for all $0\nu\beta\beta$ analyses of sterile neutrino models. Although we have focused on the isotopes ${}^{136}$Xe and $^{76}$Ge using shell-model NMEs, the same expressions can be straightforwardly extended to other isotopes and many-body approaches. We stress that the addition of potential, hard, and ultrasoft contributions only makes sense within a single nuclear many-body framework as regulators have to be applied consistently. 
\item As a byproduct of this expression, we presented up-to-date nuclear Shell Model predictions for the ${}^{136}$Xe $0\nu\beta\beta$ and ${}^{76}$Ge half-lives in the standard light-neutrino exchange mechanism for both the normal and inverted neutrino-mass hierarchy in Sec.\ \ref{sec:3+0}. Our results indicate that the current KamLAND-Zen limit reaches the bottom edge of the IH band. These predictions include potential, hard, and ultrasoft contributions. As far as we are aware, this is the first time the latter have been estimated, which we find to give $10\%$ destructive corrections.
\item The same expression applies to contributions from sterile neutrinos allowing us to apply our $0\nu\beta\beta$ formula to several simplified scenarios, such as the $3+1$ model and a case with one active and a single pseudo-Dirac $\nu_R$ pair, as well as the realistic $3+2$ model in Secs.\ \ref{sec:3+1}, \ref{sec:pseudo}, and \ref{sec:3+2}. In all cases we find significant differences in the $0\nu\beta\beta$ predictions compared to the traditional approach, which can lead to enhancements of up to two orders of magnitude depending on the neutrino mass. The formula in Eq.~\eqref{eq:fullint} can serve as the basis for the analysis of $0\nu\beta\beta$ in models of low-scale leptogenesis to update predictions in, for example, Refs.~\cite{Drewes:2016lqo, Drewes:2016jae,Hernandez:2022ivz}. 
\end{itemize}

The main difference of our approach to the traditional literature is the systematic application of $\chi$EFT and the associated power counting to identify the dominant contributions to the $0\nu\beta\beta$ amplitude for different regimes of sterile neutrino masses. We list the most important newly identified effects as well as the required input from LQCD and nuclear many-body calculations to obtain more accurate predictions:
\begin{itemize}
\item $0\nu\beta\beta$ contributions from sterile neutrinos with masses $m_i  \gtrsim $ 2 GeV cannot be obtained from naively extrapolating NME results to heavy neutrino masses through Eq.~\eqref{eq:naiveNME}. Instead, as discussed in Sec.\ \ref{heavy}, heavy sterile neutrinos must be integrated out at the quark level leading to lepton-number-violating dimension-9 operators. After evolving the effective LNV operators to low energies where QCD becomes nonperturbative, they can be matched to LNV hadronic operators without neutrinos. 
\item  For a sterile neutrino with a mass $m_i  \lesssim $ 2 GeV there are, in addition to the usually considered potential contributions, leading-order contributions associated to the exchange of hard neutrinos, with momenta $k_0\sim |{\bf k}|\sim \Lambda_\chi$, see Secs.\ \ref{sec:below_lambda_chi} and \ref{sec:below_kf}.
Hard contributions are nowadays considered in modern $0\nu\beta\beta$ calculations for the exchange of very light Majorana neutrinos, but they can play an even bigger role for massive sterile neutrinos, see the analyses in Sec.~\ref{sec:pheno}. 
\item In the final region, for light sterile neutrino masses, $m_i \lesssim k_F \sim 100$ MeV, we have identified important new contributions associated with the exchange of neutrinos with ultrasoft momenta, $k_0\sim |{\bf k}|\ll k_F$, in Sec.\  \ref{sec:below_kf}. While they formally appear at next-to-next-to-leading order in the $\chi$EFT power counting, they can become dominant in minimal seesaw models, due to the cancellation of Eq.\ \eqref{eq:cancel} affecting the leading-order terms. The inclusion of the ultrasoft contributions requires new NMEs, involving a set of  excited states of the intermediate nucleus of the decay. These ultrasoft modes can lead to an enhancement of $0\nu\beta\beta$ rates up to two orders of magnitude compared to the usually included potential modes, in parts of parameter space.
\item As always, $0\nu\beta\beta$ computations involve uncertainties which are traditionally estimated by differences between NMEs obtained with different nuclear many-body methods \cite{Engel:2016xgb,Agostini:2022zub}. The sterile neutrino contributions involve additional hadronic and nuclear matrix elements that are not always (accurately) known. The most important targets for improvements, for example through LQCD calculations, consist of the sterile neutrino mass dependence of $g_\nu^{NN}(m_i)$, and the mass-independent low-energy constants $g_1^{\pi N}$ and $g_1^{NN}$. Future determinations of these matrix elements can be directly inserted into Eq.~\eqref{eq:fullint} and would allow one to improve the accuracy of the estimates presented here. Likewise, significant reductions in the uncertainty of the NMEs could be achieved by future nuclear {\it ab-initio} calculations or by measuring processes related to $0\nu\beta\beta$~\cite{Shimizu2018,Yao:2022usd,Romeo:2021zrn}, which would allow one to update Eq.~\eqref{eq:fullint} as well.
Finally, it would be good to confirm the values of the matrix elements in Tables~\ref{tab:overlapNME} and \ref{tab:overlapNME_Ge}, associated to the ultrasoft contributions, with other nuclear many-body approaches, and to evaluate  these NMEs for additional isotopes.

\end{itemize}

\section*{Acknowledgments}

We acknowledge interesting discussions with Marco Drewes, Yannis Georis, Juraj Klaric, and Michele Lucente, and thank Pieter Mumm for the exclusion region data in the sterile neutrino oscillation phase space. JdV acknowledges support from the Dutch Research Council (NWO) in the form of a VIDI grant. WD acknowledges support from the U.S. DOE under Grant No.
DE-FG02-00ER41132. DC, JM and PS acknowledge support funded by MCIN/AEI/10.13039/501100011033 from the following grants: PID2020-118758GB-I00; RYC-2017-22781 through the ``Ram\'on y Cajal'' program funded by FSE ``El FSE invierte en tu futuro''; CNS2022-135716 funded by the ``European Union NextGenerationEU/PRTR''; and CEX2019-000918-M to the ``Unit of Excellence Mar\'ia de Maeztu 2020-2023'' award to the Institute of Cosmos Sciences.
EM is supported by the U.S. Department of Energy through the Los Alamos National Laboratory and by the Laboratory Directed Research and Development program of Los Alamos National Laboratory under project number 20230047DR. Los Alamos National Laboratory is operated by Triad National Security, LLC, for the National Nuclear Security Administration of U.S. Department of Energy (Contract No. 89233218CNA000001).

\appendix
\newpage
\section{Neutrino potential in coordinate space}\label{sec:coordinate}

Starting from the neutrino potential as given in Eq.\ \eqref{eq:potential_usoft},
which is typically used in the literature for the evaluation of sterile neutrino corrections to $0\nu\beta\beta$,
it is possible to show that the term linear in the sterile neutrino mass can be expressed 
as the NME of a potential independent of $r_{ab}$
\begin{equation}\label{eq:app.1}
 \left. \frac{d}{d m_i} \mathcal{M}(m_i) \right|_{m_i = 0}  = \frac{R_A}{2 g_A^2}
 \langle 0^+_f |  \sum_{a, b} \left( 1 - g_A^2 \boldsigma^{(a)}\cdot \boldsigma^{(b)} \right) \tau^{(a)+}\tau^{(b)+} | 0^+_i \rangle\,.
 \end{equation}
For the experimentally relevant transitions where the initial and final nuclei are in different isospin multiplets with $\Delta T=2$, the Fermi component of this NME vanishes, up to isospin-breaking corrections. The Gamow-Teller piece, proportional to the double Gamow-Teller NME~\cite{Shimizu2018}, is in general non-zero, even though it is suppressed in calculations with approximate $SU(4)$ spin-isospin symmetry.  
It can be determined by computing the normalization of the Gamow-Teller density, typically a byproduct of standard NME calculations, and therefore Eq.~\eqref{eq:app.1} offers an alternative way of subtracting the linear $m_i$ dependence from $A_\nu^{(\rm pot)}$.

To avoid the spurious linear dependence, it might be convenient to evaluate Eq.\ \eqref{eq:potential_usoft2} directly in coordinate space. 
\begin{equation}
 A_\nu^{(\rm pot)}(m_i) = A_\nu^{(\rm pot)}(0)  - \frac{m_i^2}{m_\pi^2} \left( -\frac{1}{g_A^2}\mathcal{M}^{(2)}_F
 + \mathcal{M}_{GT}^{(2)} + \mathcal{M}_T^{(2)}\right)\,,
\end{equation}
with
\begin{align}
 \mathcal{M}_F^{(2)} &= R_A  \langle 0^+_f |  \sum_{a,b} V^{(2)}_F(r_{ab}) \tau^{(a)+}\tau^{(b)+} | 0^+_i \rangle\,, \\
 \mathcal{M}_{GT}^{(2)} & = R_A \langle 0^+_f |  \sum_{a,b} V^{(2)}_{GT}(r_{ab}) \boldsigma^{(a)}\cdot \boldsigma^{(b) } \tau^{(a)+}\tau^{(b)+} | 0^+_i \rangle \,,\\
 \mathcal{M}_T^{(2)} &= R_A \langle 0^+_f |  \sum_{a,b} V^{(2)}_T(r_{ab}) S^{ab} \tau^{(a)+}\tau^{(b)+} | 0^+_i \rangle\,,
\end{align}
with $S^{ab} =  3  \boldsigma^{(a)} \cdot {\bf r}^a  \boldsigma^{(a)} \cdot {\bf r}^b  - \boldsigma^{(a)}\cdot \boldsigma^{(b) } $. 
The radial functions are given by
\begin{align}
 V^{(2)}_F(r) &=   - m_\pi  \frac{ z}{2}\,, \\
 V^{(2)}_{GT}(r) &= - m_\pi \left(  \frac{z}{2} - \frac{1}{6 z} \left( e^{- z} ( 4 + z ) - 4      \right) \right) \,,\\
 V^{(2)}_T(r) & = m_\pi \left(3+z\right)\frac{e^{-z}\left(6+4 z+z^2\right)+2z -6}{6 z^3}\,,
 \end{align}
where $z=m_\pi r$.

\section{Neutrino mass dependence of $g_{\nu}^{NN}$}
\label{app:gnuNN}

We comment here on the neutrino mass dependence of $g_{\nu}^{NN}$.
Short distance contributions to $0\nu\beta\beta$ are captured, at lowest order, by the Lagrangian 
\begin{equation}\label{eq:LagApp}
\mathcal L^{NN}_{|\Delta L| = 2} =  - \left(2 \sqrt{2} G_F V_{ud}\right)^2\, m_{\beta\beta} \bar e_L C \bar e_L^{T} \, g_{\nu}^{NN}\, \left[ \left(N^T P^{+}_{^1S_0} N\right)\, \left(N^T P^-_{^1S_0} N\right) \right] + \textrm{H.c.}  + \ldots, 
\end{equation}
where $\ldots$ denote terms with two or more pion fields, which are required by chiral symmetry, 
and  $P^{\pm}_{^1S_0} = (P^{1}_{^1S_0} \pm i P^{2}_{^1S_0} )/2$ are projectors in the $^1S_0$ wave,
\begin{equation}
P^a_{^1 S_0} = \frac{1}{\sqrt{8}} \tau^2 \tau^a \, \sigma^2.
\end{equation}
Multiple insertions of neutrino masses do not change the chiral properties of the operator, and thus can be captured by allowing $g_{\nu}^{NN}$ to depend on $m_i$.
For $m_i \ll \Lambda_\chi$, we can write
\begin{equation}
g_{\nu}^{NN}(m_i) = g_{\nu, 0}^{NN}  + m_i^2 g_{\nu, 2}^{NN} + \ldots 
\end{equation}
According to the rules of naive dimensional analysis \cite{Manohar:1983md}, the scaling of the short-distance operators is given by 
\begin{equation}
\left. g_{\nu, 0}^{NN}\right|_{\rm NDA} = \mathcal O\left(\frac{1}{\Lambda_\chi^2} \right), \qquad \left. g_{\nu, 2}^{NN}\right|_{\rm NDA} = \mathcal O\left(\frac{1}{\Lambda_\chi^4} \right), 
\end{equation}
with each additional power of $m_i$ compensated by $\Lambda_\chi$.
A $g_{\nu, 0}^{NN}$ of this size is needed to absorb divergences in loop corrections to the neutrino potential \cite{Cirigliano:2017tvr}.
Refs.\ \cite{Cirigliano:2018hja} and \cite{Cirigliano:2019vdj} pointed out that the renormalization of LO LNV scattering amplitudes require 
the promotion of $g_{\nu, 0}^{NN}$ to leading order, that is
\begin{equation}
g_{\nu, 0}^{NN} = \mathcal O \left(\frac{1}{F_\pi^2}\right).
\end{equation}
The arguments in Refs.\ \cite{Cirigliano:2018hja,Cirigliano:2019vdj} were derived keeping only linear terms in $m_i$ (leading to the factor of $m_{\bt\bt}$ in Eq.\ \eqref{eq:LagApp}), but can be generalized to a massive neutrino propagator. It is easy to see that the LO logarithmic divergence does not have $m_i$ dependence beyond an overall $m_i^1$ factor. At NLO, 
there appear additional linearly  divergent integrals  \cite{Cirigliano:2019vdj}, but again additional factors of $m_i$ only change the finite pieces and do not affect the structure of the divergence.
From these results, we can argue that $g_{\nu, 2}^{NN} < (F_\pi^3 \Lambda_\chi)^{-1}$.
A full N$^2$LO analysis of LNV scattering amplitudes in chiral or pionless EFT has so far not been performed, and is beyond the scope of this paper. 
To estimate the scaling of $g_{\nu, 2}^{NN}$ here we focus on the corrections to the neutrino potential mediated by the weak magnetic moment $g_M$
\begin{equation}\label{eq:mag}
V_{\nu}^{\rm mag} = \tau^{(a)+} \tau^{(b) + } \times (4 G_F^2 V_{ud}^2) \sum_i \frac{U_{ei} m_i}{{\bf k}^2 + m_i^2} \left(  
- g_M^2 \frac{{\bf k}^2}{6 m^2_N} \left(\boldsigma^{(a)}\cdot \boldsigma^{(b)} + \frac{1}{2} S^{ab}(\hat{\bf k}) \right)
\right),
\end{equation}
with $g_{M} = 4.7$, and $S^{ab}(\hat{\bf k})$ is the momentum-space version of the tensor operator defined in Section \ref{sec:coordinate}.
Eq.\ \eqref{eq:mag} is the only N$^2$LO contribution proportional to $g_M$, and can thus be considered in isolation, without carrying out a full N$^2$LO calculation.
The tensor component does not contribute in the $^1S_0$ channel. The GT contribution shifts the value of $g_{\nu, 0}^{NN}$
\begin{equation}
g_{\nu, 0}^{NN} \rightarrow  g_{\nu, 0}^{NN}  - \frac{g_M^2}{4 m_N^2},
\end{equation}
and gives rise to a Yukawa-like potential of the form 
\begin{equation}\label{eq:mag2}
V_{\nu}^{\rm mag} = \tau^{(a)+} \tau^{(b) + } \times (4 G_F^2 V_{ud}^2) \sum_i \frac{U_{ei} m_i}{{\bf k}^2 + m_i^2} \times 
g_M^2 \frac{m_i^2}{6 m^2_N} \left(\boldsigma^{(a)}\cdot \boldsigma^{(b)}  
\right).
\end{equation}
When evaluated on $^1S_0$ wavefunctions, this potential leads to logarithmic divergences proportional to $m_i^3$. These need to be absorbed by $g_{\nu, 2}^{NN}$, which thus scales as 
\begin{equation}\label{eq:gnu2NN}
g_{\nu, 2}^{NN} \sim g_M^2 \left(\frac{C_{^1S_0}}{4\pi}\right)^2  = \mathcal O\left(\frac{1}{F_\pi^2 \Lambda_\chi^2}\right),
\end{equation}
where $C_{^1S_0} = \mathcal O(F_\pi^{-2})$ is the leading order strong interaction short-range coupling in the $^1S_0$ channel.
Eq.\ \eqref{eq:gnu2NN} justifies the scaling assumed in the main text.

\bibliographystyle{utphysmod}
\bibliography{bibliography}

\providecommand{\href}[2]{#2}\begingroup\raggedright\begin{thebibliography}{100}

\bibitem{Fukugita:1986hr}
M.~Fukugita and T.~Yanagida, Phys. Lett. B {\bfseries 174}, 45 (1986).

\bibitem{Asaka:2005pn}
T.~Asaka and M.~Shaposhnikov, Phys. Lett. {\bfseries B620}, 17 (2005),
[\href{https://arxiv.org/abs/hep-ph/0505013}{{arXiv:hep-ph/0505013}}].

\bibitem{Dodelson:1993je}
S.~Dodelson and L.~M. Widrow, Phys. Rev. Lett. {\bfseries 72}, 17 (1994),
  [\href{https://arxiv.org/abs/hep-ph/9303287}{{arXiv:hep-ph/9303287}}].

\bibitem{Glashow:1959wxa}
S.~L. Glashow, Nucl. Phys. {\bfseries 10}, 107 (1959).

\bibitem{Salam:1959zz}
A.~Salam and J.~C. Ward, Nuovo Cim. {\bfseries 11}, 568 (1959).

\bibitem{Weinberg:1967tq}
S.~Weinberg, Phys. Rev. Lett. {\bfseries 19}, 1264 (1967).

\bibitem{Superkamiokande1998}
Y.~Fukuda {\em et~al.} [Super-Kamiokande Collaboration], Phys. Rev. Lett.
  {\bfseries 81}, 1562 (1998).

\bibitem{SNO:2001kpb}
Q.~R. Ahmad {\em et~al.} [SNO Collaboration], Phys. Rev. Lett. {\bfseries 87},
  071301 (2001),
  [\href{https://arxiv.org/abs/nucl-ex/0106015}{{arXiv:nucl-ex/0106015}}].

\bibitem{KamLAND:2002uet}
K.~Eguchi {\em et~al.} [KamLAND Collaboration], Phys. Rev. Lett. {\bfseries
  90}, 021802 (2003),
  [\href{https://arxiv.org/abs/hep-ex/0212021}{{arXiv:hep-ex/0212021}}].

\bibitem{Workman:2022ynf}
R.~L. Workman and Others [Particle Data Group], PTEP {\bfseries 2022}, 083C01
  (2022).

\bibitem{Weinberg:1979sa}
S.~Weinberg,
Phys. Rev. Lett. {\bfseries 43}, 1566 (1979).

\bibitem{Abdullahi:2022jlv}
A.~M. Abdullahi {\em et~al.}, J. Phys. G {\bfseries 50}, 020501 (2023),
  [\href{https://arxiv.org/abs/2203.08039}{{arXiv:2203.08039~[hep-ph]}}].

\bibitem{Shaposhnikov:2006nn}
M.~Shaposhnikov, Nucl. Phys. {\bfseries B763}, 49 (2007),
[\href{https://arxiv.org/abs/hep-ph/0605047}{{arXiv:hep-ph/0605047}}].

\bibitem{Shaposhnikov:2008pf}
M.~Shaposhnikov, JHEP {\bfseries 08}, 008 (2008),
  [\href{https://arxiv.org/abs/0804.4542}{{arXiv:0804.4542~[hep-ph]}}].

\bibitem{Canetti:2012kh}
L.~Canetti, M.~Drewes, T.~Frossard, and M.~Shaposhnikov, Phys. Rev. D
  {\bfseries 87}, 093006 (2013),
  [\href{https://arxiv.org/abs/1208.4607}{{arXiv:1208.4607~[hep-ph]}}].

\bibitem{Canetti:2012vf}
L.~Canetti, M.~Drewes, and M.~Shaposhnikov, Phys. Rev. Lett. {\bfseries 110},
  061801 (2013),
[\href{https://arxiv.org/abs/1204.3902}{{arXiv:1204.3902~[hep-ph]}}].

\bibitem{Drewes:2016jae}
M.~Drewes, B.~Garbrecht, D.~Gueter, and J.~Klaric, JHEP {\bfseries 08}, 018
  (2017),
  [\href{https://arxiv.org/abs/1609.09069}{{arXiv:1609.09069~[hep-ph]}}].

\bibitem{Drewes:2017zyw}
M.~Drewes, B.~Garbrecht, P.~Hernandez, M.~Kekic, J.~Lopez-Pavon, J.~Racker,
  N.~Rius, J.~Salvado, and D.~Teresi, Int. J. Mod. Phys. A {\bfseries 33},
  1842002 (2018),
  [\href{https://arxiv.org/abs/1711.02862}{{arXiv:1711.02862~[hep-ph]}}].

\bibitem{Drewes:2021nqr}
M.~Drewes, Y.~Georis, and J.~Klari\'c, Phys. Rev. Lett. {\bfseries 128}, 051801
  (2022),
  [\href{https://arxiv.org/abs/2106.16226}{{arXiv:2106.16226~[hep-ph]}}].

\bibitem{Boyarsky:2018tvu}
A.~Boyarsky, M.~Drewes, T.~Lasserre, S.~Mertens, and O.~Ruchayskiy, Prog. Part.
  Nucl. Phys. {\bfseries 104}, 1 (2019),
  [\href{https://arxiv.org/abs/1807.07938}{{arXiv:1807.07938~[hep-ph]}}].

\bibitem{Bolton:2022tds}
P.~D. Bolton, F.~F. Deppisch, M.~Rai, and Z.~Zhang,
  [\href{https://arxiv.org/abs/2212.14690}{{arXiv:2212.14690~[hep-ph]}}].

\bibitem{PIENU:2011aa}
M.~Aoki {\em et~al.} [PIENU Collaboration], Phys. Rev. D {\bfseries 84}, 052002
  (2011), [\href{https://arxiv.org/abs/1106.4055}{{arXiv:1106.4055~[hep-ex]}}].

\bibitem{Belle:2013ytx}
D.~Liventsev {\em et~al.} [Belle Collaboration], Phys. Rev. D {\bfseries 87},
  071102 (2013),
  [\href{https://arxiv.org/abs/1301.1105}{{arXiv:1301.1105~[hep-ex]}}],
  [Erratum: Phys.Rev.D 95, 099903 (2017)].

\bibitem{NA62:2020mcv}
E.~Cortina~Gil {\em et~al.} [NA62 Collaboration], Phys. Lett. B {\bfseries
  807}, 135599 (2020),
  [\href{https://arxiv.org/abs/2005.09575}{{arXiv:2005.09575~[hep-ex]}}].

\bibitem{Bryman:2019bjg}
D.~A. Bryman and R.~Shrock, Phys. Rev. {\bfseries D100}, 073011 (2019),
[\href{https://arxiv.org/abs/1909.11198}{{arXiv:1909.11198~[hep-ph]}}].

\bibitem{KATRIN:2022ith}
M.~Aker {\em et~al.} [KATRIN Collaboration], Phys. Rev. D {\bfseries 105},
  072004 (2022),
  [\href{https://arxiv.org/abs/2201.11593}{{arXiv:2201.11593~[hep-ex]}}].

\bibitem{KamLAND-Zen:2022tow}
S.~Abe {\em et~al.} [KamLAND-Zen Collaboration], Phys. Rev. Lett. {\bfseries
  130}, 051801 (2023),
  [\href{https://arxiv.org/abs/2203.02139}{{arXiv:2203.02139~[hep-ex]}}].

\bibitem{Abgrall:2017syy}
N.~Abgrall {\em et~al.} [LEGEND Collaboration], AIP Conf. Proc. {\bfseries
  1894}, 020027 (2017),
[\href{https://arxiv.org/abs/1709.01980}{{arXiv:1709.01980~[physics.ins-det]}}].

\bibitem{Albert:2017hjq}
J.~B. Albert {\em et~al.} [nEXO Collaboration], Phys. Rev. {\bfseries C97},
  065503 (2018),
[\href{https://arxiv.org/abs/1710.05075}{{arXiv:1710.05075~[nucl-ex]}}].

\bibitem{Agostini:2022zub}
M.~Agostini, G.~Benato, J.~A. Detwiler, J.~Men\'endez, and F.~Vissani, Rev.
  Mod. Phys. {\bfseries 95}, 025002 (2023),
  [\href{https://arxiv.org/abs/2202.01787}{{arXiv:2202.01787~[hep-ex]}}].

\bibitem{Adams:2022jwx}
C.~Adams {\em et~al.},
  [\href{https://arxiv.org/abs/2212.11099}{{arXiv:2212.11099~[nucl-ex]}}].

\bibitem{Blennow:2010th}
M.~Blennow, E.~Fernandez-Martinez, J.~Lopez-Pavon, and J.~Men\'endez, JHEP
  {\bfseries 07}, 096 (2010),
[\href{https://arxiv.org/abs/1005.3240}{{arXiv:1005.3240~[hep-ph]}}].

\bibitem{Mitra:2011qr}
M.~Mitra, G.~Senjanovi\'c, and F.~Vissani, Nucl. Phys. {\bfseries B856}, 26
  (2012),
[\href{https://arxiv.org/abs/1108.0004}{{arXiv:1108.0004~[hep-ph]}}].

\bibitem{Li:2011ss}
Y.~F. Li and S.-s. Liu, Phys. Lett. {\bfseries B706}, 406 (2012),
[\href{https://arxiv.org/abs/1110.5795}{{arXiv:1110.5795~[hep-ph]}}].

\bibitem{deGouvea:2011zz}
A.~de~Gouv\^{e}a and W.-C. Huang, Phys. Rev. {\bfseries D85}, 053006 (2012),
[\href{https://arxiv.org/abs/1110.6122}{{arXiv:1110.6122~[hep-ph]}}].

\bibitem{Faessler:2014kka}
A.~Faessler, M.~Gonz\'alez, S.~Kovalenko, and F.~\v{S}imkovic, Phys. Rev. D
  {\bfseries 90}, 096010 (2014),
  [\href{https://arxiv.org/abs/1408.6077}{{arXiv:1408.6077~[hep-ph]}}].

\bibitem{Barea:2015zfa}
J.~Barea, J.~Kotila, and F.~Iachello, Phys. Rev. {\bfseries D92}, 093001
  (2015),
[\href{https://arxiv.org/abs/1509.01925}{{arXiv:1509.01925~[hep-ph]}}].

\bibitem{Giunti:2015kza}
C.~Giunti and E.~M. Zavanin, JHEP {\bfseries 07}, 171 (2015),
[\href{https://arxiv.org/abs/1505.00978}{{arXiv:1505.00978~[hep-ph]}}].

\bibitem{Asaka:2011pb}
T.~Asaka, S.~Eijima, and H.~Ishida, JHEP {\bfseries 04}, 011 (2011),
[\href{https://arxiv.org/abs/1101.1382}{{arXiv:1101.1382~[hep-ph]}}].

\bibitem{Asaka:2013jfa}
T.~Asaka and S.~Eijima, PTEP {\bfseries 2013}, 113B02 (2013),
[\href{https://arxiv.org/abs/1308.3550}{{arXiv:1308.3550~[hep-ph]}}].

\bibitem{Asaka:2016zib}
T.~Asaka, S.~Eijima, and H.~Ishida, Phys. Lett. {\bfseries B762}, 371 (2016),
[\href{https://arxiv.org/abs/1606.06686}{{arXiv:1606.06686~[hep-ph]}}].

\bibitem{Dekens:2023iyc}
W.~Dekens, J.~de~Vries, E.~Mereghetti, J.~Men\'endez, P.~Soriano, and G.~Zhou,
  Phys. Rev. C {\bfseries 108}, 045501 (2023),
  [\href{https://arxiv.org/abs/2303.04168}{{arXiv:2303.04168~[hep-ph]}}].

\bibitem{deGouvea:2005er}
A.~de~Gouv\^{e}a, Phys. Rev. {\bfseries D72}, 033005 (2005),
[\href{https://arxiv.org/abs/hep-ph/0501039}{{arXiv:hep-ph/0501039}}].

\bibitem{Beneke:1997zp}
M.~Beneke and V.~A. Smirnov, Nucl. Phys. B {\bfseries 522}, 321 (1998),
  [\href{https://arxiv.org/abs/hep-ph/9711391}{{arXiv:hep-ph/9711391}}].

\bibitem{Doi:1985dx}
M.~Doi, T.~Kotani, and E.~Takasugi, Prog. Theor. Phys. Suppl. {\bfseries 83}, 1
  (1985).

\bibitem{Cirigliano:2018hja}
V.~Cirigliano, W.~Dekens, J.~De~Vries, M.~L. Graesser, E.~Mereghetti,
  S.~Pastore, and U.~Van~Kolck, Phys. Rev. Lett. {\bfseries 120}, 202001
  (2018),
[\href{https://arxiv.org/abs/1802.10097}{{arXiv:1802.10097~[hep-ph]}}].

\bibitem{Cirigliano:2019vdj}
V.~Cirigliano, W.~Dekens, J.~De~Vries, M.~L. Graesser, E.~Mereghetti,
  S.~Pastore, M.~Piarulli, U.~Van~Kolck, and R.~B. Wiringa, Phys. Rev.
  {\bfseries C100}, 055504 (2019),
[\href{https://arxiv.org/abs/1907.11254}{{arXiv:1907.11254~[nucl-th]}}].

\bibitem{Cirigliano:2022rmf}
V.~Cirigliano {\em et~al.}, J. Phys. G {\bfseries 49}, 120502 (2022),
  [\href{https://arxiv.org/abs/2207.01085}{{arXiv:2207.01085~[nucl-th]}}].

\bibitem{Davoudi:2020gxs}
Z.~Davoudi and S.~V. Kadam, Phys. Rev. Lett. {\bfseries 126}, 152003 (2021),
  [\href{https://arxiv.org/abs/2012.02083}{{arXiv:2012.02083~[hep-lat]}}].

\bibitem{Cirigliano2020May}
V.~Cirigliano, W.~Detmold, A.~Nicholson, and P.~Shanahan, Prog. Part. Nucl.
  Phys. {\bfseries 112}, 103771 (May, 2020).

\bibitem{Davoudi2022May}
Z.~Davoudi and S.~V. Kadam, Phys. Rev. D {\bfseries 105}, 094502 (May, 2022).

\bibitem{Cirigliano:2021qko}
V.~Cirigliano, W.~Dekens, J.~de~Vries, M.~Hoferichter, and E.~Mereghetti, JHEP
  {\bfseries 05}, 289 (2021),
  [\href{https://arxiv.org/abs/2102.03371}{{arXiv:2102.03371~[nucl-th]}}].

\bibitem{Cirigliano:2020dmx}
V.~Cirigliano, W.~Dekens, J.~de~Vries, M.~Hoferichter, and E.~Mereghetti, Phys.
  Rev. Lett. {\bfseries 126}, 172002 (2021),
  [\href{https://arxiv.org/abs/2012.11602}{{arXiv:2012.11602~[nucl-th]}}].

\bibitem{Richardson:2021xiu}
T.~R. Richardson, M.~R. Schindler, S.~Pastore, and R.~P. Springer, Phys. Rev. C
  {\bfseries 103}, 055501 (2021),
  [\href{https://arxiv.org/abs/2102.02184}{{arXiv:2102.02184~[nucl-th]}}].

\bibitem{Kotila:2012zza}
J.~Kotila and F.~Iachello, Phys. Rev. {\bfseries C85}, 034316 (2012),
[\href{https://arxiv.org/abs/1209.5722}{{arXiv:1209.5722~[nucl-th]}}].

\bibitem{Horoi:2017gmj}
M.~Horoi and A.~Neacsu,
[\href{https://arxiv.org/abs/1706.05391}{{arXiv:1706.05391~[hep-ph]}}].

\bibitem{Cirigliano:2017djv}
V.~Cirigliano, W.~Dekens, J.~de~Vries, M.~L. Graesser, and E.~Mereghetti, JHEP
  {\bfseries 12}, 082 (2017),
[\href{https://arxiv.org/abs/1708.09390}{{arXiv:1708.09390~[hep-ph]}}].

\bibitem{Cirigliano:2017tvr}
V.~Cirigliano, W.~Dekens, E.~Mereghetti, and A.~Walker-Loud, Phys. Rev.
  {\bfseries C97}, 065501 (2018),
  [\href{https://arxiv.org/abs/1710.01729}{{arXiv:1710.01729~[hep-ph]}}],
[Erratum: Phys. Rev.C100,no.1,019903(2019)].

\bibitem{Pastore:2017ofx}
S.~Pastore, J.~Carlson, V.~Cirigliano, W.~Dekens, E.~Mereghetti, and R.~B.
  Wiringa, Phys. Rev. {\bfseries C97}, 014606 (2018),
[\href{https://arxiv.org/abs/1710.05026}{{arXiv:1710.05026~[nucl-th]}}].

\bibitem{Wirth:2021pij}
R.~Wirth, J.~M. Yao, and H.~Hergert, Phys. Rev. Lett. {\bfseries 127}, 242502
  (2021),
  [\href{https://arxiv.org/abs/2105.05415}{{arXiv:2105.05415~[nucl-th]}}].

\bibitem{Weiss:2021rig}
R.~Weiss, P.~Soriano, A.~Lovato, J.~Men\'endez, and R.~B. Wiringa, Phys. Rev. C
  {\bfseries 106}, 065501 (2022),
  [\href{https://arxiv.org/abs/2112.08146}{{arXiv:2112.08146~[nucl-th]}}].

\bibitem{Belley:2023btr}
A.~Belley, T.~Miyagi, S.~R. Stroberg, and J.~D. Holt,
  [\href{https://arxiv.org/abs/2307.15156}{{arXiv:2307.15156~[nucl-th]}}].

\bibitem{Dekens:2020ttz}
W.~Dekens, J.~de~Vries, K.~Fuyuto, E.~Mereghetti, and G.~Zhou, JHEP {\bfseries
  06}, 097 (2020),
  [\href{https://arxiv.org/abs/2002.07182}{{arXiv:2002.07182~[hep-ph]}}].

\bibitem{Bolton:2019pcu}
P.~D. Bolton, F.~F. Deppisch, and P.~S. Bhupal~Dev, JHEP {\bfseries 03}, 170
  (2020),
  [\href{https://arxiv.org/abs/1912.03058}{{arXiv:1912.03058~[hep-ph]}}].

\bibitem{Fang:2021jfv}
D.-L. Fang, Y.-F. Li, and Y.-Y. Zhang, Phys. Lett. B {\bfseries 833}, 137346
  (2022),
  [\href{https://arxiv.org/abs/2112.12779}{{arXiv:2112.12779~[hep-ph]}}].

\bibitem{Prezeau:2003xn}
G.~Pr\'ezeau, M.~Ramsey-Musolf, and P.~Vogel, Phys. Rev. {\bfseries D68},
  034016 (2003),
[\href{https://arxiv.org/abs/hep-ph/0303205}{{arXiv:hep-ph/0303205}}].

\bibitem{Graesser:2016bpz}
M.~L. Graesser, JHEP {\bfseries 08}, 099 (2017),
[\href{https://arxiv.org/abs/1606.04549}{{arXiv:1606.04549~[hep-ph]}}].

\bibitem{Buras:2000if}
A.~J. Buras, M.~Misiak, and J.~Urban, Nucl. Phys. {\bfseries B586}, 397 (2000),
[\href{https://arxiv.org/abs/hep-ph/0005183}{{arXiv:hep-ph/0005183}}].

\bibitem{Buras:2001ra}
A.~J. Buras, S.~J\"{a}ger, and J.~Urban, Nucl. Phys. {\bfseries B605}, 600
  (2001),
[\href{https://arxiv.org/abs/hep-ph/0102316}{{arXiv:hep-ph/0102316}}].

\bibitem{Cirigliano:2018yza}
V.~Cirigliano, W.~Dekens, J.~de~Vries, M.~L. Graesser, and E.~Mereghetti, JHEP
  {\bfseries 12}, 097 (2018),
[\href{https://arxiv.org/abs/1806.02780}{{arXiv:1806.02780~[hep-ph]}}].

\bibitem{Nicholson:2018mwc}
A.~Nicholson {\em et~al.}, Phys. Rev. Lett. {\bfseries 121}, 172501 (2018),
  [\href{https://arxiv.org/abs/1805.02634}{{arXiv:1805.02634~[nucl-th]}}].

\bibitem{Detmold:2020jqv}
W.~Detmold and D.~Murphy [NPLQCD Collaboration],
  [\href{https://arxiv.org/abs/2004.07404}{{arXiv:2004.07404~[hep-lat]}}].

\bibitem{Detmold:2022jwu}
W.~Detmold, W.~I. Jay, D.~J. Murphy, P.~R. Oare, and P.~E. Shanahan, Phys. Rev.
  D {\bfseries 107}, 094501 (2023),
  [\href{https://arxiv.org/abs/2208.05322}{{arXiv:2208.05322~[hep-lat]}}].

\bibitem{MPinedo}
E.~Caurier, G.~Martinez-Pinedo, F.~Nowacki, A.~Poves, and A.~P. Zuker, Rev.
  Mod. Phys. {\bfseries 77}, 427 (2005).

\bibitem{Engel:2016xgb}
J.~Engel and J.~Men\'endez, Rept. Prog. Phys. {\bfseries 80}, 046301 (2017),
[\href{https://arxiv.org/abs/1610.06548}{{arXiv:1610.06548~[nucl-th]}}].

\bibitem{Jokiniemi:2022yfr}
L.~Jokiniemi, B.~Romeo, C.~Brase, J.~Kotila, P.~Soriano, A.~Schwenk, and
  J.~Men\'endez, Phys. Lett. B {\bfseries 838}, 137689 (2023),
  [\href{https://arxiv.org/abs/2211.03764}{{arXiv:2211.03764~[nucl-th]}}].

\bibitem{Menendez:2017fdf}
J.~Men\'endez,
J. Phys. {\bfseries G45}, 014003 (2018).

\bibitem{Caurier08}
E.~Caurier, J.~Men\'endez, F.~Nowacki, and A.~Poves, Phys. Rev. Lett.
  {\bfseries 100}, 052503 (2008).

\bibitem{FNowacki}
E.~Caurier and F.~Nowacki, Acta Physica Polonica B {\bfseries 30}, 705 (1999).

\bibitem{Jokiniemi:2021qqv}
L.~Jokiniemi, P.~Soriano, and J.~Men\'endez, Phys. Lett. B {\bfseries 823},
  136720 (2021),
  [\href{https://arxiv.org/abs/2107.13354}{{arXiv:2107.13354~[nucl-th]}}].

\bibitem{Cirigliano:2019jig}
V.~Cirigliano, Z.~Davoudi, T.~Bhattacharya, T.~Izubuchi, P.~E. Shanahan,
  S.~Syritsyn, and M.~L. Wagman [USQCD Collaboration], Eur. Phys. J. {\bfseries
  A55}, 197 (2019),
[\href{https://arxiv.org/abs/1904.09704}{{arXiv:1904.09704~[hep-lat]}}].

\bibitem{Cirigliano:2022oqy}
V.~Cirigliano {\em et~al.},
  [\href{https://arxiv.org/abs/2203.12169}{{arXiv:2203.12169~[hep-ph]}}].

\bibitem{Tuo:2022hft}
X.-Y. Tuo, X.~Feng, and L.-C. Jin, Phys. Rev. D {\bfseries 106}, 074510 (2022),
  [\href{https://arxiv.org/abs/2206.00879}{{arXiv:2206.00879~[hep-lat]}}].

\bibitem{Jokiniemi:2022ayc}
L.~Jokiniemi, B.~Romeo, P.~Soriano, and J.~Men\'endez, Phys. Rev. C {\bfseries
  107}, 044305 (2023),
  [\href{https://arxiv.org/abs/2207.05108}{{arXiv:2207.05108~[nucl-th]}}].

\bibitem{Yao2019Aug}
J.~M. Yao, B.~Bally, J.~Engel, R.~Wirth, T.~R. Rodr\'\i{}guez, and H.~Hergert,
  Phys. Rev. Lett. {\bfseries 124}, 232501 (2020),
  [\href{https://arxiv.org/abs/1908.05424}{{arXiv:1908.05424~[nucl-th]}}].

\bibitem{Belley2020Aug}
A.~Belley, C.~G. Payne, S.~R. Stroberg, T.~Miyagi, and J.~D. Holt, Phys. Rev.
  Lett. {\bfseries 126}, 042502 (2021),
  [\href{https://arxiv.org/abs/2008.06588}{{arXiv:2008.06588~[nucl-th]}}].

\bibitem{Novario2020Aug}
S.~Novario, P.~Gysbers, J.~Engel, G.~Hagen, G.~R. Jansen, T.~D. Morris,
  P.~Navr\'atil, T.~Papenbrock, and S.~Quaglioni, Phys. Rev. Lett. {\bfseries
  126}, 182502 (2021),
  [\href{https://arxiv.org/abs/2008.09696}{{arXiv:2008.09696~[nucl-th]}}].

\bibitem{Belley:2023lec}
A.~Belley {\em et~al.},
  [\href{https://arxiv.org/abs/2308.15634}{{arXiv:2308.15634~[nucl-th]}}].

\bibitem{Muto94}
K.~Muto, Nucl. Phys. A {\bfseries 577}, 415C (1994).

\bibitem{Senkov13}
R.~A. Sen'kov and M.~Horoi, Phys. Rev. C {\bfseries 88}, 064312 (2013).

\bibitem{Senkov16}
R.~A. Sen'kov and M.~Horoi, Phys. Rev. C {\bfseries 93}, 044334 (2016).

\bibitem{GERDA:2020xhi}
M.~Agostini {\em et~al.} [GERDA Collaboration], Phys. Rev. Lett. {\bfseries
  125}, 252502 (2020),
  [\href{https://arxiv.org/abs/2009.06079}{{arXiv:2009.06079~[nucl-ex]}}].

\bibitem{Ichimura:2022kvl}
K.~Ichimura [KamLAND-Zen Collaboration], PoS {\bfseries NOW2022}, 067 (2023).

\bibitem{Abada:2018qok}
A.~Abada, A.~Hern\'andez-Cabezudo, and X.~Marcano, JHEP {\bfseries 01}, 041
  (2019),
  [\href{https://arxiv.org/abs/1807.01331}{{arXiv:1807.01331~[hep-ph]}}].

\bibitem{Asaka:2020wfo}
T.~Asaka, H.~Ishida, and K.~Tanaka, Phys. Rev. D {\bfseries 103}, 015014
  (2021),
  [\href{https://arxiv.org/abs/2012.12564}{{arXiv:2012.12564~[hep-ph]}}].

\bibitem{Asaka:2020lsx}
T.~Asaka, H.~Ishida, and K.~Tanaka, PTEP {\bfseries 2021}, 063B01 (2021),
  [\href{https://arxiv.org/abs/2012.13186}{{arXiv:2012.13186~[hep-ph]}}].

\bibitem{Asaka:2021hkg}
T.~Asaka, H.~Ishida, and K.~Tanaka, JHEP {\bfseries 07}, 062 (2023),
  [\href{https://arxiv.org/abs/2101.12498}{{arXiv:2101.12498~[hep-ph]}}].

\bibitem{Barinov:2021asz}
V.~V. Barinov {\em et~al.}, Phys. Rev. Lett. {\bfseries 128}, 232501 (2022),
  [\href{https://arxiv.org/abs/2109.11482}{{arXiv:2109.11482~[nucl-ex]}}].

\bibitem{DANSS:2018fnn}
I.~Alekseev {\em et~al.} [DANSS Collaboration], Phys. Lett. B {\bfseries 787},
  56 (2018),
  [\href{https://arxiv.org/abs/1804.04046}{{arXiv:1804.04046~[hep-ex]}}].

\bibitem{Gariazzo:2018mwd}
S.~Gariazzo, C.~Giunti, M.~Laveder, and Y.~F. Li, Phys. Lett. B {\bfseries
  782}, 13 (2018),
  [\href{https://arxiv.org/abs/1801.06467}{{arXiv:1801.06467~[hep-ph]}}].

\bibitem{Dentler:2018sju}
M.~Dentler, A.~Hern\'andez-Cabezudo, J.~Kopp, P.~A.~N. Machado, M.~Maltoni,
  I.~Martinez-Soler, and T.~Schwetz, JHEP {\bfseries 08}, 010 (2018),
  [\href{https://arxiv.org/abs/1803.10661}{{arXiv:1803.10661~[hep-ph]}}].

\bibitem{Giunti:2019aiy}
C.~Giunti and T.~Lasserre, Ann. Rev. Nucl. Part. Sci. {\bfseries 69}, 163
  (2019),
  [\href{https://arxiv.org/abs/1901.08330}{{arXiv:1901.08330~[hep-ph]}}].

\bibitem{Mohapatra:1986bd}
R.~N. Mohapatra and J.~W.~F. Valle, Phys. Rev. D {\bfseries 34}, 1642 (1986).

\bibitem{Mohapatra1986Feb}
R.~N. Mohapatra, Phys. Rev. Lett. {\bfseries 56}, 561 (Feb., 1986).

\bibitem{Nandi1986Feb}
S.~Nandi and U.~Sarkar, Phys. Rev. Lett. {\bfseries 56}, 564 (Feb., 1986).

\bibitem{Dev:2012sg}
P.~S.~B. Dev and A.~Pilaftsis, Phys. Rev. D {\bfseries 86}, 113001 (2012),
  [\href{https://arxiv.org/abs/1209.4051}{{arXiv:1209.4051~[hep-ph]}}].

\bibitem{BhupalDev:2012jvh}
P.~S. Bhupal~Dev and A.~Pilaftsis, Phys. Rev. D {\bfseries 87}, 053007 (2013),
  [\href{https://arxiv.org/abs/1212.3808}{{arXiv:1212.3808~[hep-ph]}}].

\bibitem{Hernandez:2022ivz}
P.~Hernandez, J.~Lopez-Pavon, N.~Rius, and S.~Sandner, JHEP {\bfseries 12}, 012
  (2022),
  [\href{https://arxiv.org/abs/2207.01651}{{arXiv:2207.01651~[hep-ph]}}].

\bibitem{PIENU:2017wbj}
A.~Aguilar-Arevalo {\em et~al.} [PIENU Collaboration], Phys. Rev. D {\bfseries
  97}, 072012 (2018),
  [\href{https://arxiv.org/abs/1712.03275}{{arXiv:1712.03275~[hep-ex]}}].

\bibitem{Friedrich:2020nze}
S.~Friedrich {\em et~al.}, Phys. Rev. Lett. {\bfseries 126}, 021803 (2021),
  [\href{https://arxiv.org/abs/2010.09603}{{arXiv:2010.09603~[nucl-ex]}}].

\bibitem{Borexino:2013bot}
G.~Bellini {\em et~al.} [Borexino Collaboration], Phys. Rev. D {\bfseries 88},
  072010 (2013),
  [\href{https://arxiv.org/abs/1311.5347}{{arXiv:1311.5347~[hep-ex]}}].

\bibitem{Barouki:2022bkt}
R.~Barouki, G.~Marocco, and S.~Sarkar, SciPost Phys. {\bfseries 13}, 118
  (2022),
  [\href{https://arxiv.org/abs/2208.00416}{{arXiv:2208.00416~[hep-ph]}}].

\bibitem{DELPHI:1996qcc}
P.~Abreu {\em et~al.} [DELPHI Collaboration], Z. Phys. C {\bfseries 74}, 57
  (1997), [Erratum: Z.Phys.C 75, 580 (1997)].

\bibitem{Davidson:2002qv}
S.~Davidson and A.~Ibarra, Phys. Lett. B {\bfseries 535}, 25 (2002),
  [\href{https://arxiv.org/abs/hep-ph/0202239}{{arXiv:hep-ph/0202239}}].

\bibitem{Casas:2001sr}
J.~A. Casas and A.~Ibarra, Nucl. Phys. B {\bfseries 618}, 171 (2001),
  [\href{https://arxiv.org/abs/hep-ph/0103065}{{arXiv:hep-ph/0103065}}].

\bibitem{Esteban:2020cvm}
I.~Esteban, M.~C. Gonzalez-Garcia, M.~Maltoni, T.~Schwetz, and A.~Zhou, JHEP
  {\bfseries 09}, 178 (2020),
  [\href{https://arxiv.org/abs/2007.14792}{{arXiv:2007.14792~[hep-ph]}}].

\bibitem{NuFIT}
``{NuFIT 5.2 (2022)}.'' \url{www.nu-fit.org}.

\bibitem{deVries:2024rfh}
J.~de~Vries, M.~Drewes, Y.~Georis, J.~Klari\'c, and V.~Plakkot,
  [\href{https://arxiv.org/abs/2407.10560}{{arXiv:2407.10560~[hep-ph]}}].

\bibitem{Drewes:2016lqo}
M.~Drewes and S.~Eijima, Phys. Lett. B {\bfseries 763}, 72 (2016),
  [\href{https://arxiv.org/abs/1606.06221}{{arXiv:1606.06221~[hep-ph]}}].

\bibitem{Shimizu2018}
N.~Shimizu, J.~Men\'{e}ndez, and K.~Yako, Phys. Rev. Lett. {\bfseries 120},
  142502 (2018).

\bibitem{Yao:2022usd}
J.~M. Yao, I.~Ginnett, A.~Belley, T.~Miyagi, R.~Wirth, S.~Bogner, J.~Engel,
  H.~Hergert, J.~D. Holt, and S.~R. Stroberg, Phys. Rev. C {\bfseries 106},
  014315 (2022),
  [\href{https://arxiv.org/abs/2204.12971}{{arXiv:2204.12971~[nucl-th]}}].

\bibitem{Romeo:2021zrn}
B.~Romeo, J.~Men\'endez, and C.~Pe\~na Garay, Phys. Lett. B {\bfseries 827},
  136965 (2022),
  [\href{https://arxiv.org/abs/2102.11101}{{arXiv:2102.11101~[nucl-th]}}].

\bibitem{Manohar:1983md}
A.~Manohar and H.~Georgi,
Nucl. Phys. {\bfseries B234}, 189 (1984).

\end{thebibliography}\endgroup

\end{document}